\documentclass[useAMS,usenatbib]{mn2e}

\usepackage{amssymb,amsmath,graphicx}
\usepackage{textcomp}
\usepackage{graphicx}
\usepackage{longtable}[1]
\usepackage{lscape}
\usepackage{color}
\usepackage{xcolor}
\usepackage{rotating}
\usepackage{cite}
\usepackage{ulem}
\usepackage{natbib}
\usepackage{journal_names}
\usepackage{multicol,float}
\title[YMCs in the interacting LIRG Arp\,299]{Young massive clusters in the interacting LIRG Arp\,299:
evidence for the dependence of star cluster formation and evolution on environment}
\author[Z. Randriamanakoto et al.]{Z. Randriamanakoto$^{1,\,2}$\thanks{E-mail:
zara@saao.ac.za}, P. V\"ais\"anen$^{1,\,3}$, S.\,D. Ryder$^{4,\,5}$ and P. Ranaivomanana$^{6}$\\
$^{1}$South African Astronomical Observatory, P.O. Box 9 Observatory, Cape Town, South Africa\\
$^{2}$Department of Astronomy, University of Cape Town, Private Bag X3, Rondebosch 7701, South Africa\\
$^{3}$Southern African Large Telescope, P.O. Box 9 Observatory, Cape Town, South Africa\\
$^{4}$Australian Astronomical Observatory, 105 Delhi Road, North Ryde, NSW 2113, Australia\\
$^{5}$Department of Physics \& Astronomy, Macquarie University, NSW 2109, Australia\\
$^{6}$Department of Physics, University of Antananarivo, P.O. Box 906, Antananarivo, Madagascar
}

\newcommand{\lsun}{L_{\odot}}
\newcommand{\msun}{M_{\odot}}
\newcommand{\lir}{{L}_{\rm IR}}
\newcommand{\arp}{Arp\,299}
\newcommand{\p}{$+$}
\newcommand{\m}{$-$}

\begin{document}
\normalem
\date{Accepted 2018 October 17. Received 2018 October 17; in original form 2018 June 15}

\pagerange{\pageref{firstpage}--\pageref{lastpage}} \pubyear{2018}

\maketitle

\label{firstpage}

\begin{abstract}

Archival WFC3/UVIS imaging of Arp 299 (NGC\,3690E\,+\,NGC\,3690W) 
is retrieved to investigate the young massive cluster (YMC) population of
this ongoing merger. We extract 2182 cluster candidates, 
including 1323 high confidence photometric sources. Multiband photometry 
is matched with {\tt Yggdrasil} models to estimate the age, mass, and extinction
of each cluster. A Schechter fit of the truncated cluster mass function
results in a characteristic mass ${\rm M_{\star} = 1.6 \times 10^6\,\msun}$. 
Our results confirm that intensely star-forming galaxies
such as Arp\,299 host more massive clusters than quiescent dwarf
and normal spirals. In the case of NGC\,3690E, we find that the 
cluster masses decrease with an increasing galactocentric radius 
likely due to the gas density distribution. On the other hand,
the fitted age distributions of a mass-limited sample
suggest that YMCs of the western component undergo stronger disruption
than those hosted by the eastern galaxy. This is in agreement with the properties
of the underlying cluster luminosity functions: a clear
truncation at high luminosities with slopes generally shallower by $\sim\,0.3$\,dex
than the ones of the NGC\,3690E.  
Finally, the derived cluster formation efficiency, $\Gamma \sim 19$\,percent, 
indicates that \arp~has
$\sim 3-5$ times more star formation happening in bound clusters compared to 
the cases of gas-poor spirals like NGC\,2997 and NGC\,4395. The merger generally follows the $\Gamma - $ 
star formation rate density relation from the literature. The YMC photometric
study of Arp\,299 thus reveals that both formation and disruption mechanisms
of the star cluster population are most likely environment-dependent.
\end{abstract}


\section{Introduction}

\arp~is a nearby luminous infrared galaxy (LIRG) system (log\,${\rm \lir/\lsun = 11.88}$,
\citealp{2003AJ....126.1607S}) in an early merging stage, at a luminosity
distance of $\approx$\,45\,Mpc where $1\,{\rm kpc} \sim 4.6$\,arcsec. 
The interacting system consists of a pair of two irregular
galaxies: NGC\,3690E\footnote{NGC\,3690E is commonly known as source A in the literature.}
in the eastern part, and NGC\,3690W\footnote{NGC\,3690W is commonly known as sources
B\p C.} the western component. 
While the former is possibly the remnant of a retrograde spiral galaxy, the latter 
has a more disturbed morphology with multiple nuclei. With a projected nuclear separation 
of around 4.7\,kpc and the presence of a disk overlap region (known as ${\rm C+C'}$), 
the interaction between the two galaxies is believed to have 
started at least 750\,Myr ago \citep{1999AJ....118..162H}. 

\citet{2016ApJ...825..128L} classified \arp~as a major merger with visible tidal tails where they
found that the system contains a molecular gas fraction of 31\,percent. In addition, extensive 
multi-wavelength observations by \citet{2000ApJ...532..845A,2002AJ....124..166A,2009ApJ...697..660A,2013ApJ...779L..14A}
concluded that the interacting LIRG hosts extreme star formation (SF) activity with
a recent episode of massive SF mainly enshrouded in the dusty nuclear starburst regions.
Relatively high numbers of supernovae (SNe) and optically hidden core
collapse SNe (CCSNe) have also been recorded in 
the circumnuclear regions of \arp~\citep[e.g.][]{2004ApJ...611..186N,2011MNRAS.415.2688R,2012A&A...539A.134B,
2012ApJ...756..111M,2014MNRAS.440.1052K}. It is important to study SNe/CCSNe as they could be spatially 
associated with the young massive star 
clusters (YMCs), also known as super star clusters (SSCs) and proven to be good tracers of recent
massive star formation \citep{2014MNRAS.440.1052K}. 

(Near)-infrared observations by \citet{1999A&A...351..834L}, \citet{2000ApJ...532..845A,2002AJ....124..166A,2009ApJ...697..660A}
and \citet{2013MNRAS.431..554R} revealed that a large population of YMCs reside in the dusty nuclear starbursts 
and star-forming regions of \arp. Alternatively,  \citet{2011PhDT.........8V} and \citet{2017ApJ...843...91L}
made use of archival HST/ACS data as part of the GOALS survey to probe the photometric properties of the YMC candidates. 
They found that the interacting system has very young and extinguished star clusters with
a mass range between $\approx 10^4 -10^7\,{\rm \msun}$. In particular, \citet{2017ApJ...843...91L} suggested that
the extreme environments of merging LIRGs favor both the formation of the very massive star clusters 
(with masses ${\rm M}\,>\,10^6\,{\rm \msun}$) and the rapid dissolution of its cluster population. 
They also reported evidence of mass-independent disruption mechanisms by interpreting the mass distribution of the YMCs. 
Note, however, that only 53 cluster candidates hosted by \arp~were included in their analyses which use a combined catalogue drawn
from a sample of 22 LIRGs. It is therefore worth revisiting the effects of the galactic environments and any other 
factors on the star cluster formation,
evolution and disruption mechanisms happening in \arp.

It is important to investigate the influence(s) of both internal and external effects on the cluster properties since 
its universality is being actively questioned and debated \citep[see~e.g.][]{2010ARA&A..48..431P, 
2010AJ....140...75W,2012MNRAS.419.2606B,2012ApJ...752...96F,2014ApJ...786..117F,2015ApJ...810....1C,2016MNRAS.460.2087H,
2017ApJ...839...78J,2018ASSL..424...91A}. {\it Are the characteristics of the star cluster population
tightly related to the environmental properties of the host galaxy? Are internal processes such as stellar evolution and two-body 
relaxation strong enough to completely dissolve very 
massive and dense YMCs? Or do they need help from environment and/or mass-dependent external effects (e.g. strong
tidal fields, dense giant molecular clouds) at some stages?} The cluster mass function (CMF) is a
powerful tool commonly used to help answer such questions. 

Previously known to be well represented by a power-law function, \citet{2009A&A...494..539L} and
\citet{2009MNRAS.394.2113G} introduced another form of the CMF that is suggested to be more consistent 
with the data at both low and high-mass ends: a Schechter distribution. It is of the form
${\rm {dN} \sim M^{-\beta_{\star}}\,{\rm exp}(-M/M_{\star})\,dM}$, where $\beta_{\star} \sim 2$ is the slope and ${\rm M_{\star}}$ 
the truncated characteristic mass varying with the cluster 
environments. A review by \citet{2010ARA&A..48..431P} compared the Schechter MFs of different galaxies
including the LMC, a cluster-poor spiral galaxy, a cluster-rich one, and the disturbed Antennae galaxies
to explore the possibility of a mass and/or environment-dependent cluster initial mass function (CIMF).
The results indicated that the location of the truncation is not the same for the different galaxies:
quiescent normal galaxies have ${\rm M_{\star} \approx 2 \times 10^{5}\,M_{\odot}}$ while strongly interacting galaxies
are associated with higher values of ${\rm M_{\star} \approx 2 \times 10^{6}\,M_{\odot}}$.
Observational works by \citet{2012MNRAS.419.2606B}, \citet{2013AJ....145..137K}, and \citet{2015MNRAS.452..246A} also noticed 
that the inner and outer field of the host galaxy have two distinct Schechter characteristic masses: 
e.g.\,${\rm \sim 1.6 \times 10^{5}\,M_{\odot}}$ and ${\rm \sim 0.5 \times 10^{5}\,M_{\odot}}$, respectively. 
Such a difference arises because of the change in the maximum mass of the giant molecular clouds (GMCs). Furthermore,
the high gas pressures of the extreme environments\textcolor{black}{, seen in galaxy mergers,} and the galactic nuclear regions enhance
the formation of the most massive cluster candidates, and hence, result in a relatively high truncated ${\rm M_{\star}}$ of their corresponding 
Schechter CMF. \textcolor{black}{Star cluster analyses of the Legacy ExtraGalactic UV Survey (LEGUS) sample
also revealed strong evidence of a Schechter-type CMF with a slope $\beta_{\star} \sim 2$ and a truncation mass
 ${\rm M_{\star} \approx 10^{5}\,M_{\odot}}$ \citep{2017ApJ...841..131A, 2018MNRAS.473..996M}}.
In contrast, \citet{2014ApJ...786..117F} maintain that pure power-law mass distributions fit the data better 
than a Schechter one, though they also agree that the CMF may vary with respect to the host environment. 
\textcolor{black}{Any turnover or high truncation in the CMF is believed to be a mere reflection of a constant disruption triggered by 
internal mechanisms (e.g.\,infant mortality, stellar evolution) during the cluster evolutionary processes
\citep[e.g.][]{2007AJ....133.1067W, 2010AJ....140...75W, 2014AJ....147...78W}}. 
\citet{2014ApJ...787...17C,2017ApJ...849..128C}, \citet{2016ApJ...826...32M}, and 
\textcolor{black}{\citet{2018arXiv180611192M}} also
emphasize the quasi-universality of the CIMF and report that systematics alter the real feature of the mass function.

\textcolor{black}{To further assess the formation conditions of YMCs,} \citet{2008MNRAS.390..759B} refers to the fraction of star formation happening in
bound stellar clusters as cluster formation efficiency (CFE or $\Gamma$). The author derives 
the value of such a fraction by using the following expression:
\begin{equation}
\Gamma (\%) = \frac{\rm CFR}{\rm SFR} \times 100
\label{eqn:CFE}
\end{equation}
where CFR is the cluster formation rate and SFR the host galaxy star formation rate.
The former parameter is the total mass formed in clusters at a certain age
interval $\Delta t$ divided by the duration of time of such an interval. 
In the solar neighborhood, the value of $\Gamma$ has been found to remain relatively constant:
$\Gamma \simeq $\,5\,\% by \citet{2003ARA&A..41...57L}, $\Gamma \simeq $\,7\,\% by 
\citet{2008ASPC..388..367L}, and $\Gamma \simeq $\,3\,\% for the SMC by
\citet{2008A&A...482..165G}. However, subsequent YMC extragalactic studies
ruled out the concept of a constant parameter and suggested instead an environmentally-dependent
CFE \citep{2010MNRAS.405..857G,2011MNRAS.417.1904A,2015MNRAS.452..246A}. Such arguments were supported
with the predictions by e.g.\,\citet{2003MNRAS.338..665B}, \citet{2012MNRAS.420.1503P}
and \citet{2012MNRAS.426.3008K}. In fact, the fraction has been found to go beyond
40\,\% in high-SFR luminous blue compact galaxies \citep{2011MNRAS.417.1904A}.
A new version of the CFE - SFR surface density (${\rm \Sigma _{SFR}}$) relation was recently published by 
\textcolor{black}{\citet{2016ApJ...827...33J}} and \citet{2018MNRAS.473..996M}. The correlation between the
two parameters is thought to be a mere reflection of the CFE - gas density relation which means that high SF efficiency environments
produce more GMCs. Hence, they are expected to have more stars forming in bound stellar clusters.  
The use of a diverse type of galaxies (normal spirals, starbursts, etc) to draw the ${\rm \Gamma - \Sigma _{SFR}}$ 
relation, however, has raised some doubts. An inhomogeneous sample combined with low spatial resolution could affect
the upper trend of the relation by showing a spurious increase of $\Gamma$ at high SFR densities \citep{2017ApJ...849..128C}.

It has also been reported that the host environment and the cluster mass regulate
the rate at which the cluster population get dissolved \citep[e.g.][]{2009MNRAS.394.2113G,2011MNRAS.417L...6B,2012MNRAS.419.2606B,2017ApJ...843...91L}.
The intrinsic shape and slope of
both the cluster luminosity function (CLF) and the CMF as well as the star cluster age distribution are usually analysed
to define the intensity of the disruption.
The latter distribution of the form ${\rm dN/d\tau \sim \tau^{- \delta}}$ is a mass-limited cluster sample 
plotted per time interval, where  $\delta$ is the power-law slope. 
Such a diagram is also \textcolor{black}{useful for producing} the cluster formation history of the 
galaxy. \textcolor{black}{While \citet{2005ApJ...631L.133F} and \citet{2007AJ....133.1067W} interpret the age distribution
$\tau^{-1}$ for the clusters in the Antennae galaxies to provide evidence for a mass independent disruption scenario 
(e.g.\,a relaxation-driven cluster dissolution model)}, 
\citet{2005A&A...441..117L}, \citet{2013MNRAS.430..676B} and \citet{2014MNRAS.440L.116S} 
have suggested that a steeper slope of ${\rm dN/d\tau}$ is an imprint of
a stronger cluster dissolution rate from strong tidal forces and GMCs of high surface density. In such a scenario, the massive cluster candidates 
(${\rm M}\,>\,10^5\,{\rm \msun}$) and those hosted by a relatively weak tidal environment are likely
to remain gravitationally bound after a gradual cluster mass-loss. These YMCs are expected to have a long-term survival chance and could
potentially become the present day globular clusters (see \citealt{2016EAS....80....5B} for a recent review).

The goal of this paper is to investigate whether the host galaxy environment along
with other physical effects play a major role in defining the physical properties and 
the evolution of its YMC population. The study focuses on the star clusters of Arp\,299 
since the target has a large sample of YMCs and has been recently
imaged by the HST/WFC3 camera. The cluster age, mass, and extinction will be
derived by fitting high-resolution photometric data covering from the UV to the near-infrared (NIR)
bands with carefully chosen single stellar population (SSP) models.
The paper is organized as follows. In Section\,\ref{obs-sec}, we describe the data and observations.
Object detection, aperture photometry as well as SSC candidate selection are reported in Section\,\ref{phot-sec}. 
Section\,\ref{age-sec} presents the cluster age and mass modelling. We analyse and discuss the star cluster photometric properties 
and the cluster formation histories in Sections\,\ref{sec:ccd} \& \ref{analysis-sec}. 
In Section\,\ref{end-sec}, we summarize our work and then draw our conclusions.

\section{Data and observations}\label{obs-sec}
This work mainly uses multiband observations of \arp~imaged with the HST WFC3/UVIS camera.
Such a wealth of data will enable us to study the interacting galaxies to an 
unprecedented magnitude limit, hopefully detecting a significant number of new star
clusters that also includes the low-mass candidates. To investigate whether there 
are NIR excess star clusters (see Section\,\ref{subsec:K-selected}), we will combine the optical
datasets with some observations taken with the Gemini/NIRI adaptive optics (AO) systems.

\subsection{HST WFC3/UVIS data}\label{wfc3-data}
\begin{figure*}
\centering
\resizebox{1.0\hsize}{!}{\rotatebox{0}{\includegraphics{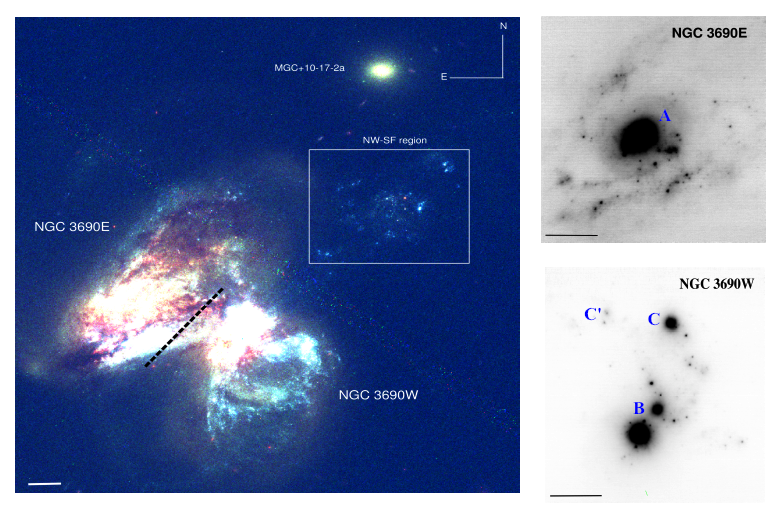}}}
\caption{\small {\em Left:} WFC3/UVIS three-color image of Arp\,299~in a 1.90
by 1.85\,arcmin field: F336W ($U$-band, blue), 
F438W ($B$-band, green), and F814W ($I$-band, red). The dashed line, \textcolor{black}{going through the overlap region,} 
marks the boundary used 
in this work to separate YMC candidates hosted by NGC\,3690E (East) from the ones located 
in the westen component NGC\,3690W. The galaxy companion 
MCG+10-17-2a and a star-forming region (inside the box) $\sim$\,10\,kpc northwest to the system are also
imaged in the field. North-East orientation and a length scale of 1\,kpc
(the horizontal line) are indicated. {\em Right:} $K$-band
images of the individual components
taken with Gemini/NIRI AO systems in a 22\,$\times$\,22\,arcsec field: NGC\,3690E (top)
and NGC\,3690W (bottom). \textcolor{black}{The letters A, B, C, and ${\rm C^{'}}$ denote the major radio sites in Arp\,299.}
The horizontal lines represent a 1\,kpc scale.}
\label{arp299-rgb}
\end{figure*}

 We retrieved already-processed WFC3/UVIS images of \arp~from the publicly
available Hubble Legacy Archive (HLA\footnote{http://hla.stsci.edu}) in the
following broad-band filters: F336W ($U$, 790\,s), F438W ($B$, 740\,s),
and F814W ($I$, 740\,s). The plate scale is 0.039 arcsec\,pixel$^{-1}$ and the 
point-spread function (PSF) values for the point sources vary between
0.07$-$0.09\,arcsec depending
on the filter. Table\,\ref{details-obs} lists a summary of the observation
log  on June 24, 2010 (PI: Bond, PID: 12295). The left panel of  Fig.\,\ref{arp299-rgb} 
shows the three-color image of the pair of interacting galaxies from the $UBI$ exposures
in a $1.85 \times 1.90$ arcmin field of view (FoV). The northwest spheroidal MCG+10-17-2a 
and a star-forming region located 10\,kpc away from the disk overlap were also covered
during the HST imaging. Compared to previous HST instruments such as WFPC2 and WFC/ACS,
the WFC3/UVIS camera provides sharper images with a more stable PSF. Furthermore, the use
of deep $U$-band data will help to constrain the cluster age-extinction degeneracy and
to develop a more accurate approach on how to interpret results from $BIK$-band YMC 
analyses.

\subsection{Gemini/NIRI data}
Arp\,299 is amongst the targets observed in an ongoing survey dubbed 
SUNBIRD (SUperNovae and starBursts in the InfraReD or 
Supernovae UNmasked By InfraRed Detection) to search for dust-obscured CCSNe in 
starbursts and LIRGs using NIR AO systems \citep{2014mysc.conf..185V,2018MNRAS.473.5641K}.
The $K_S$-band (hereafter, referred to as $K$-band) AO 
imaging of the interacting LIRG were obtained
with the ALTAIR/NIRI instrument mounted on the Gemini-North telescope. 
Final science images of the two individual components, as shown in the right
panel of Fig.\,\ref{arp299-rgb},  are the results of a
multi-epoch survey during the period $2007-2009$ (PI: Ryder). 
The galaxy components of \arp~were
observed separately due to the small FoV of 22 arcsec of the Gemini/NIRI
camera with a plate scale of 0.022 arcsec pixel$^{-1}$. The total integration times per
galaxy  are 1260\,s (NGC\,3690E) and 2192\,s (NGC\,3690W), respectively. The AO correction using
laser guide stars resulted in a PSF of 0.1\,arcsec for the point sources. 
The $K$-band SSC catalogues were compiled using a criterion based on 
the concentration index vs.\,FWHM plot of the detected sources, with 81 candidates
selected in each component. Further details on the observations, data reduction,
as well as the selection criteria are reported in \citet{2013MNRAS.431..554R}.

\begin{table}
\begin{scriptsize}
\begin{center}
\caption{UV and optical archival data of \arp~obtained with the HST WFC3/UVIS 
camera (PI: Bond, PID: 12295).}
\label{details-obs}
\begin{tabular}{lcccc}
\hline 
   \noalign{\smallskip}
   
  Filter &  Pix.size & Mean wav.  & Exp.time  & PSF/FWHM \\
                                & (arcsec)   & (\AA)                        &  (sec) & (arcsec)  \\
     (1) &      (2)  &     (3)  &     (4)         &        (5)     \\      
   \noalign{\smallskip}
\hline \hline
   \noalign{\smallskip}
F336W ($U$)  & 0.039 & 3345.8 & 790   & 0.09 \\
F438W ($B$)  &  0.039 & 4300.5  & 740   & 0.08 \\
F814W ($I$)  &  0.039 & 8174.5  &  740  &  0.07 \\
\noalign{\smallskip}
   \hline
\noalign{\smallskip}
\multicolumn{5}{@{} p{8.5cm} @{}}{\footnotesize{\textbf{Notes. }Column 1: broad-band filter used; 
Column 2: the camera plate scale; Column 3: mean wavelength of the filter; 
Column 4: exposure time; Column 5: FWHM of bright isolated stars in the original image.}}

\end{tabular}
\end{center}
\end{scriptsize}
\end{table}

\section{Photometry and source selection}\label{phot-sec}

\subsection{Source extraction and aperture photometry}\label{detect-sec}
Automated source detection was performed on the unsharp-masked version
of the combined $BI$-image using {\tt SExtractor} \citep{1996A&AS..117..393B}.
Such a frame is chosen to gain a higher signal-to-noise ratio (SNR), 
and hence to facilitate the extraction of faint and extended objects. 
Two sets of the software critical parameters were adopted to optimally detect
SSC candidates in the outer and inner regions of the galaxies. Besides the minimum 
object area of 8 pixels, a minimum detection limit of 4$\sigma$
combined with a background mesh width equal to 85 were used to extract
the sources in the outer field. In contrast, a higher threshold of 9$\sigma$
with a relatively low background mesh of 20
were necessary to reduce the number of spurious objects from
the inner field.

\begin{figure}
\centering
\resizebox{\hsize}{!}{\rotatebox{0}{\includegraphics{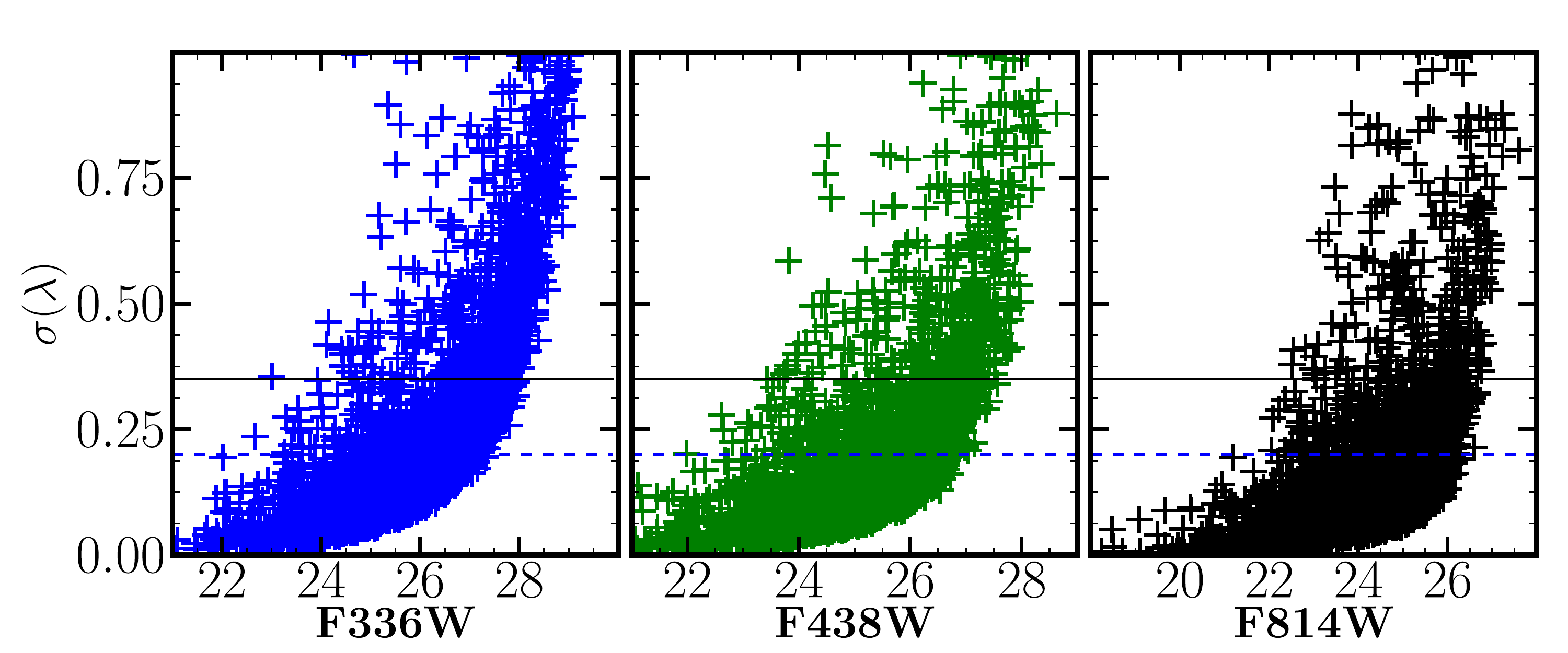}}}
\caption{\small $UBI$-photometric errors of the detected sources plotted against 
the instrumental magnitudes. Both uncertainty range and scatter of the data
points vary from filter to filter. The dashed and solid lines represent cutoff errors
at $\sigma_m = 0.20$\,mag and $\sigma_m = 0.35$\,mag, respectively.}
\label{merr-arp}
\end{figure}

We then used {\tt IRAF/DAOPHOT} package to perform aperture photometry on
the resulting coordinates from {\tt SExtractor}. Matched-aperture photometry
was applied in all three $UBI$-images with a fixed aperture
radius of 2.5\,pixels ($\sim$\,0.1\,arcsec)
and sky background annuli from 4 to 6\,pixels (0.08\,arcsec wide).
Figure\,\ref{merr-arp} shows the multiband instrumental magnitudes plotted against their
respective photometric uncertainties. Such plots help in deciding the value of the
photometric error cutoff $\sigma_m$ (see Section\,\ref{select-arp}). Growth curves
of bright and isolated sources were drawn up to 1\,arcsec to calibrate 
$B$- and $I$-band aperture photometry. The value of the aperture 
correction was obtained from the HST/WFC3 data manuals for the $U$-band data due
to the noisy background and the lack of isolated sources.
The VEGAMAG photometric zeropoints were also retrieved from the same manuals.
These values along with the foreground Galactic extinction of
each filter are listed in Table\,\ref{ubihk-arp}. The Vega-based absolute magnitudes 
are estimated to have uncertainties of $\approx 0.1 - 0.3$\,mag;  the 
relatively high errors are associated with the $U$-band data. 

\begin{table}
\begin{scriptsize}
\caption[The sample]{Photometric measurements of Arp\,299 and the number of optically-selected star cluster
candidates of each filter.}
\label{ubihk-arp}
\begin{center}
\begin{tabular}{lccccc}
\hline 
   \noalign{\smallskip}
   
Filter 	& $A_{\lambda}$ & $a_c, m_0$  & $\bar \sigma_{M}$ & Nb.SSCs & Mag.limit\\
	&   (mag)	&  (mag) & (mag)   &  ($\sigma \leq \sigma_m$) & (Vega-mag)	\\
(1)     &   (2)		& (3)	 & (4)	& (5)	 & (6)	  	\\
   \noalign{\smallskip}
\hline \hline
   \noalign{\smallskip}
$U$	&	   0.07	& \m0.41, 23.46 &0.17 &1491  & 25.9  \\
$B$	&	  0.06	& \m0.40, 24.98 &0.13 & 1879  & 27.0 \\
$I$	&	  0.03	& \m0.52, 24.67 & 0.12 & 1913 & 25.8 \\
\noalign{\smallskip}
\hline
\noalign{\smallskip}
\multicolumn{6}{@{} p{8.5cm} @{}}{\footnotesize{\textbf{Notes. }Column 1: broad-band filter used;
 Column 2: foreground Galactic reddening from NED Database; Column 3:  aperture correction
 and zeropoint magnitude; Column 4: mean uncertainty of the cluster absolute magnitude; Column 5: number of 
 SSCs with $\sigma \leq 0.20$\,mag in the  WFC3/UVIS field; Column 6: the corresponding 
 Vega-based magnitude limit.}}
 
\end{tabular}
\end{center}
\end{scriptsize}
\end{table}

\subsection{The optically-selected star cluster catalogue}\label{select-arp}

The selection of star cluster candidates remains similar
to the philosophy adopted by \citet{2013ApJ...775L..38R, 2013MNRAS.431..554R} to draw
their catalogues. The only difference lies on a cross-identification
of the {\tt SExtractor} coordinates in all three filters to extract the common 
sources instead of a criterion based on the concentration index
vs.\,FWHM plot. As in \citet{2013ApJ...775L..38R, 2013MNRAS.431..554R}, preliminary
steps prior to the source cross-matching include rigorous visual inspection to
clean the catalogue from false detections, cosmic rays and hot pixels, foreground sources,
and any contamination from the galaxy nuclei. 

Only objects with ${\rm SNR} \geq$\,5 in all three filters were initially considered
in order to output a catalogue of 1323 high confidence photometric
SSC candidates. This number comes from cross-matching 1491, 1879, and 1913
objects that satisfy the conditions in the individual fields of $U$-, $B$-, and $I$-band, respectively. 
The corresponding Vega-magnitude limits are listed in the last column of Table\,\ref{ubihk-arp}.
\textcolor{black}{Both the visual inspection and data point distribution} in Fig.\,\ref{merr-arp}, however,
indicate that a significant number of obvious SSCs would be rejected by applying 
such criteria in the photometric errors. Therefore, to draw our final $UBI$-band catalogue of 
2182 star cluster candidates, we set a cutoff of $\sigma_m = 0.35$\,mag as a trade-off and 
also included 116 extra candidates clearly
visible in all three filters but with a relatively high uncertainty in their $U$-band
magnitudes ($\sigma_{U} \leq$\,0.50\,mag). To strengthen our analyses, the resulting age and
extinction of objects with magnitude uncertainties of $\sigma \leq$\,0.20\,mag
will be checked and evaluated separately. 
\textcolor{black}{Nevertheless, we investigated the effects of photometric biases using the final
catalogue which turned out
to be insignificant.} The $UBI$-band apparent magnitudes with their 
uncertainties are available online in a machine-readable format as sampled in 
Table\,\ref{source-tab}. Note that 74 of the selected cluster
candidates are hosted by the northwest star-forming region. This subset of the catalogue
will be analysed separately in Section\,\ref{sec:SSCs-NWreg}. 

\begin{figure}
\centering
 \begin{tabular}{c}
\resizebox{1.\hsize}{!}{\rotatebox{0}{\includegraphics[trim= 0cm 0cm 0cm 0cm, clip]{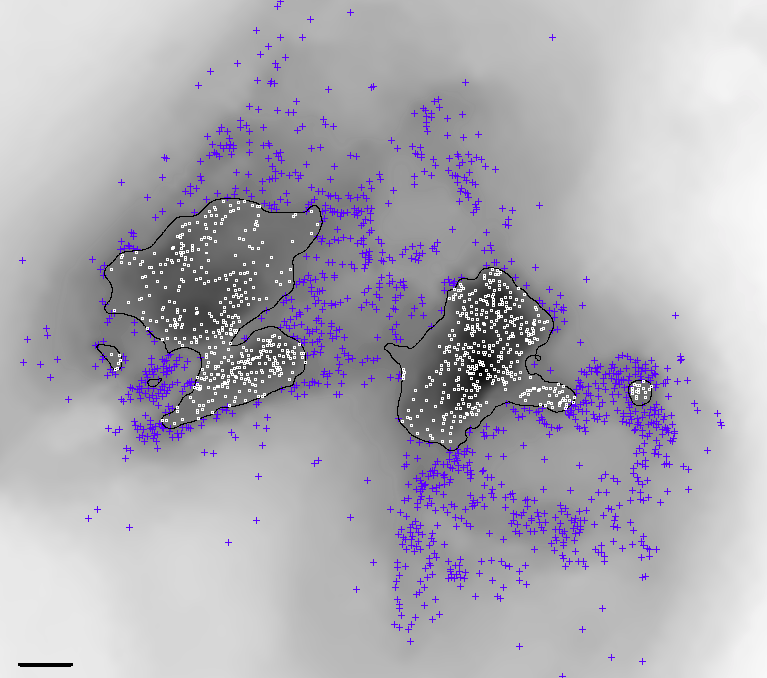}}}\\
\resizebox{1.\hsize}{!}{\rotatebox{0}{\includegraphics[trim= 0cm 0cm 0cm 0cm, clip]{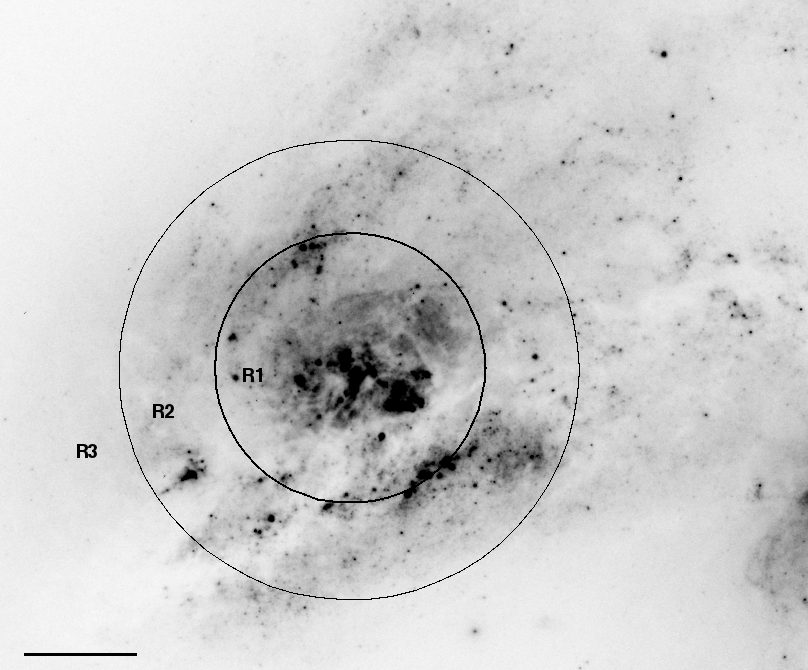}}}
\end{tabular}
\caption{\small \textcolor{black}{{\em Upper:} The background map of Arp\,299 in a logarithmic scale. 
The black solid line marks the edge of the contour level used to split the star cluster candidates depending on their surrounding
background levels: white squares are sources hosted by the nuclear regions and purple crosses
represent YMCs in the outer field. {\em Lower:} NGC\,3690E in the $I$-band. The concentric circles divide the 
eastern component into three radial bins of an equal number of 330 clusters. The horizontal lines indicate 
a length scale of 1\,kpc.}}
\label{fig:split-sscs}
\end{figure}

To study the influence of any environment- and/or mass-dependent effects on the
star cluster formation and disruption mechanisms, SSC candidates were divided in two
different ways:
\begin{enumerate}
 \item as a function of the host galaxy component, i.e. NGC\,3690E\,(A) vs. NGC\,3690W\,(B+C).
Setting a boundary (the dashed line in Fig.\,\ref{arp299-rgb}) between the two galaxy 
components is not trivial due to the ongoing disk overlaps in the area ${\rm C+C'}$.
The surface brightness distribution and \textcolor{black}{the visual overlap region} were, therefore,
used to split the catalogue for convenience.
A number of 988 and 1120 star cluster candidates were recovered
for NGC\,3690E and NGC\,3690W, respectively. With a relatively large SSC population, NGC\,3690W
also hosts candidates with brighter
magnitudes than the ones in the eastern component. For instance, the star clusters of 
NGC\,3690W have $B$-band
absolute magnitudes between $-15.2$\,mag and $-6.5$\,mag. This range becomes
$-13.1 \lesssim M_B \lesssim -6.5$\,mag in the field of NGC\,3690E.

\item depending on their surrounding background levels, i.e. outer vs. inner regions.
A logarithmic scale of the background map was used to define these two regions
\textcolor{black}{as shown in the upper panel of Fig.\,\ref{fig:split-sscs}}.
Obscured nuclear regions of the interacting LIRG host 836\, star clusters
while we counted a population of 1272 SSC candidates in the outer field of the system. 

\end{enumerate}
The SSC population of NGC\,3690E was also split into three categories by defining radial bins 
containing approximately an equal number of 330\,clusters 
\textcolor{black}{(see lower panel of Fig.\,\ref{fig:split-sscs})}.
The bins range from 0 to 1.38\,kpc (${\rm R_1}$), 1.38$-$2.25\,kpc (${\rm R_2}$), 2.25\,kpc and beyond (${\rm R_3}$) 
the galaxy detectable optical emission in the frame. This eastern component is chosen 
as a laboratory to further investigate any environmental effects on the cluster properties
due to its morphological feature resembling that of a spiral galaxy.

\subsection{Completeness of the catalogue}\label{comp-arp}
\begin{figure}
\centering
\resizebox{1.1\hsize}{!}{\rotatebox{0}{\includegraphics{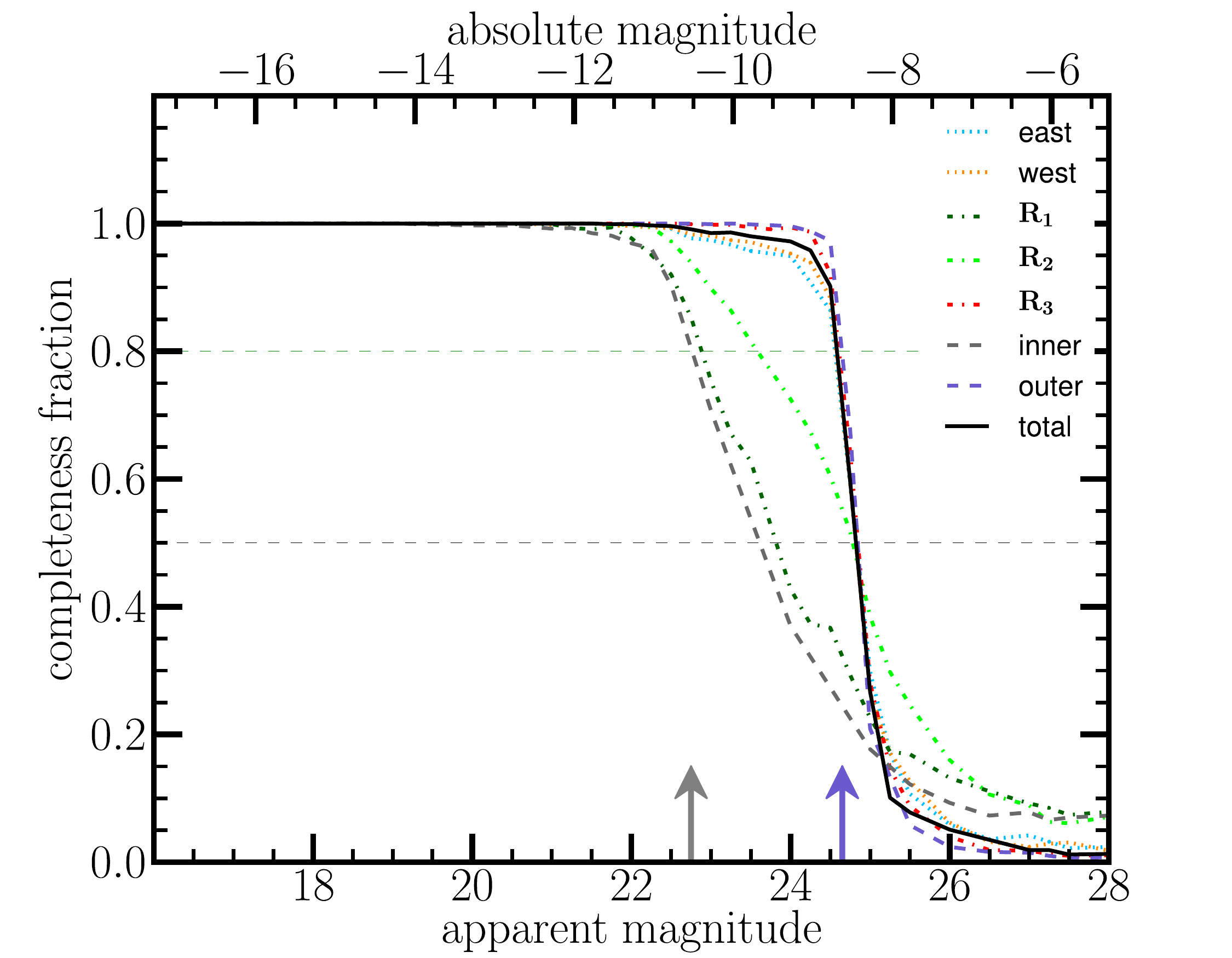}}}\\
\caption{\small The completeness rates per magnitude 
\textcolor{black}{of Arp\,299 (solid line) and its different regions which were defined as a function of
 the galaxy components (dotted), the radial bins of NGC\,3690E (dash-dotted), and the background levels (dashed).}
The horizontal lines mark the 50\,percent and 80\,percent completeness limits.
\textcolor{black}{The grey and purple arrows respectively indicate the magnitudes at 
which the inner and the outer field are 80\,percent complete.}}
\label{fig:comp-frac}
\end{figure}

We performed Monte-Carlo completeness simulations to estimate the fraction of the sources 
being missed due to the detection limits and our source extraction methods. The same
procedures as in \citet{2013MNRAS.431..554R} were adopted and applied on the combined
$BI$-frame between 16 to 28 magnitude range and considering
\textcolor{black}{the whole interacting system and different regions described in Section\,\ref{select-arp}}.
Figure\,\ref{fig:comp-frac} plots the recovered completeness fractions as a function of magnitude
in \textcolor{black}{Arp\,299 (black solid line), the inner (dashed grey) and outer (dashed purple) field. 
The other lines represent the recovery rates in the two galaxy components (dotted) and the three radial bins of NGC\,3690E (dash-dotted).} Based on the simulations, 
our catalogue is believed to be 80\,percent complete down to $m_B \sim 22.8$ and $\sim 24.7$
magnitudes, in the inner and outer regions, respectively. Note that core nuclei were excluded
from the simulation because of their relatively high background levels.

\begin{table*}

\caption{\small Catalogue of the SSC candidates in the field of Arp\,299. 
A full table of 2182 star clusters is available online.}

\centering
\scalebox{0.85}{
  \begin{tabular}{ccccccccc}
  \hline
  \hline
Source name           &  RA  & DEC & F336W  & F438W  & F814W & A$_V$ & log($\tau)$  & log(${\rm M}/{\rm M_{\odot}}$)  \\
                      &   [deg] & [deg]      & [mag]   & [mag] & [mag] & [mag] &  & \\
(1)   &  (2)  & (3)   &   (4)    &   (5)  & (6) & (7) & (8) & (9) \\ \hline

1 & 172.105729167 & 58.5713222222 & 23.23 $\pm$  0.07 & 24.56 $\pm$ 0.06 & 24.19 $\pm$ 0.07 & 0.35 & 6.00 & 3.74 \\
2 & 172.106341667 & 58.5724000000 & 24.61 $\pm$  0.15 & 25.34 $\pm$ 0.10 & 24.94 $\pm$ 0.14 & 0.01 & 7.15 & 3.61 \\
3 & 172.106708333 & 58.5724972222 & 23.67 $\pm$  0.08 & 24.83 $\pm$ 0.08 & 25.27 $\pm$ 0.16 & 0.00 & 6.48 & 3.09 \\
4 & 172.106895833 & 58.5725333333 & 25.05 $\pm$  0.24 & 26.01 $\pm$ 0.15 & 25.05 $\pm$ 0.11 & 0.25 & 6.95& 3.48  \\
5 & 172.107008333 & 58.5723416667 & 21.55 $\pm$  0.03 & 22.94 $\pm$ 0.03 & 22.45 $\pm$ 0.03 & 0.25 & 6.00 & 4.36 \\
6 & 172.107150000 & 58.5728333333 & 23.86 $\pm$  0.11 & 24.94 $\pm$ 0.09 & 23.95 $\pm$ 0.07 & 0.25 & 6.95 & 3.96 \\
7 & 172.107154167 & 58.5722750000 & 23.29 $\pm$  0.08 & 24.58 $\pm$ 0.07 & 23.78 $\pm$ 0.05 & 0.07 & 7.18 & 4.20 \\
8 & 172.107204167 & 58.5726777778 & 21.36 $\pm$  0.03 & 22.57 $\pm$ 0.03 & 21.82 $\pm$ 0.03 & 0.00 & 7.18 & 4.92 \\
9 & 172.109154167 & 58.5704083333 & 24.97 $\pm$  0.18 & 25.23 $\pm$ 0.10 & 24.38 $\pm$ 0.07 & 0.63 & 8.30 & 5.29 \\
10& 172.109408333 & 58.5703333333 & 23.59 $\pm$  0.09 & 24.87 $\pm$ 0.09 & 23.88 $\pm$ 0.06 & 0.13 & 6.95 & 3.93\\

\hline
\multicolumn{9}{@{} p{17.2cm} @{}}{\footnotesize{\textbf{Notes. }Column 1: source identification; Columns 2 \& 3: right ascencion \& declination in J2000 coordinates;
Columns 4 $-$ 6: Vega-based apparent magnitudes with their uncertainties in $U$, $B$ and $I$-band, respectively; Column 7: the resulting extinction value; 
Columns 8 \& 9: the cluster age and mass in logarithmic base.}}
\end{tabular}
}
\label{source-tab}
\end{table*}

\subsection{Comparisons with other catalogues}\label{sec:compare-cat}
The number of the optically-selected SSC candidates is quite high compared 
to that recovered by \citet{2013MNRAS.431..554R} in their $K$-band AO images.
Different selection methods and a larger FoV but especially, the capability of
the WFC3/UVIS camera to deliver sharper images enabled many more objects to be
detected. The observed characteristics of the $K$-band star clusters are discussed 
in Section\,\ref{subsec:K-selected}.

Previous works by \citet{2011PhDT.........8V} and \citet{2017ApJ...843...91L}
have used WFC/ACS images (PI: Evans) in their star cluster analyses. 
The latter combined their optical data with small FoV far-UV images (a field
35 times smaller than that imaged by WFC3/UVIS camera) and hence,
only generated a cross-matched catalogue of 53 candidates with a Vega-magnitude 
range of $-14.9 < M_B < -8.9$\,mag in the $B$-band.
A further comparison of the cluster ages and masses will be done after performing 
spectral energy distribution (SED) fitting as described in Section\,\ref{age-sec}.

Our $BI$-band \textcolor{black}{cross-matched catalogue is composed of 1646 cluster candidates
with $\sigma \leq 0.20$\,mag}. The data are consistent with the detection and 
photometry algorithms \textcolor{black}{of 1321 sources} derived by \citet{2011PhDT.........8V} which only use
these two filters in their work. The high quality data from the UVIS camera,
however, include more cluster detection at fainter magnitude levels as expected. 
Table\,\ref{ACS-WFC3} summarizes comparison between the two datasets: the 
magnitude ranges (assuming $D_L = 45.3$\,Mpc) are in agreement within the
cluster photometric uncertainties.

\begin{table}
\begin{scriptsize}
\begin{center}
\caption{Comparing our data with the \citet{2011PhDT.........8V} catalogue.} 

\label{ACS-WFC3}
\begin{tabular}{lcccc}
\hline 
   \noalign{\smallskip}
   
  Instrument & Pix.scale  & Nb.SSCs & F438W & Reference \\
             & (arcsec)   & ($\sigma \leq \sigma_m$) &  (mag) &   \\
     ~~~~(1) &      (2)  &     (3)  &     (4)         &        (5)     \\      
   \noalign{\smallskip}
\hline \hline
   \noalign{\smallskip}
WFC/ACS   & 0.05 &  1321 & -15.6 & Valvikin$+$11 \\
WFC3/UVIS & 0.04 &  1646 & -15.2 & This work \\
\noalign{\smallskip}
   \hline
\noalign{\smallskip}
\multicolumn{5}{@{} p{8.5cm} @{}}{\footnotesize{\textbf{Notes. }Column 1: HST instrument used; 
Column 2: the camera plate scale; Column 3: the number of SSCs with $\sigma \leq 0.20$\,mag
by cross-matching $B$- and $I$-band catalogues; 
Column 4: the $B$-band magnitude of the brightest star cluster; Column 5:  reference from literature.
}}

\end{tabular}
\end{center}
\end{scriptsize}
\end{table}

\section{The cluster age and mass fitting}\label{age-sec}

To derive the star cluster properties of \arp, we fit our $UBI$-band photometric catalogue to one of the most up-to-date
SSP models: {\tt Yggdrasil} by \citet{2011ApJ...740...13Z}. Such models are already used extensively
in the study of YMCs to output reliable fits of their ages $\tau$ and extinction $A_V$
\citep[e.g][]{2012MNRAS.419.2606B,2014AJ....148...33R,2015MNRAS.452..246A,
2016MNRAS.460.2087H,2018MNRAS.473..996M}. In contrast to the well-known {\tt Starburst99\,(SB99)} code
by \citet{1999ApJS..123....3L,2014ApJS..212...14L}, {\tt Yggdrasil} has options to define the gas covering factor, $f_{cov}$, 
 \textcolor{black}{at low and indermediate redshifts to model unresolved stellar populations like galaxies or star clusters}. 
The treatment of photoionized gas from nebular emission and continuum is crucial to output 
robust physical properties of the very young stellar clusters. 

In this work, the models were retrieved assuming an instantaneous burst, a solar metallicity, a Kroupa IMF, 
and a Padova-AGB stellar evolutionary model. With a factor of $f_{cov} = 1$ and a low redshift coverage, 
the evolutionary track has an age range between 1\,Myr and 10\,Gyr. A metallicity of ${\rm Z} = 0.02$  (i.e. Solar) 
was chosen to align with the
SALT/RSS\footnote{The Robert Stobie Spectograph (RSS) is a spectrograph mounted
on the 10-meter class Southern African Large Telescope (SALT).} spectroscopic analyses of the  
SUNBIRD sample, which show typically such metallicities 
(Ramphul \& V{\"a}is{\"a}nen, private communication).  Finally, a starburst attenuation curve was adopted
to estimate the extinction coefficient of each filter ($R_V = 4.05$, \citealt{2000ApJ...533..682C}).

Note, however, that using appropriate SSP models alone is not enough  to estimate values
of the cluster age and mass. 
We therefore constructed
an extinction map of the galactic field (see Section\,\ref{sec:range-Av})
in order to apply constraints on the extinction range prior to $\chi^2$ minimization.
The fitting algorithm is expressed as follows:
\begin{equation}
\chi^{2}(\tau, A_{V}) = \sum_{\lambda}W_{\lambda} \left( m^{obs}_{\lambda} - m^{mod}_{\lambda}\right)^{2}
\end{equation}
where $\lambda = U, B, {\rm and}~I$-band, $m^{obs}$ and $m^{mod}$ are respectively the observed and the synthetic 
magnitudes while $W_{\lambda}$ is a normalization factor.

\subsection{Constraining the extinction ranges}\label{sec:range-Av}

\begin{figure}
\centering
\resizebox{1.\hsize}{!}{\rotatebox{0}{\includegraphics[trim= 0.cm 1.5cm 0cm 0cm, clip]{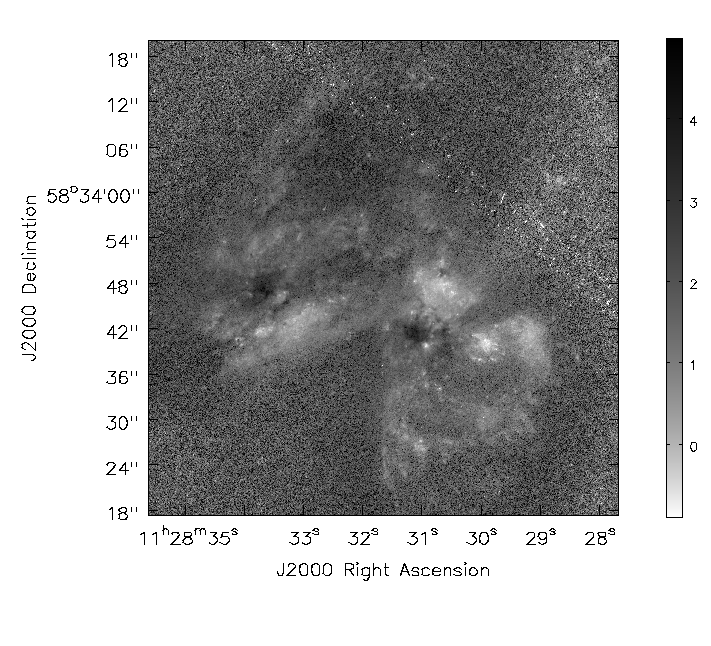}}}
\caption{\small A broad-band $U - I$ color map of \arp. The colorbar represents a linear scale of the
colors where darker shades are associated with highly extinguished regions.}
\label{Avmap-arp}
\end{figure}

For each cluster to have its own range of extinction 
as an input to the $\chi^2$ fitting, an extinction map of \arp~was produced based on a smoothed background version
of a broad-band $U - I$ color map as displayed in Fig.\,\ref{Avmap-arp}. An intrinsic color of $U - I \approx 0.35$\,mag to
the starburst galaxy \citep{1996ApJ...467...38K} at a zeroth-order approximation 
along with a starburst attenuation law were adopted. Based on the derived map, high extinction values of $A_V \gtrsim 3$\,mag
are associated with the nuclear starburst
regions of the galaxies: $A_V \sim$\,3.8\,mag and $A_V \sim$\,3.6\,mag in the case of NGC\,3690E and NGC\,3690W, respectively.
\citet{2000ApJ...532..845A} reported similar extinction ranges using optical spectroscopy.
They also estimated the extinction to the gas in the nuclei with NIR HST/NICMOS observations and by using an 
aperture of 2 by 2 arcsec, they found the following ranges: $A_V = 5 - 6$\,mag for NGC\,3690E
and $A_V = 3-4$\,mag for NGC\,3690W.
The higher extinction range in the case of NGC\,3690E is expected because
estimates from optical data usually reflect the dust screen only. Measurements based on NIR observations, however, probe
deeper into the interstellar medium. These comparisons demonstrate that our derived extinction
map should provide at least a crude approximation to the extinction of the cluster.

An area of 5 by 5 pixels centered at the spatial coordinates of any given SSC was used to calculate an average
value $A_V ^0$ for its initial extinction estimate. The cluster's range of extinction for the SED fitting
was then defined as follows:
\begin{equation}
A_V (A_V^0) = \left \{
\begin{array}{lcl}
 \verb*=[=0,\,0.25]       & \, {\rm if} & A_V^0 < 0.25 \\
 \verb*=[=A_V^0 - A_V^{\star},\,A_V^0 + A_V^{\star}]  & \, {\rm if} & A_V^0  \geq 0.25 \\

\end{array}
\right .
\end{equation}
\begin{equation}
{\rm where}~A_V^{\star} = \left \{ 
 \begin{array}{lcl}
 0.25 & \, {\rm if} & 0.25 \leq A_V^0 \leq 0.75 \\
 0.75 & \, {\rm if} & A_V^0 > 0.75 \\
 \end{array}
 \right .
 \label{eqn:Avinit}
\end{equation}
\textcolor{black}{The higher the value of the initial extinction $A_V^0$ is, the wider the range in which $A_V$ can vary.
This is to ensure that $A_V^0$ is optimally used to constrain each cluster's range of extinction.}
Such constraints are essential and they were all suitably chosen after comparing the derived cluster ages output
from different sets of extinction ranges. 
\textcolor{black}{In fact, we performed the following tests}:
\begin{enumerate}
 \item \textcolor{black}{by estimating the resulting extinctions and ages of datasets associated with less patchy dust distributions ($A_V^0 < 0.25$\,mag).
These are expected to be young objects ($\tau \lesssim 100$\,Myr) which are almost free of dust based on their colors. 
We varied the range of extinction between $0 - 0.15$\,mag, then $0 - 0.25$\,mag, and finally $0 - 0.35$\,mag while 
performing the SED fitting. Results from all three sets are similar and consistent with the distribution of the datapoints 
in the color-color plots (see Section\,\ref{sec:ccd}). Adopting a range of $A_V = [0,0.25]$ for star clusters with $A_V^0 < 0.25$\,mag 
is therefore reasonable and hence, was used to derive the final results;} 

\item \textcolor{black}{by considering objects with relatively high initial extinctions 
of $1.2 < A_V^0\,{\rm (mag)} < 2$. Based on their positions in the color-color plots, the datasets either represent highly extinguished young clusters or
old clusters with low extinction values. This degeneracy between age and extinction should be reduced 
once UV-data is incorporated. Various ranges were again used for comparison,
i.e.\,$A_V = A_V^0 \pm A_V^{\star}$ where $A_V^{\star} = 0.25, 0.5, 0.75, 1$\,mag. The fitted results
are generally consistent with the prediction, regardless of the extinction ranges;}

\item \textcolor{black}{by adopting a range of $A_V = [0,A_V^{nucl}]$ for the whole catalogue, where $A_V^{nucl} \sim 3.8\,{\rm mag}$. 
Compared with the results from the previous methods, we find that 
around 58 - 71 percent of the clusters have similar ages within 0.25 - 0.5 dex. Such a percentage
is expected since the constraints on $A_V$ are only governed by the maximum extinction value associated
with the nuclear regions.} 
\end{enumerate}
\textcolor{black}{Based on the results from these tests,} \textcolor{black}{the chosen
values of $A_V^{\star}$ in Eq.\,\ref{eqn:Avinit}
are therefore expected to output reliable estimates of the star cluster properties.}

\begin{figure}
\centering
\resizebox{1.1\hsize}{!}{\rotatebox{0}{\includegraphics{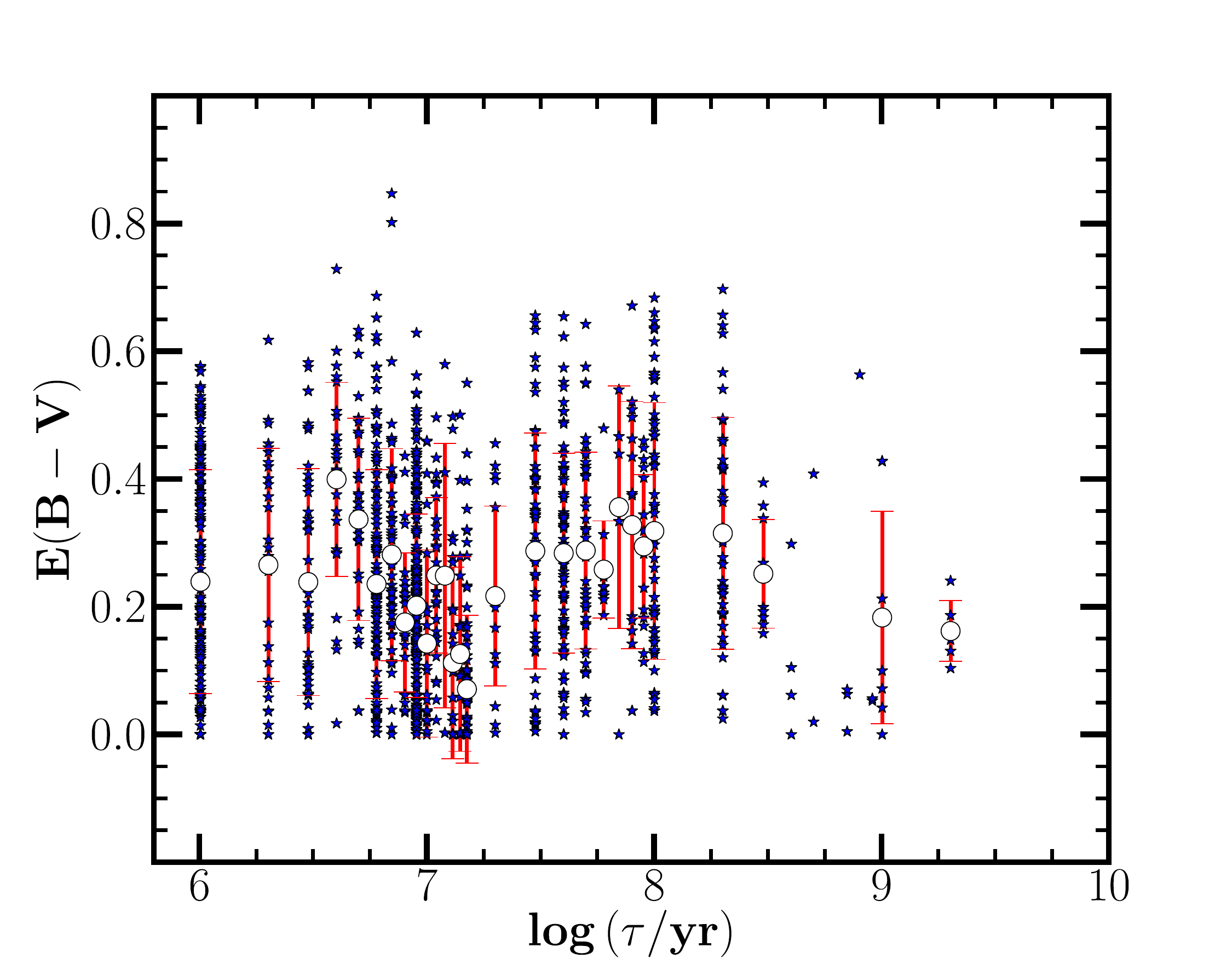}}}\\
\caption{\small The cluster visual extinction plotted against its age. Only SSCs with $\sigma \leq 0.20$\,mag
are included for clarity. The open circles and the error bars represent the mean extinction values and the associated
$1\sigma$ uncertainties at each age step, respectively.}
\label{fig:Ebv-age}
\end{figure}

We should also note that the use of the extinction map does not bias our results:
\textcolor{black}{{\it i)} we constructed an extinction map based on a $B - I$ color and then derived the corresponding value  
$A_V^0$ of each cluster. The estimated ages were compared with results considering 
$U - I$ color map. We find that 80\,percent of the clusters have similar ages within 0.5\,dex.
SED fitting that consider $A_V^0$ retrieved from a $U - I$ color map output clusters with slightly older
ages as expected;} \textcolor{black}{{\it ii)} we also varied $A_V$ between the initial extinction estimates $A_V^0$
derived from the $U - I$ and $B - I$ color maps. We find that about 70 - 90 percent of the clusters have similar ages
within 0.25 - 0.5 dex compared with results using $U -I$ color map.}
\textcolor{black}{These tests emphasize the fact that the color map used to constrain the extinction does not bias
the derived properties of the clusters.} In addition, as shown in
Fig.\,\ref{fig:Ebv-age}, where the open circles represent the mean extinction values at each age step,
there is no clear correlation between the derived extinction and the cluster age.
\textcolor{black}{The cluster mass, however, increases with an increasing extinction. This is expected
since highly extinguished nuclear regions are reported to be potential birthsites of the very massive star
cluster candidates (see Section\,\ref{sec:distr-arp}).}

\begin{figure}
\centering
 \begin{tabular}{c}
\resizebox{1.\hsize}{!}{\rotatebox{0}{\includegraphics[trim= .7cm 0cm 1.5cm 0cm, clip]{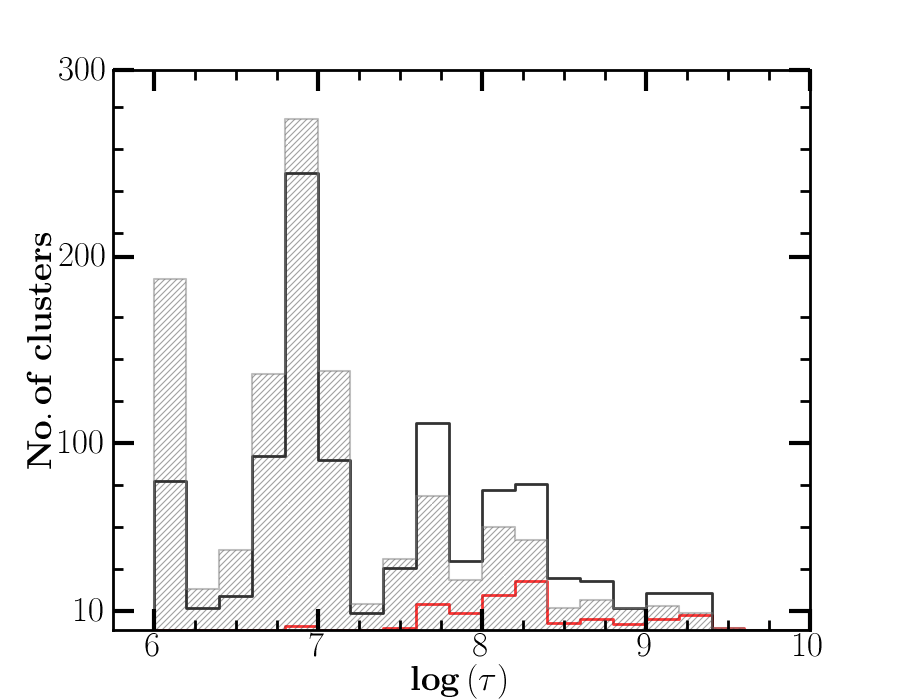}}}\\
\resizebox{1.\hsize}{!}{\rotatebox{0}{\includegraphics[trim= .7cm 0cm 1.5cm 0cm, clip]{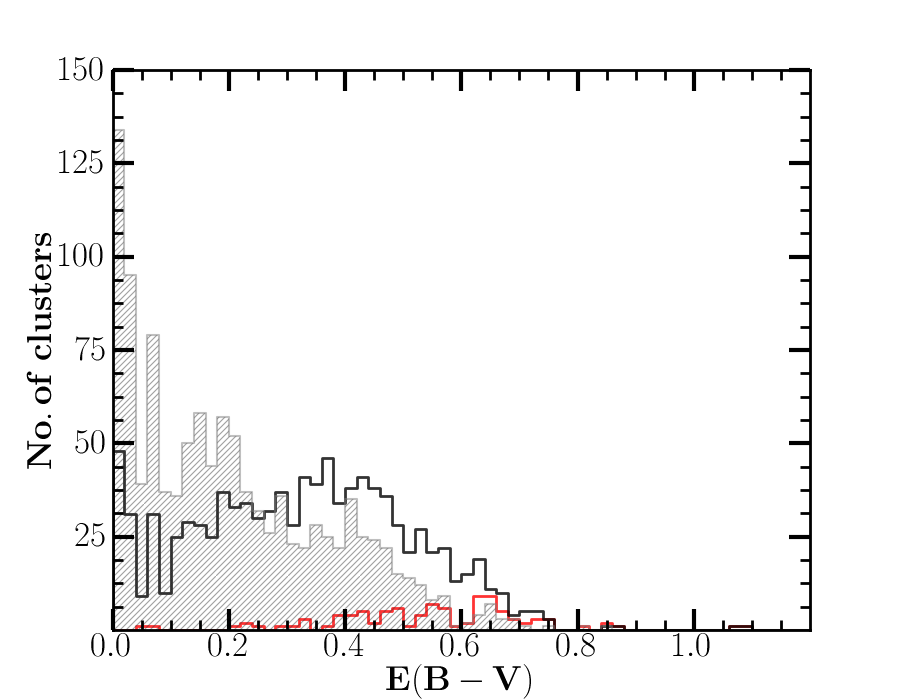}}}
\end{tabular}
\caption{\small The histograms representing the age ({\em top}) and extinction ({\em bottom}) distributions
of the SSC candidates in Arp\,299.
The solid line histograms correspond to the datasets of NGC\,3690E while the hatched ones represent those of NGC\,3690W. 
The red distributions correspond to cluster candidates with masses \,$ > 10^7\,{\rm M_{\odot}}$ in the system. 
More than half of the clusters have ages younger than $\sim$\,15\,Myr old.
There is a significant number of SSC candidates with low extinction values in the\ field of NGC\,3690W.}
\label{fig:age-hist}
\end{figure}

\subsection{The resulting properties of the star clusters}\label{subsec:result-fits}
\begin{figure}
\centering
\resizebox{1.\hsize}{!}{\rotatebox{0}{\includegraphics[trim= 1.3cm 0cm 1.8cm 0cm, clip]{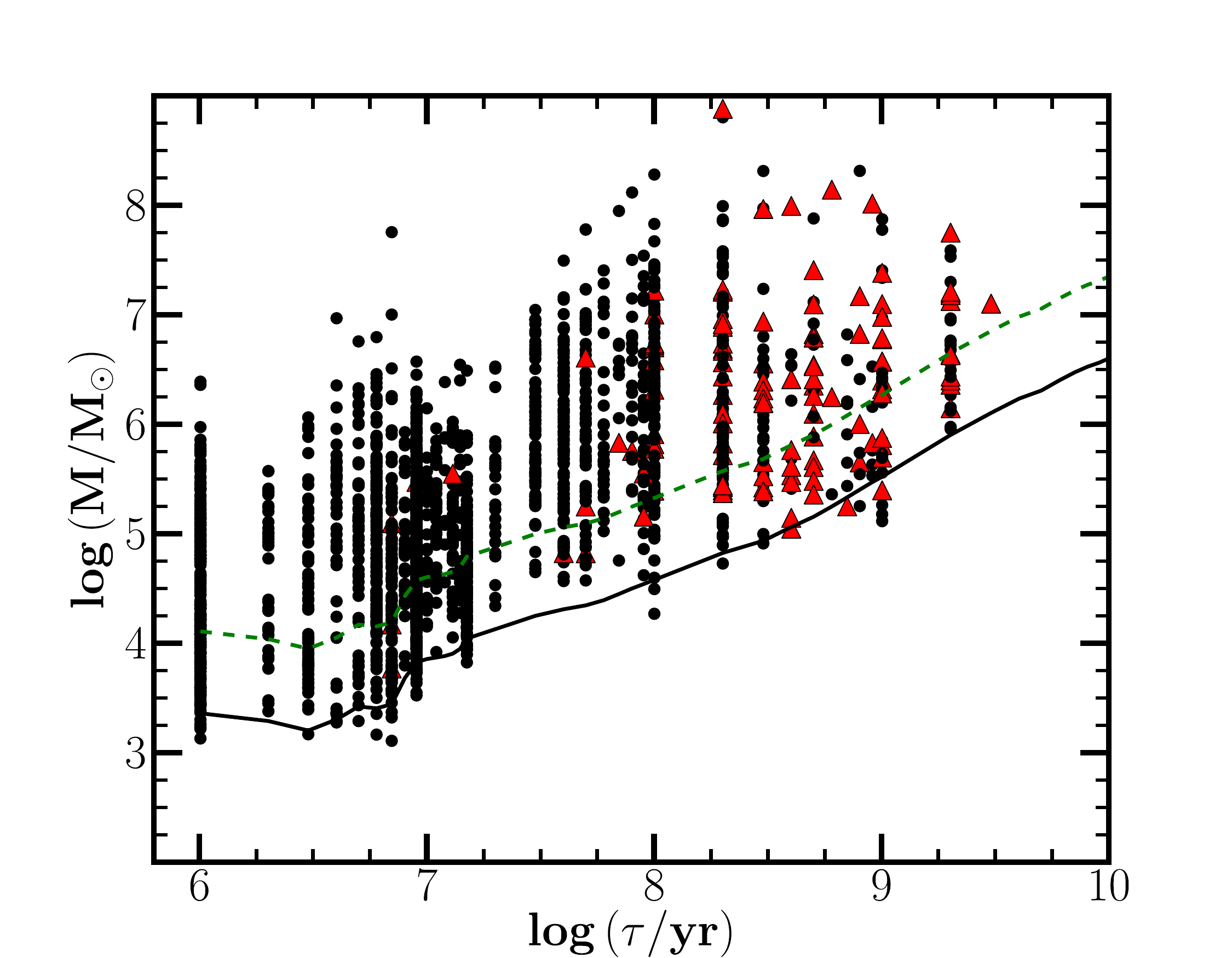}}}\\
\resizebox{1.\hsize}{!}{\rotatebox{0}{\includegraphics[trim= 1.3cm 0cm 1.8cm 0cm, clip]{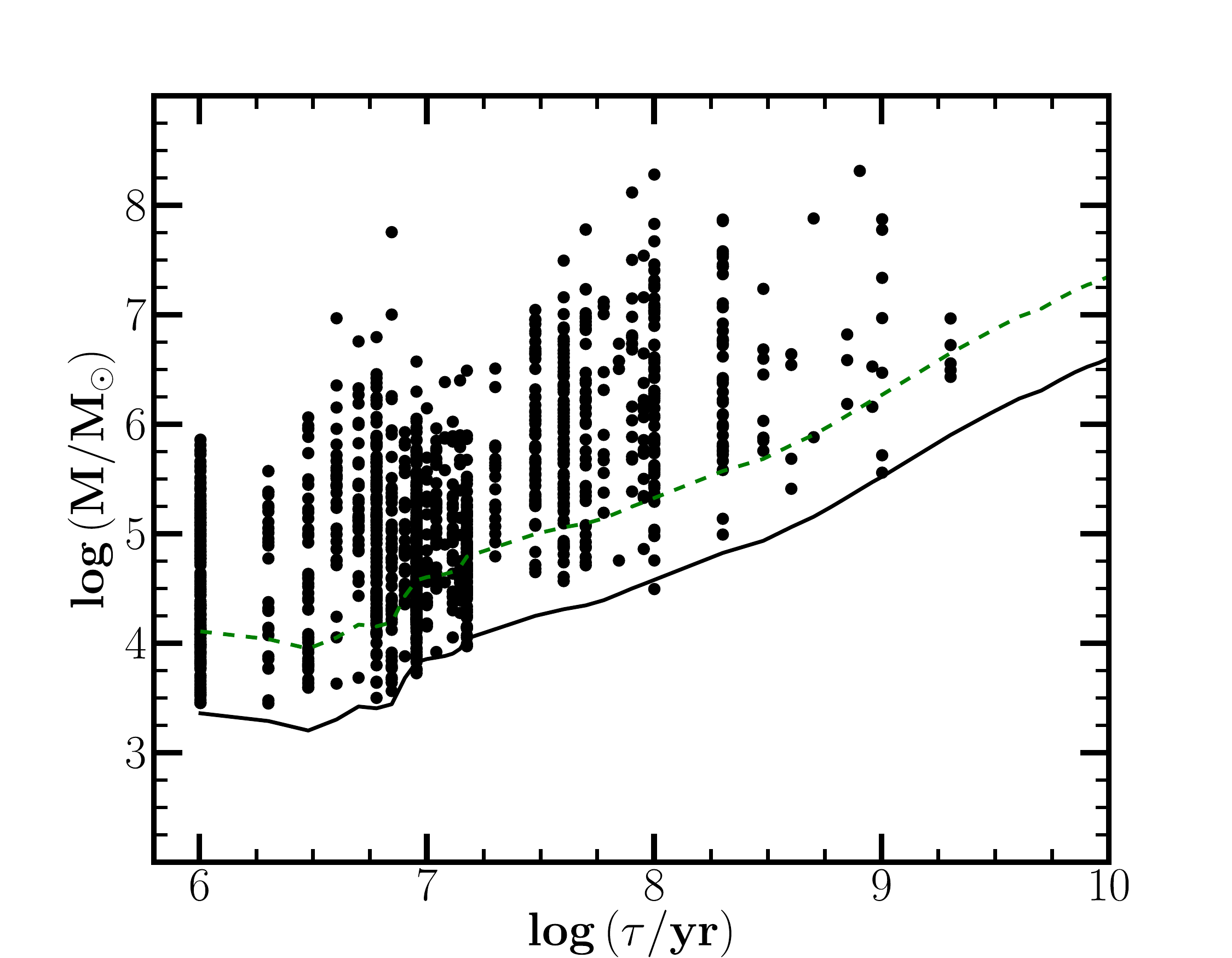}}}\\
\caption{\small The cluster age-mass plane of Arp\,299 from SED-fitting with {\tt Yggdrasil} models. 
The solid and dashed lines denote the 80\,percent photometric
detection limits for the outer ($M_B = -8.6$\,mag) and inner ($M_B = -10.5$\,mag) regions,
respectively.  Chimneys are clearly observed at around 1\,Myr and other age
steps. {\em Upper:} All selected candidates are considered where the red points correspond to clusters
with a high magnitude error in the $U$-band. {\em Lower:} Same distribution but 
including only sources with ${\rm S/N \geq 5}$. }
\label{fig:age-mass}
\end{figure}

Outputs from our $UBI$-fit indicate that \arp~hosts a very young population of  
star clusters: 62 percent of the SSCs have ages $\tau \lesssim 15$\,Myr old.
The upper panel of Fig.\,\ref{fig:age-hist} overplots
\textcolor{black}{the number of star clusters found per age bin} in NGC\,3690E (black solid line) and NGC\,3690W (hatched)
where $\sim$\,60\,percent of the SSCs
in the western component are younger than $\tau \lesssim 10$\,Myr old.
Figure\,\ref{fig:age-mass}, on the other hand, shows the resulting distributions in the age-mass plane, 
considering all candidates (top panel) 
and those with high confidence photometry only (bottom). 
Clusters with $0.35\,{\rm mag} < \sigma_{U} \leq$\,0.50\,mag are labelled as red triangles.
They are mostly old SSCs with $\tau > 200$\,Myr which are consistent with their $U$-band magnitude errors.
The solid and dashed lines represent 
the evolutionary fading of the SSP models associated with 80\,percent completeness limits for the outer 
($M_B = -8.6$\,mag) and inner regions ($M_B = -10.5$\,mag), respectively. 
Concentrations, referred to as \textgravedbl chimneys\textacutedbl, are present at ages around
$\tau \approx $\,1, 10, 50 and 200\,Myr. Poor fitting
of the youngest star clusters explains the chimney at log\,$\tau = $\,6 which corresponds to the lower
age limit of the SSP models. Concentrations at log\,$\tau \approx $\,7 and log\,$\tau \approx$\,8.3 
also exist because red super giants (RSGs) and asymptotic-giant branch stars of the star clusters become apparent 
at these ages, respectively \citep[e.g][]{2005A&A...431..905B,2013MNRAS.431.2917D}. The real age-mass distribution
is thus believed to spread into a wider range at these ages. The peak at log\,$\tau \approx$\,7.7
for both components, however, should be investigated further. {\it Could it be a signature of a past episode of intense star
formation in the interacting system}? Section\,\ref{subsec:dNdt} reports possible explanations of the trend.

SSC masses lie in the range of $\approx 10^{3} - 10^8 {\rm \msun}$. Even though chimneys exist in age modelling,
the mass of the clusters is much less affected by those uncertainties. Furthermore, the trends seen in 
Fig.\,\ref{fig:age-mass} clearly demonstrate that the population of star clusters decrease with
time because of various cluster disruption mechanisms \citep[e.g.][]{2003A&A...397..473B}. 
Nevertheless, SSCs with a derived mass below $10^4{\rm \msun}$ may be affected by stochastic effects while some
datapoints with ${\rm M > 10^7 \msun}$ (red distributions in Fig.\,\ref{fig:age-hist}) 
could be complexes of star clusters and hence, are analysed separately in Section\,\ref{sec:massive-SSC}.

The lower panel of Fig.\,\ref{fig:age-hist} shows the resulting extinction distributions: the visual 
extinction $A_V$ of the star clusters varies between 0 to 4.5\,mag, i.e. ${\text{E}(\text{B}-\text{V})=0-1.1}$.
NGC\,3690W (hatched) has a significant number of star clusters with a low extinction value of ${\text{E}(\text{B}-\text{V})<\,0.1}$
compared to the population of SSCs in NGC\,3690E (black solid line). This is not surprising
since Fig.\,\ref{Avmap-arp} shows that NGC\,3690W has a much larger white area (indicating $A_V \sim 0$) than NGC\,3690E.

\subsection{Consistency checks} \label{subsec:testfit}

At least four filters, covering a large broad-band that
includes Balmer lines, are usually to be considered to secure SED fitting results with a high precision
 \citep{2004MNRAS.347..196A}.  Nevertheless, $\chi^2$ minimization using three filters only should be 
reliable if an UV-filter is included to break the cluster age-extinction degeneracy, especially if
constraints are applied on the extinction and metallicity. To assess the robustness of our results,
various sanity checks were performed by only considering sources with low photometric uncertainties 
(i.e.\,$\sigma \leq 0.20$\,mag).

\subsubsection{SED fitting using SB99}

\begin{figure}
\centering
\resizebox{1.\hsize}{!}{\rotatebox{0}{\includegraphics[trim= 1.3cm 0cm 1.8cm 0cm, clip]{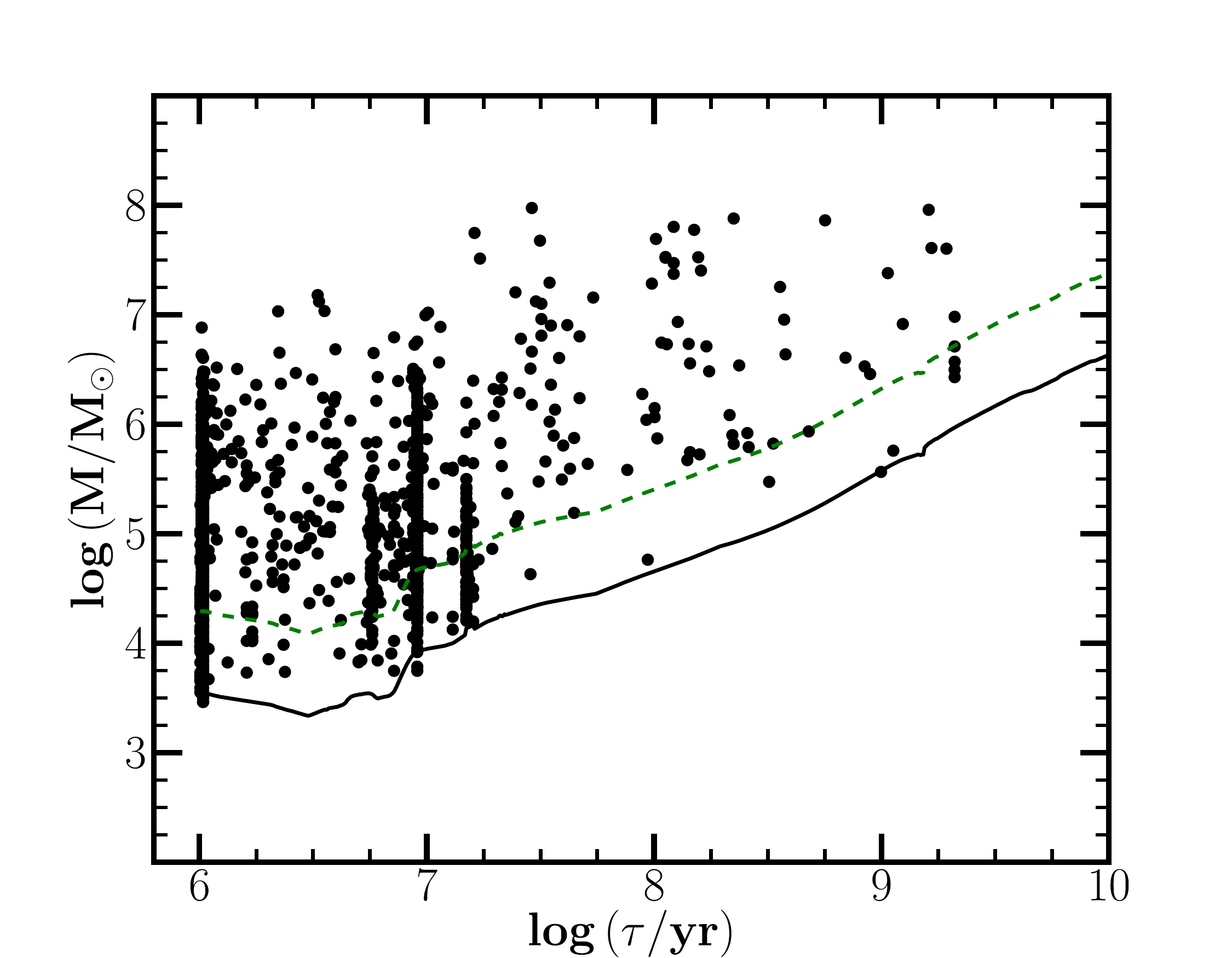}}}
\caption{\small The age-mass plane when performing $\chi^{2}$\,fitting with {\tt SB99} models. 
Same labels as in Fig.\,\ref{fig:age-mass}. Clusters tend to have younger ages compared with results from
{\tt Yggdrasil} SED fitting.}
\label{fig:age-mass-SB99}
\end{figure}

We performed SED fitting of the observed data with {\tt SB99} for comparison: the SSP models 
tend to output clusters of younger ages (see Fig.\,\ref{fig:comp-SB99} in the Appendix). 
In fact, half of the population are younger than 5\,Myr old while 
the number of clusters with ages between 10 and 100\,Myr double-up when using
{\tt Yggdrasil} SSP models. The distribution
in the age-mass plane is shown in Fig.\,\ref{fig:age-mass-SB99}: the resulting ages from {\tt SB99}
mostly converge and form more pronounced chimneys at log\,$\tau \approx 6$ and log\,$\tau \approx 7$.
This is in contrast to the trend seen in Fig.\,\ref{fig:age-mass} where the ages from {\tt Yggdrasil} 
spread into a wider time interval (despite the presence of chimneys). One main reason of the discrepancies
is that {\tt SB99} models do not account for the effects 
of nebular emission and hence do a poor fitting of the emission lines.
We therefore chose not to use the results from {\tt SB99}
but rather focused our analyses based on the physical properties derived with {\tt Yggdrasil}.

\subsubsection{Comparison with Linden age results}\label{subsec:Linden-ages}

\begin{figure}
\centering
\resizebox{1.1\hsize}{!}{\rotatebox{0}{\includegraphics{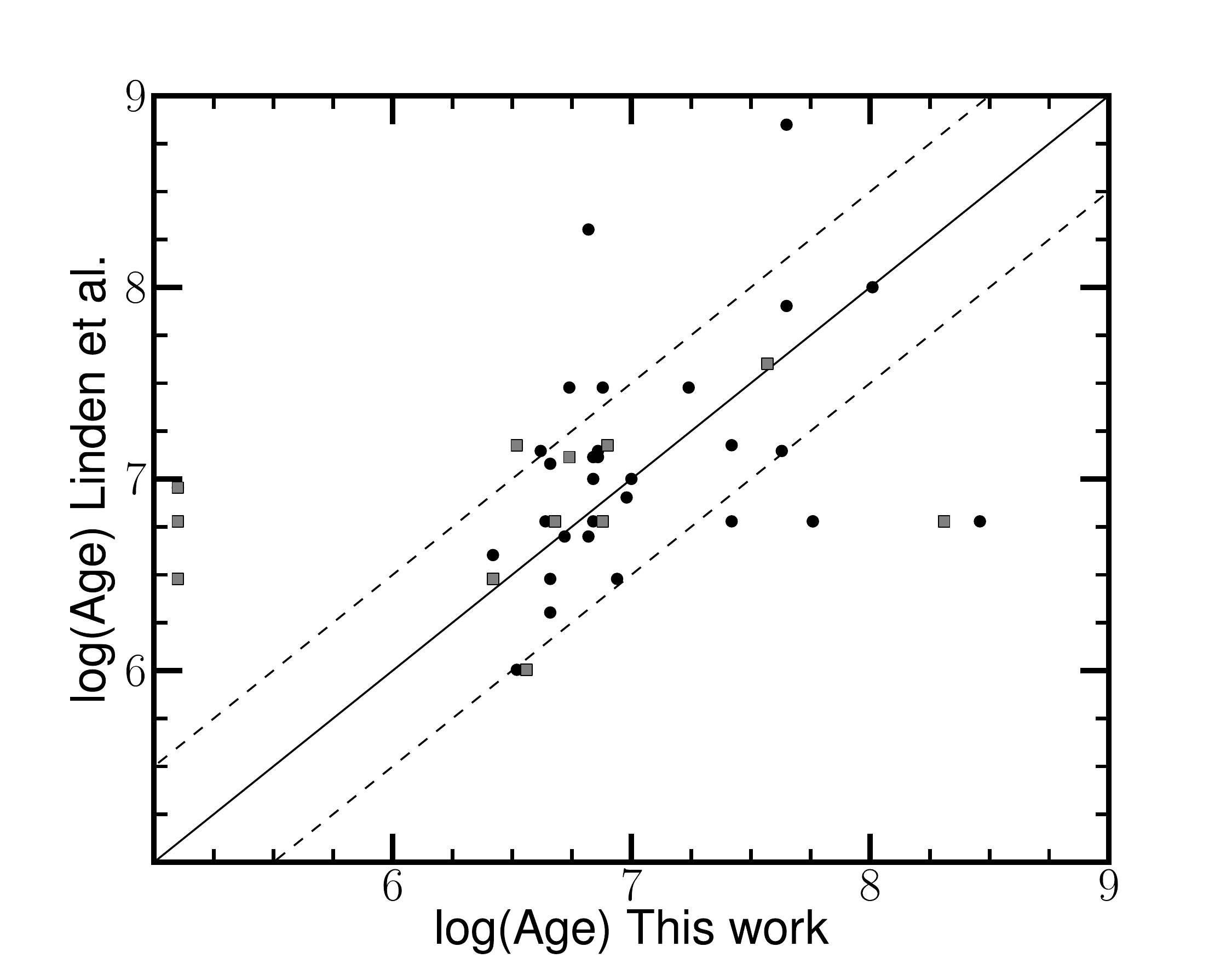}}}
\caption{\small Comparison between the derived ages from \citet{2017ApJ...843...91L} and our 
$UBI$-band {\tt Yggdrasil} SED fitting. A match radius of 0.01\,arcsec recovers 31 common sources (black circles).
The grey squares, however, are matched sources identified within $0.01-0.1$\,arcsec radii. The solid line 
illustrates a linear correlation, whereas the dashed lines are within 0.5\,dex of the relation.}
\label{fig:Linden-cat}
\end{figure}

As already mentioned in Section\,\ref{sec:compare-cat},  \citet{2017ApJ...843...91L} have studied the physical
properties of 53 SSC candidates using WFC/ACS and far-UV images. Figure\,\ref{fig:Linden-cat} compares the 
derived ages in this work with the ones from \citet{2017ApJ...843...91L} who considered
{\tt GALAXEV}\footnote{{\tt GALAXEV} is a library of SSP models
built from the isochrone synthesis code of \citet{2003MNRAS.344.1000B}.}. A match radius of 0.1\,arcsec was 
adopted to recover 43 common sources: the black circles denote sources identified within 0.01\,arcsec, 
whereas the other sources are labelled as grey squares. Our results are in agreement with the values found by \citet{2017ApJ...843...91L}:
approximately 72\,percent of the matched sources are consistent in  their ages
within 0.5\,dex. This work also generates less massive clusters (see Fig.\,\ref{fig:comp-Linden}).
Contamination is expected to affect more the Linden catalogue since it was drawn from 
\textcolor{black}{WFC/ACS datasets with a slightly coarser pixel scale of 0.05 arcsec/pixel.} 

\section{The color diagrams}\label{sec:ccd}
\begin{figure}
\centering
 \begin{tabular}{c}
\resizebox{1.\hsize}{!}{\rotatebox{0}{\includegraphics[trim= 0.4cm 0cm 1.8cm 0cm, clip]{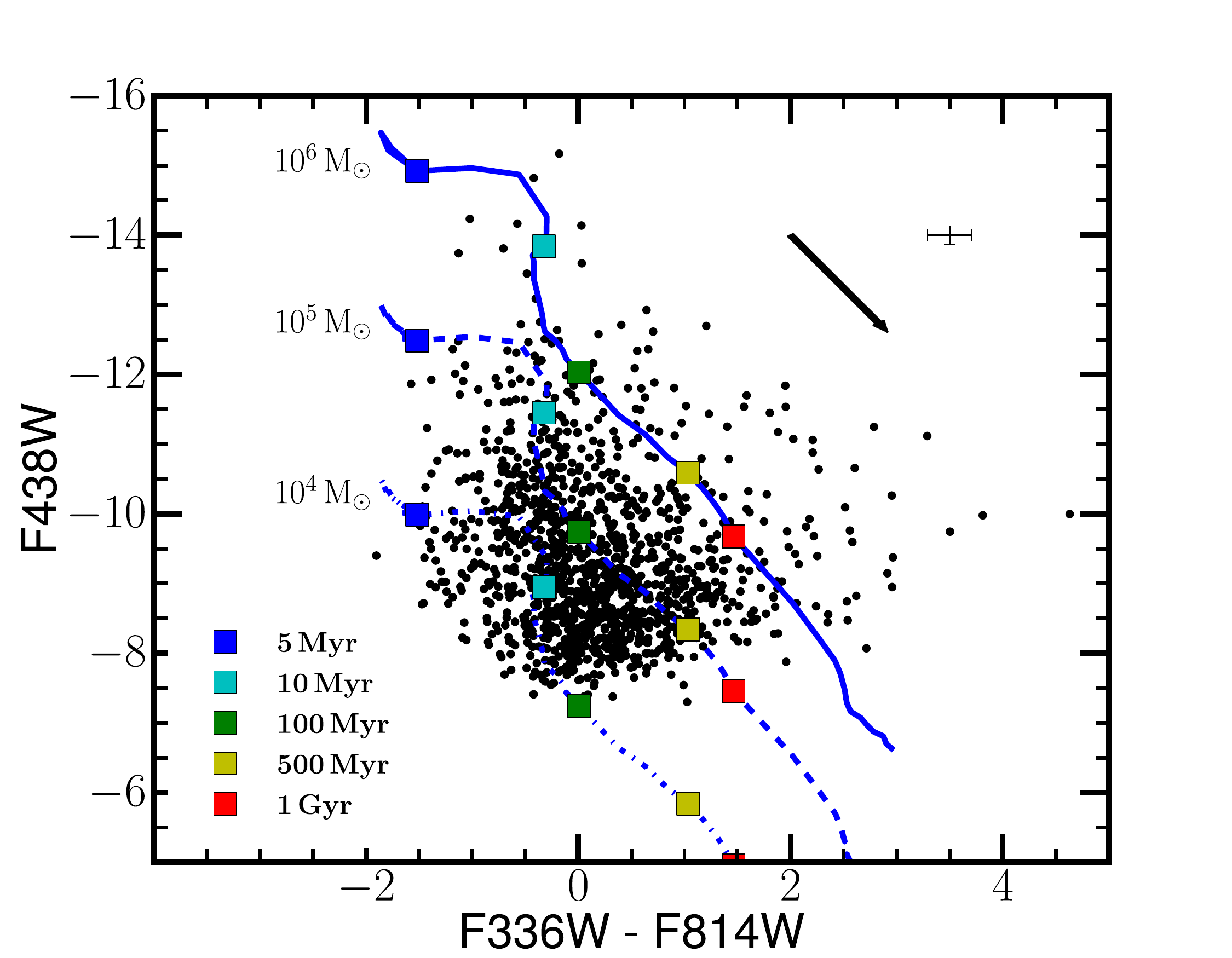}}}\\
\resizebox{1.\hsize}{!}{\rotatebox{0}{\includegraphics[trim= 0.4cm 0cm 1.8cm 0cm, clip]{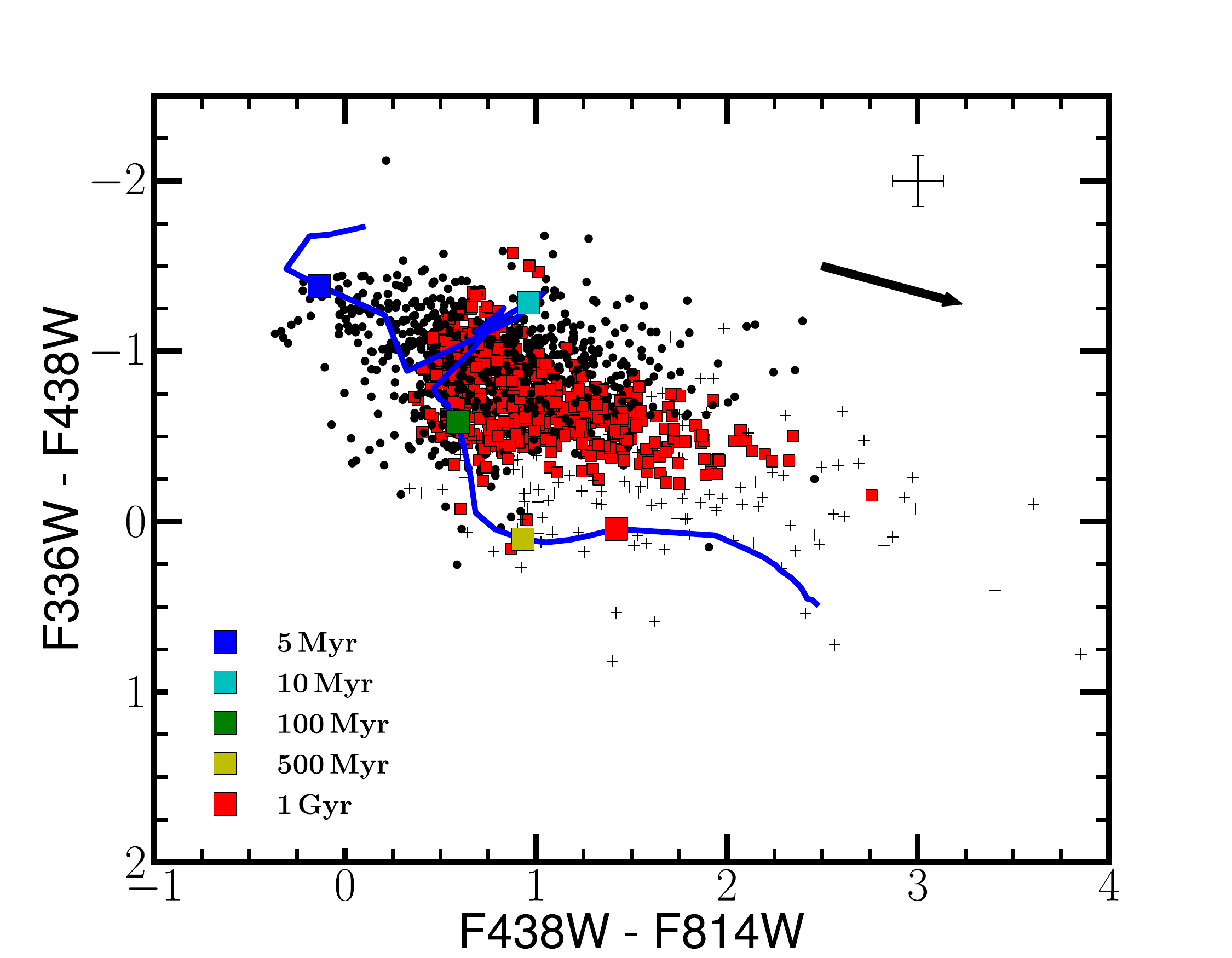}}}

 \end{tabular}
\caption{\small CMD and CCD for star clusters with $\sigma \leq $\,0.20\,mag.
The {\tt Yggdrasil} SSP models are plotted over the datapoints as solid and dashed lines. Blue, cyan, green, yellow and red squares,
respectively, mark 5, 10, 100, 500 and 1000\,Myr in the evolutionary tracks. The arrows 
indicate a reddening of $E(B-V) = 0.25$. {\em Upper:} $B$-band magnitude plotted against $U - I$ color. The model is plotted
with a synthetic cluster mass of $10^4,\,10^5$, and\,$10^6\,{\rm \msun}$. {\em Lower:} $U - B$ versus $B - I$ colors. Datapoints are labelled as a function of their
derived ages: black circles for $\tau \leq 10$\,Myr, red squares for $10 < \tau \leq 100$\,Myr, and black crosses for $\tau > 100$\,Myr.}
\label{cmds-arp}
\end{figure}

The color-magnitude diagram (CMD) and the color-color diagram (CCD) for SSCs with $\sigma \leq 0.20$\,mag are respectively
presented in the upper and lower panels of Fig.\,\ref{cmds-arp}. The solid and dashed lines denote {\tt Yggdrasil}
models where different ages of the synthetic cluster are marked with multicolor squares in the evolutionary track. 
The distribution of the datapoints in the CMD indicates that a significant number of the SSCs are predicted to have masses
between $\approx 10^4 - 10^7 {\rm \msun}$. This is in agreement with our results where 88\,percent of the clusters  
have derived masses within that range. 

The CCD suggests that the merging system mainly hosts 
young SSCs. Since the cluster population is concentrated in the field that corresponds to $1-100$\,Myr of the 
evolutionary track, most of the sources are likely to be associated with ages younger than 100\,Myr old. 
Again, such predictions
are consistent with the results reported in Section\,\ref{subsec:result-fits}. The cluster age distribution in the CCD can clearly be seen by
color-coding each datapoint as follows. Black circles,
red squares and black crosses correspond to $\tau \leq 10$\,Myr, ages between $10-100$\,Myr, and $\tau > 100$\,Myr, respectively.

\section{Interpreting the YMC properties}\label{analysis-sec}
\subsection{Investigating the cluster spatial distributions} \label{sec:distr-arp}
\begin{figure*}
\centering
 \begin{tabular}{c}
\resizebox{.8\hsize}{!}{\rotatebox{0}{\includegraphics[trim= 1.cm 0.5cm 2.5cm 1cm, clip]{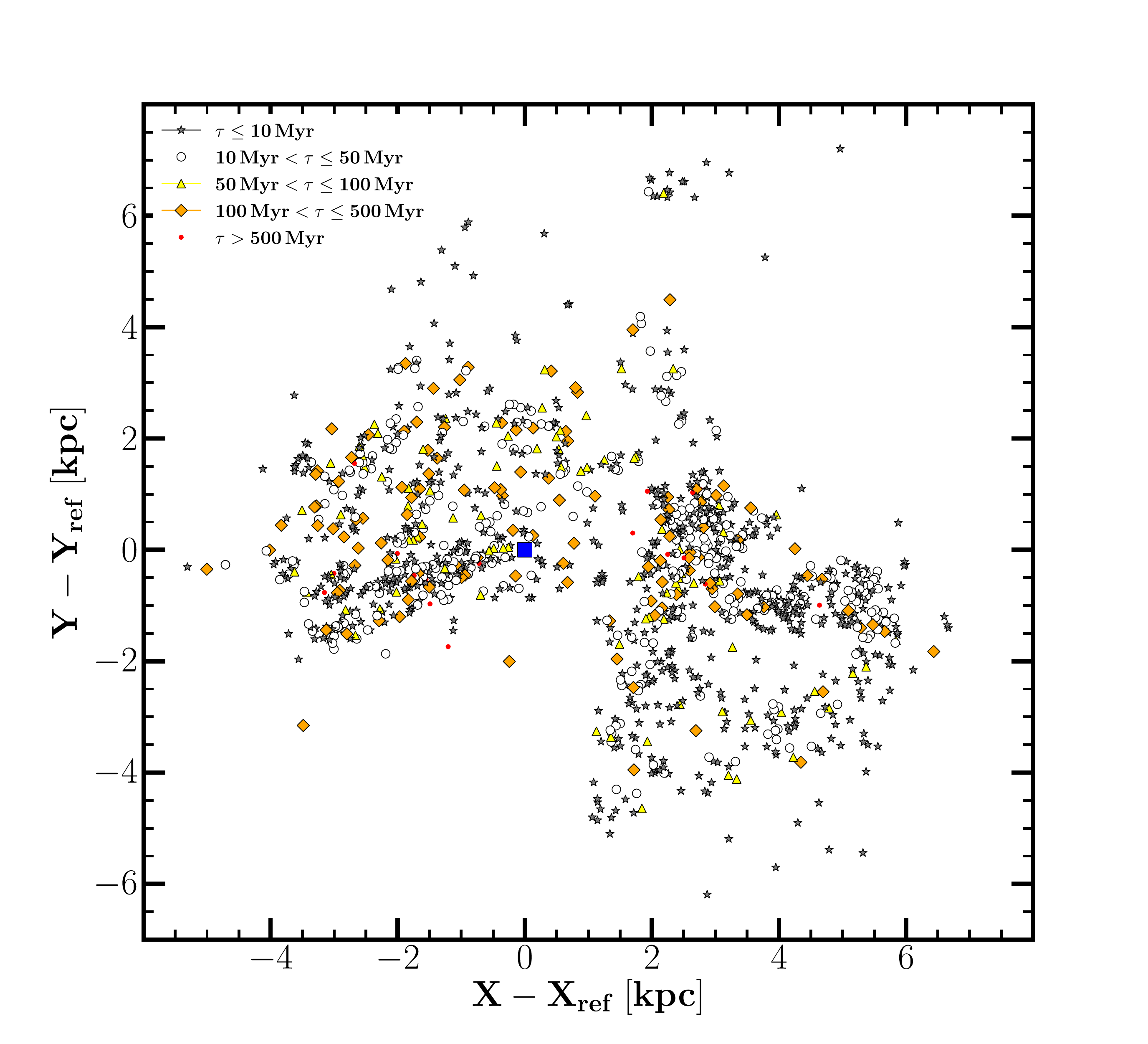}}}\\ 
\resizebox{.5\hsize}{!}{\rotatebox{0}{\includegraphics{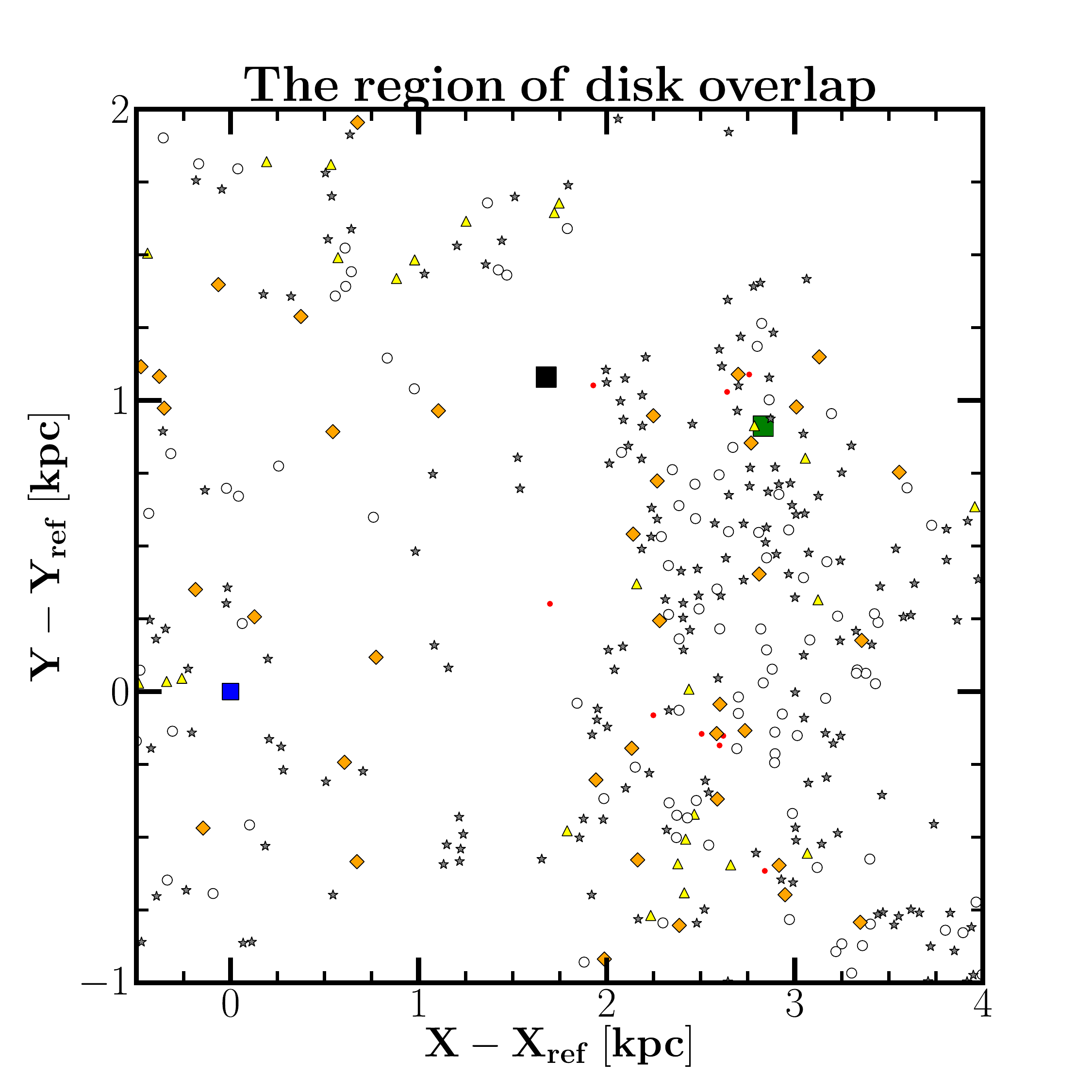}}}
\resizebox{.5\hsize}{!}{\rotatebox{0}{\includegraphics{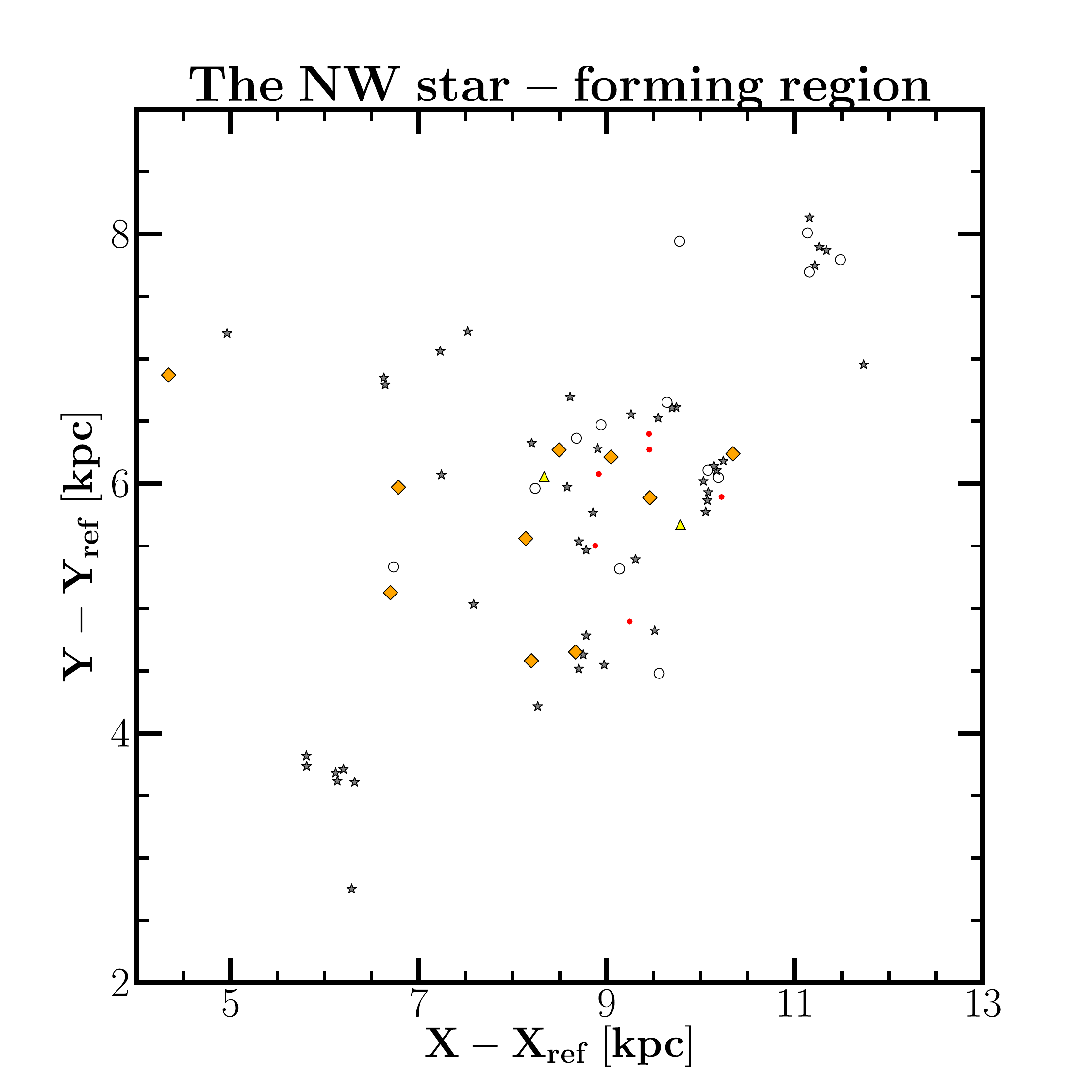}}}
\end{tabular}
\caption{{\em Upper: } The age spatial distribution of SSCs with $\sigma \leq 0.20$\,mag. The blue square
marks the intersection of the distance between the galaxy nuclei
with the boundary line used to separate the YMCs with respect to their galaxy hosts.
{\em Lower: } Zoomed-in versions looking at the disk overlap region ({\em left}) and the star-forming region
northwest of \arp~({\em right}). The different labels correspond to different age ranges as displayed in 
the legend.}

\label{fig:age-spatial-distr}
\end{figure*}

\begin{figure*}
\centering
 \begin{tabular}{c}
\resizebox{.8\hsize}{!}{\rotatebox{0}{\includegraphics[trim= 1.cm 0.5cm 2.5cm 1cm, clip]{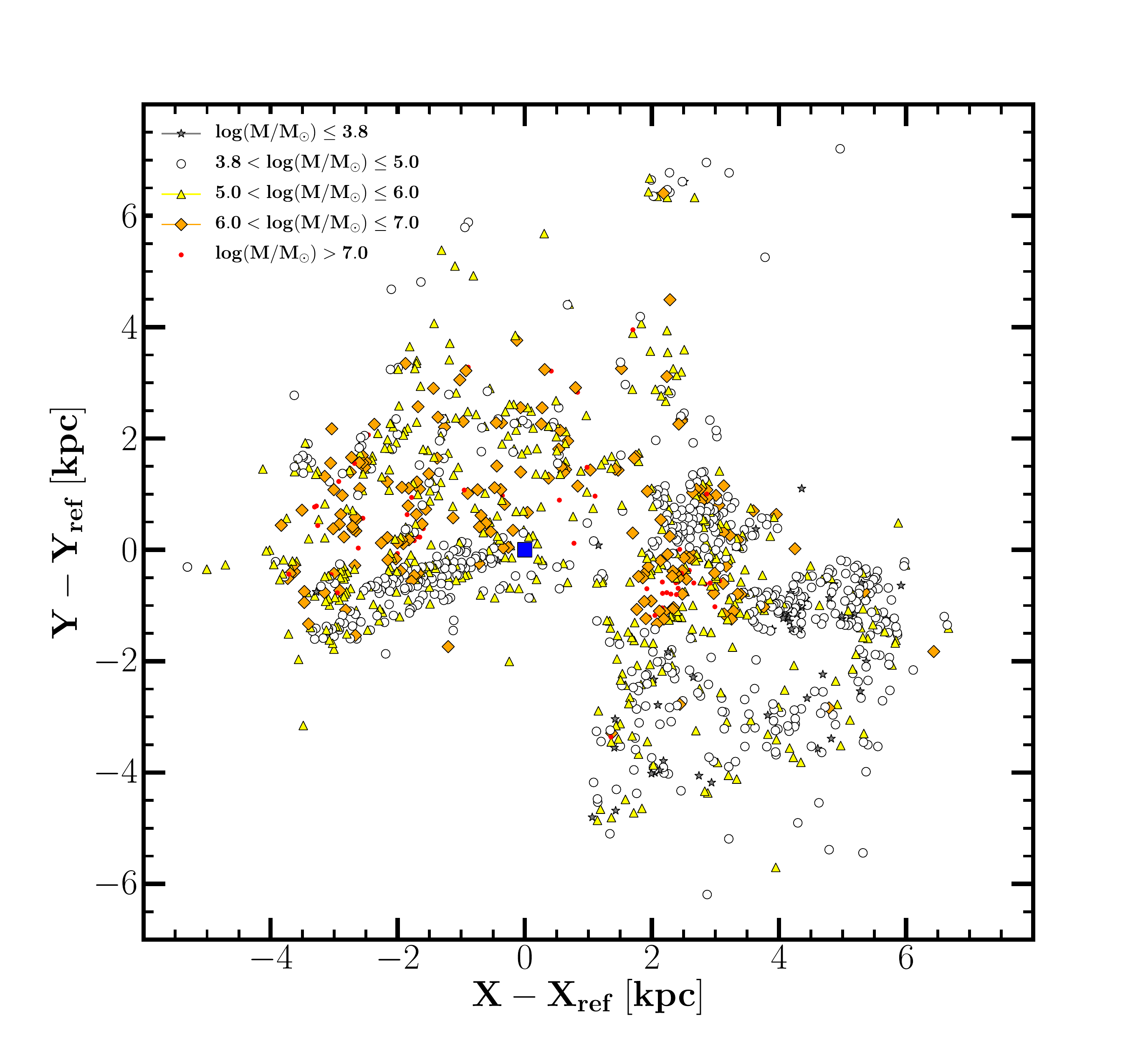}}}\\ 
\resizebox{.5\hsize}{!}{\rotatebox{0}{\includegraphics{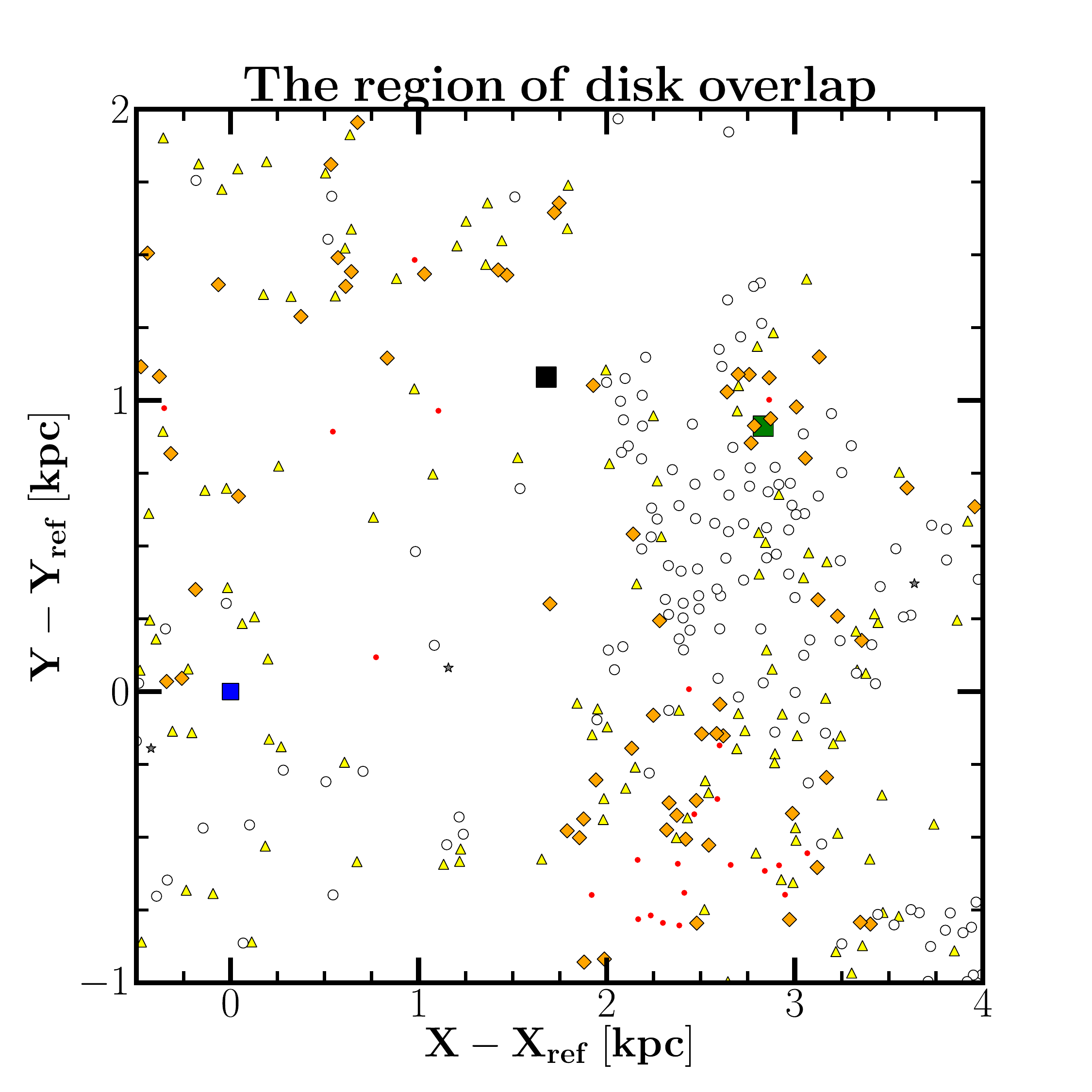}}}
\resizebox{.5\hsize}{!}{\rotatebox{0}{\includegraphics{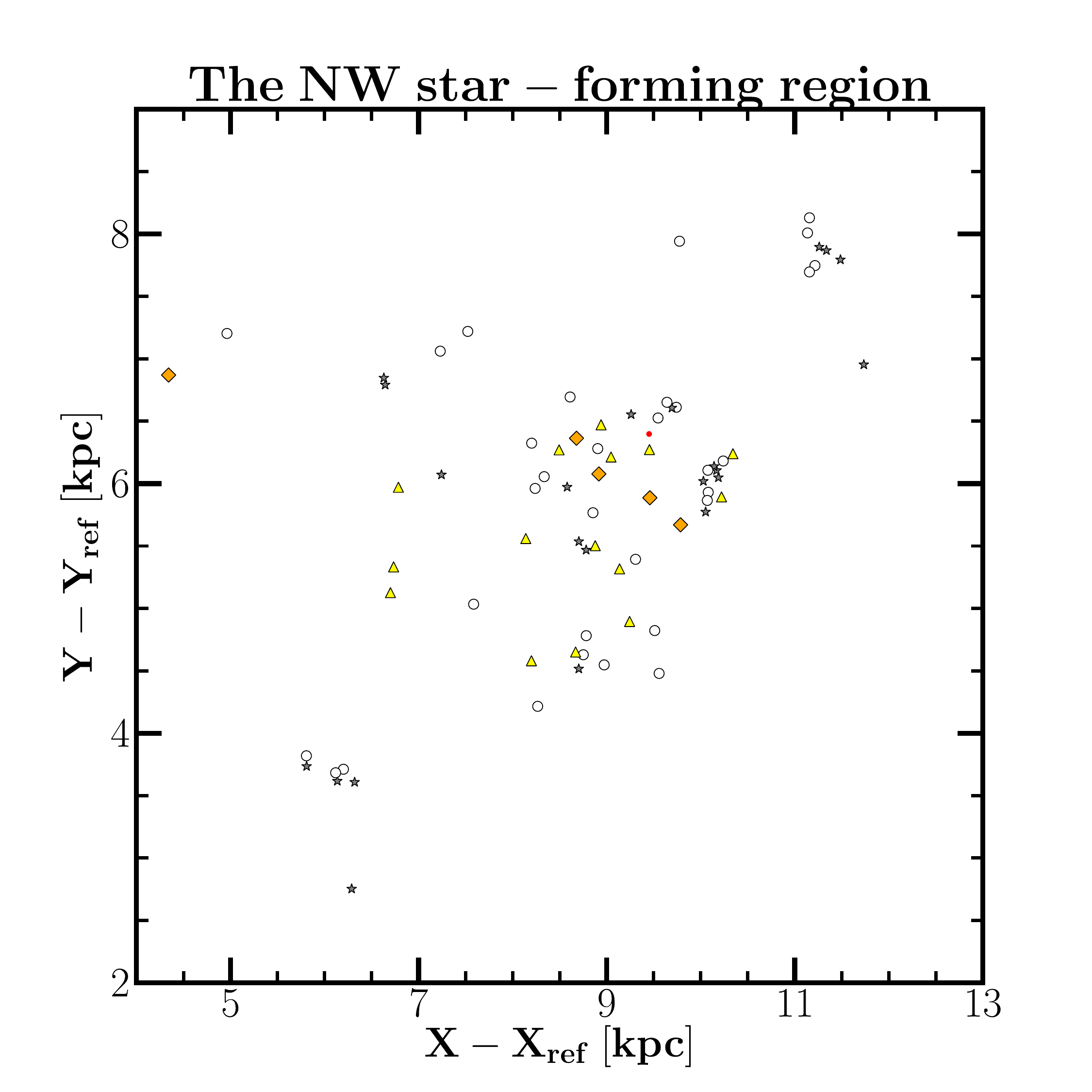}}}
\end{tabular}
\caption{Same as in Fig.\,\ref{fig:age-spatial-distr} but with the datapoints labelled as a function of the cluster
mass. The most massive SSCs tend to be located near the nuclear regions of the interacting system.}
\label{fig:mass-spatial-distr}
\end{figure*}

\begin{figure*}
\centering
 \begin{tabular}{c}
\resizebox{.8\hsize}{!}{\rotatebox{0}{\includegraphics[trim= 1.cm 0.5cm 2.5cm 1cm, clip]{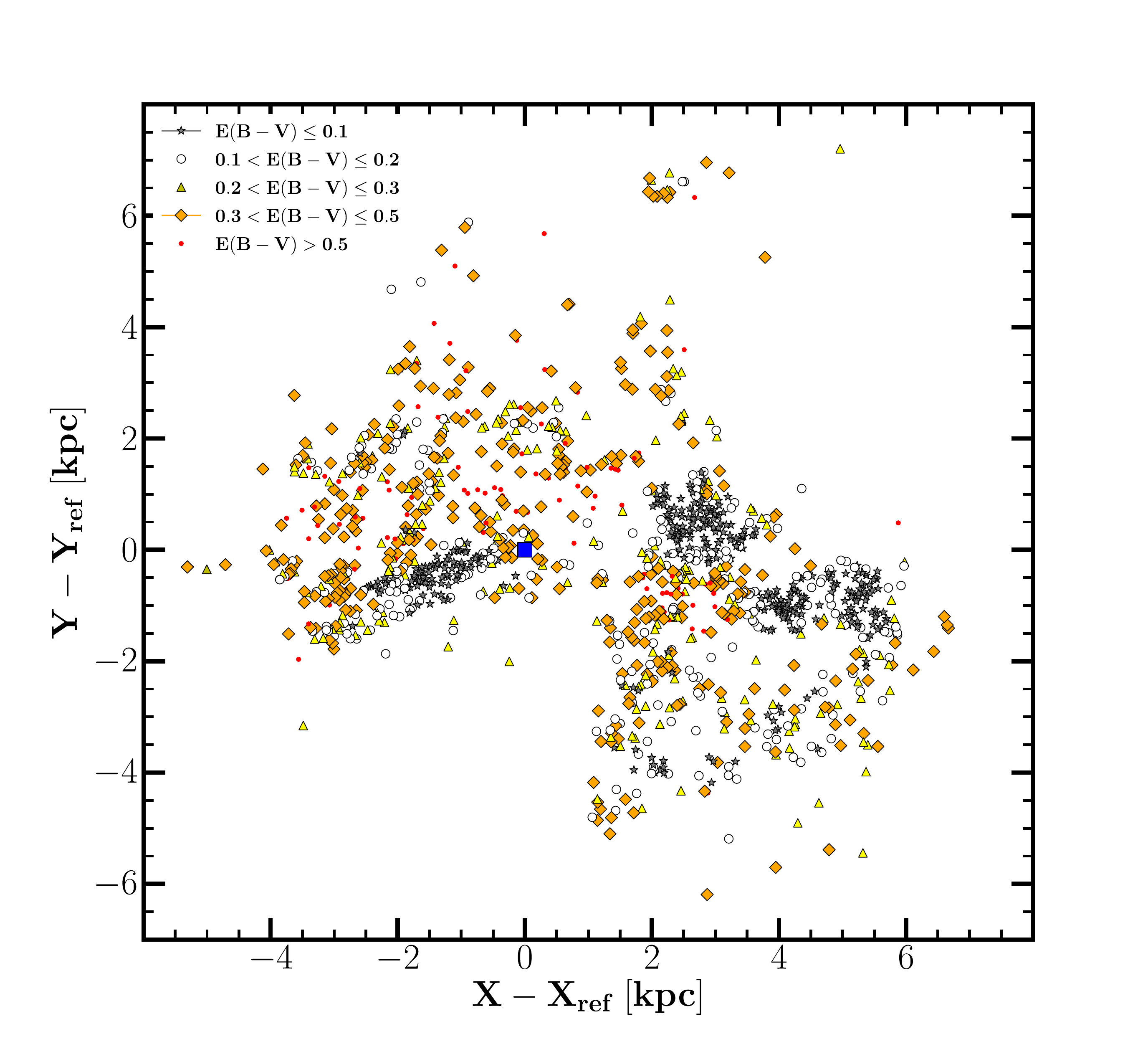}}}\\ 
\resizebox{.5\hsize}{!}{\rotatebox{0}{\includegraphics{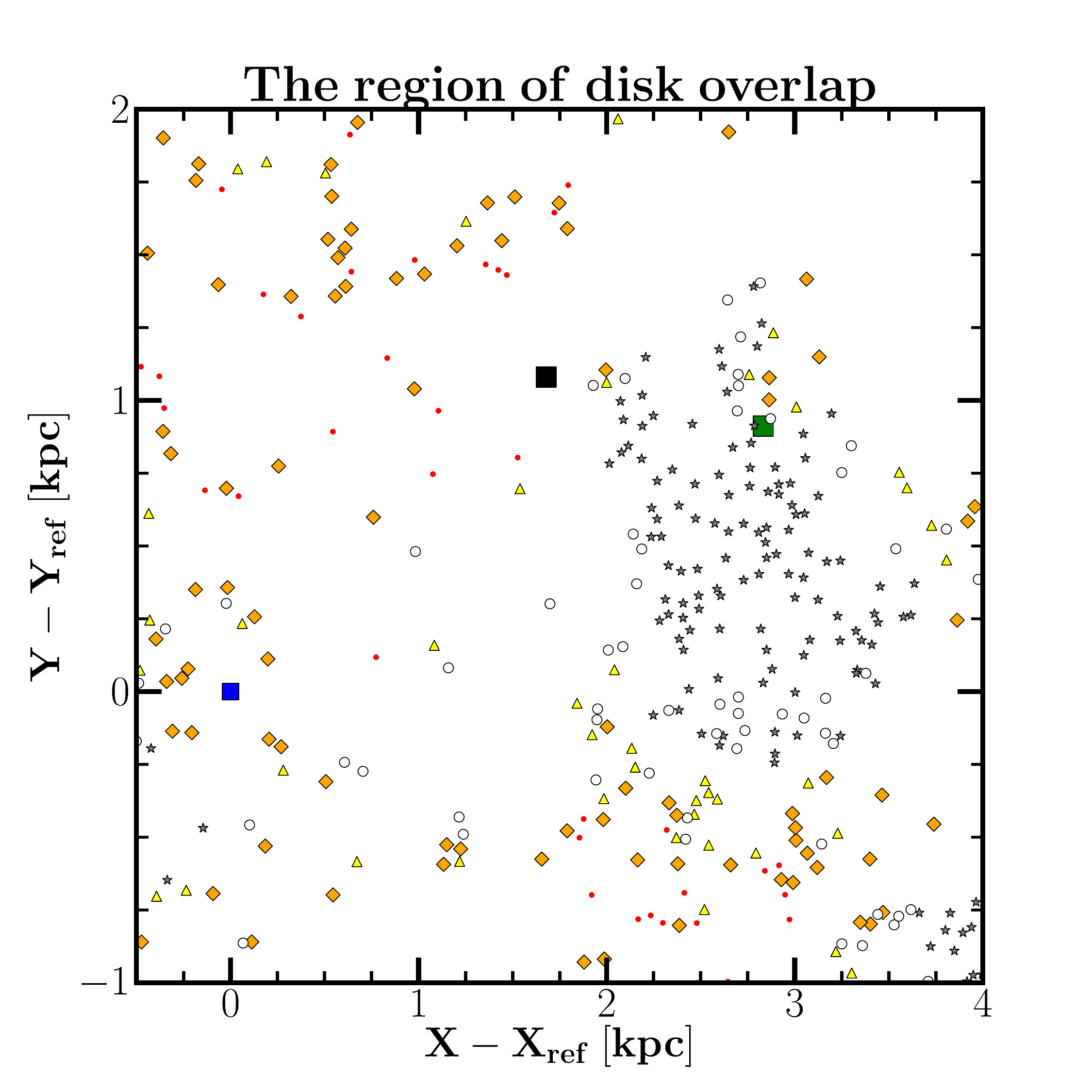}}}
\resizebox{.5\hsize}{!}{\rotatebox{0}{\includegraphics{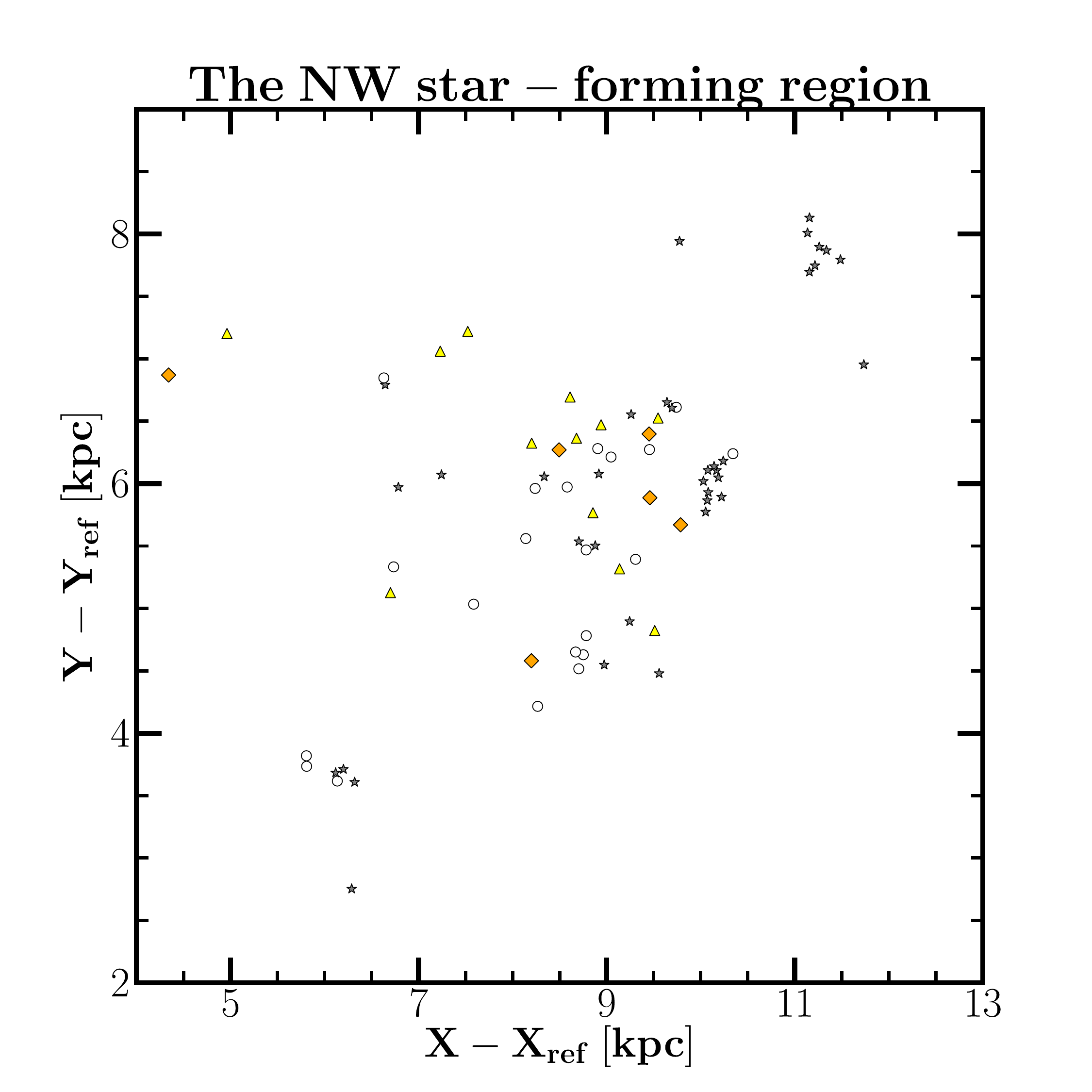}}}
\end{tabular}
\caption{Same as in Fig.\,\ref{fig:age-spatial-distr} but with the datapoints labelled as a function of the cluster
reddening. Regions such as the southeast spiral arm of NGC\,3690E, the disk overlap 
${\rm C+C'}$ and the HII regions northwest of NGC\,3690W are almost free of extinction.}

\label{fig:Ebv-spatial-distr}
\end{figure*}

Figure\,\ref{fig:age-spatial-distr}, Fig.\,\ref{fig:mass-spatial-distr} 
and Fig.\,\ref{fig:Ebv-spatial-distr} display the spatial distributions
of the SSCs split into different age, mass and extinction bins, respectively. Only clusters 
with high confidence photometry are included for clarity.
The blue square denotes the intersection of the distance between the two galaxy nuclei
with the boundary line used to separate their YMC catalogues. Insets in the lower panel zoom two regions of the interacting 
system: the disk overlap region (left) and the northwest star forming region (right). The latter is
analysed further in Section\,\ref{sec:SSCs-NWreg}.

In the case of Fig.\,\ref{fig:age-spatial-distr}, different age bins tend to overlap spatially. Such a random distribution 
may be an imprint of ongoing cluster migration due to
the strong tidal shocks generated through the merging of NGC\,3690E with NGC\,3690W. Nevertheless, 
the region of disk overlap
${\rm C+C'}$ as well as the southern spiral arm of NGC\,3690E host a siginificant number of clusters younger than 50\,Myr, some of them as young 
as $5-7$\,Myr. This is consistent with the mid-IR spectral features of these star-forming regions 
which also host supergiant HII regions \citep{2000ApJ...532..845A,2009ApJ...697..660A}.

In contrast, a clear pattern is observed in the spatial distribution of the cluster population
with respect to their masses. YMCs with masses ${\rm M > 10^{6} \msun}$
tend to be located near the galaxy nuclei but as we move away from the inner regions,
we start to observe the less massive cluster candidates. Though detection
of the faint and low-mass clusters is likely to be affected by the relatively
high-background of the nuclear regions, such a
pattern is also in agreement with \citet{1991ApJ...366L...1S} previous findings: they have recorded large 
quantities of dense molecular gas in the inner field of \arp. Such a concentration is reported to 
be an ideal birthsite of the very massive YMCs. Simulations by \citet{1996ApJ...471..115B}
also predict that collisions between gas-rich galaxies displace significant amounts of molecular
gas into the nuclear regions of the interacting \arp.

Finally, the southeast spiral arm of NGC\,3690E, the disk overlap region ${\rm C+C'}$ as well as the HII regions 
northwest of NGC\,3690W are populated by SSCs with relatively low visual extinction, i.e.\,${\rm E(B-V) \leq 0.1}$. 
A significant number of these clusters have ages between $\approx 7-50$\,Myr old. They survived 
the gas expulsion phase where the surrounding dust and ionized gas have been blown away, thus allowing them
to become optically visible. The majority of the most extinguished star clusters, however, are within the obscured
nuclear starburst regions. Such a distribution is consistent
with the properties of the derived extinction map in Fig.\,\ref{Avmap-arp}.

\begin{figure}
\centering
 \begin{tabular}{c}
\resizebox{1.\hsize}{!}{\rotatebox{0}{\includegraphics[trim= 0.2cm 0cm 1cm 0cm, clip]{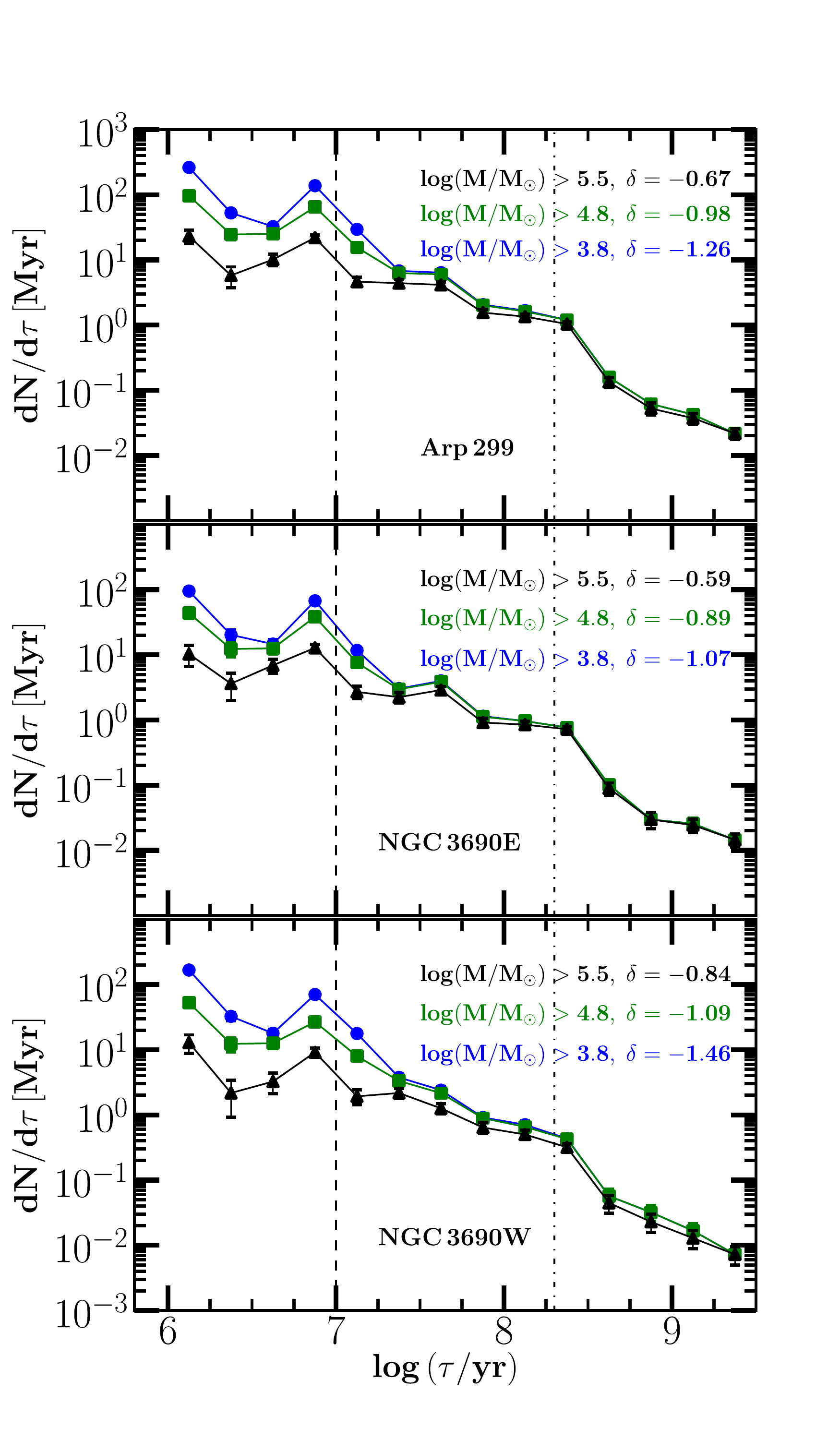}}}
\end{tabular}
\caption{\small The cluster age distribution dN/d$\tau$ as a function of age for \arp~({\em top}),
NGC\,3690E\,({\em middle}) and NGC\,3690W ({\em bottom}). Three different lower mass limits
were considered to derive various formation/dissolution rates of the observed clusters. 
Blue circles, green squares and black triangles
correspond respectively to the age distributions of star clusters more massive than
${\rm 10^{3.8}\,\msun,\,10^{4.8}\,\msun,\, and\,10^{5.5}\,\msun}$.
The vertical lines represent the borderline of the three different age ranges as 
discussed in Section\,\ref{sec:Cfqcy}. The values of the slope $\delta$ from fitting a power-law 
distribution to the $10 - 200$\,Myr age range are also included.
}
\label{fig:CFH-arp}
\end{figure}

\subsection{Age distributions of the star clusters}\label{sec:Cfqcy}
Figure\,\ref{fig:CFH-arp} shows the number of clusters per time interval
with the data binned by 0.25 in log$\,\tau$.
The upper panel depicts the dN/d$\tau$ shape 
for the entire system of \arp. The middle and bottom panels
represent the age distributions for NGC\,3690E and NGC\,3690W, respectively.
Lower mass limits of ${\rm 10^{3.8}\,\msun}$\,(blue circles), ${\rm 10^{4.8}\,\msun}$\,(green squares),
\,and\,${\rm 10^{5.5}\,\msun}$\,(black triangles) were adopted to establish cluster formation rates
of three different mass-limited catalogues. Dashed and dotted-dashed vertical lines mark
the timescale at 10\,Myr and 200\,Myr, respectively.

\subsubsection{Star cluster evolution during the first 10\,Myr}

For clusters younger than log$\,\tau = 7$, the sample is complete down to a mass cutoff 
of ${\rm \approx 10^{3.8}\,\msun}$. By omitting the peak around log\,$\tau \sim 6.8$,
the age distribution rapidly decreases by $\approx 1$\,dex 
during this first 10\,Myr of the
cluster lifetime. This trend reflects the
impact of gas expulsion (infant mortality, \citealt{2003ARA&A..41...57L}) combined with any
contamination from gravitationally unbound associations \citep{2011MNRAS.410L...6G}. The bump
is ignored as it is less likely to be real due to
the appearance of RSGs around that timescale. We also notice that an increase in the value of
the lower mass limit results in a flatter distribution
of the CFR. 
Mass-limited catalogues should be interpreted carefully since there is a trade-off between 
excluding less massive clusters and analysing a complete sample for a certain age range.
Nevertheless, the flat distribution may also indicate
that SSCs with masses ${\rm \gtrsim 10^6\,\msun}$ are less affected by the early disruption
mechanisms of the embedded phase.

Figure\,\ref{fig:CFH-arp} also suggests that the
CFR has increased \textcolor{black}{continuously} over the last $5-10$\,Myr. With a current high SFR of 
\textcolor{black}{$\sim$\,86\,${\rm M_{\odot}\,yr^{-1}}$ \citep{2017MNRAS.471.1634H}},
the merging process between NGC\,3690E and NGC\,3690W
keeps inducing violent and extreme
starburst episodes. It is thus highly probable that both cluster and star formation rates will gradually
increase as the two galaxies collide \citep{2009ApJ...701..607B}.
\subsubsection{dN/d$\tau$ analysis of the 10 - 200\,Myr age range}\label{subsec:dNdt}
We have a complete sample between $10 - 200\,{\rm Myr}$ by applying a mass cutoff 
of ${\rm \approx 10^{4.8}\,\msun}$ to the catalogue. There is a weak peak around log$\,\tau \approx 7.7$ ($\tau \sim 50$\,Myr)
that persists even when we vary the logarithmic age-binning or we use variable-size bins 
recommended by \citet{2005ApJ...629..873M} to draw
the age distributions.
The upper panel of Fig.\,\ref{fig:age-hist} exhibits the same trend around that age.
The bump is mostly prominent in the dN/d$\tau$ distribution of NGC\,3690E and
could be the imprint of an intense starburst activity that may have occured $\approx$\,50\,Myr ago.
Such a value is consistent with the starburst ages derived by \citet{2012ApJ...756..111M} from 
the SED fits for \arp: 45\,Myr and 55\,Myr for NGC\,3690E and NGC\,3690W, respectively. A close encounter 
between the galaxy pairs of the merging system could have displaced/disturbed
a large amount of gas and dust and thus had favored the formation of
YMCs. Nevertheless, \citet{2005A&A...431..905B} cautioned that SED fitting models adopted to output the 
cluster ages can 
also introduce artefacts in the age distributions of the clusters. We should therefore consider studying
the kinematics of gas and dust of \arp~over its lifetime to further investigate the origin of
such a peak. 

The sample is  likely
to be free of gravitationally unbound systems above 10\,Myr (Adamo et al. 2015). 
Hence, a power-law fit should help
in accurately deriving the cluster dissolution rate. \textcolor{black}{The values of $\delta$  
as a function of the mass-limited samples are given in the legend of each panel in
Fig.\ref{fig:CFH-arp}. We focus our analysis
using} the high mass cutoff
of ${\rm 10^{5.5}\,\msun}$ to avoid completeness bias introduced by
SSCs from the inner regions (see the dashed line in Fig.\,\ref{fig:age-mass}).
The slopes $\delta$ are respectively equal to $-0.59$ and $-0.84$ in the case of NGC\,3690E and NGC\,3690W. 
\textcolor{black}{By adopting a variable binning, the values of $\delta$ are $-0.69$ and $-0.86$, respectively.}
We record $\delta = - 0.67$ by considering the whole system. Such values are consistent with 
those of \citet{2011A&A...529A..25S} and \citet{2014MNRAS.440L.116S} who also considered mass-limited samples
but with a slightly longer time interval ($10 - 300/500\,{\rm Myr}$).
The steeper age distribution of 
the western component suggests that its cluster population endures stronger disruption mechanisms
than the YMCs hosted by NGC\,3690E. Studies of the CLFs/CMFs will be used for consistency checks.

\subsubsection{A rapid decline after 200\,Myr}
Apart from the significant evolutionary fading of old clusters, observational incompleteness becomes an issue at older ages
\citep[e.g.][]{2013MNRAS.430..676B, 2018ASSL..424...91A}. 
The rapid decline just after 200\,Myr is thus not surprising and most probably independent of the environment. 
This is the reason why we decided not to fit a power-law to the overall age distribution but rather restricted our analysis to the 
$10 - 200$\,Myr time interval.

\subsection{$UBI$-band cluster luminosity functions}\label{sec:CLF}
\begin{figure*}
\centering
\resizebox{.95\hsize}{!}{\rotatebox{0}{\includegraphics{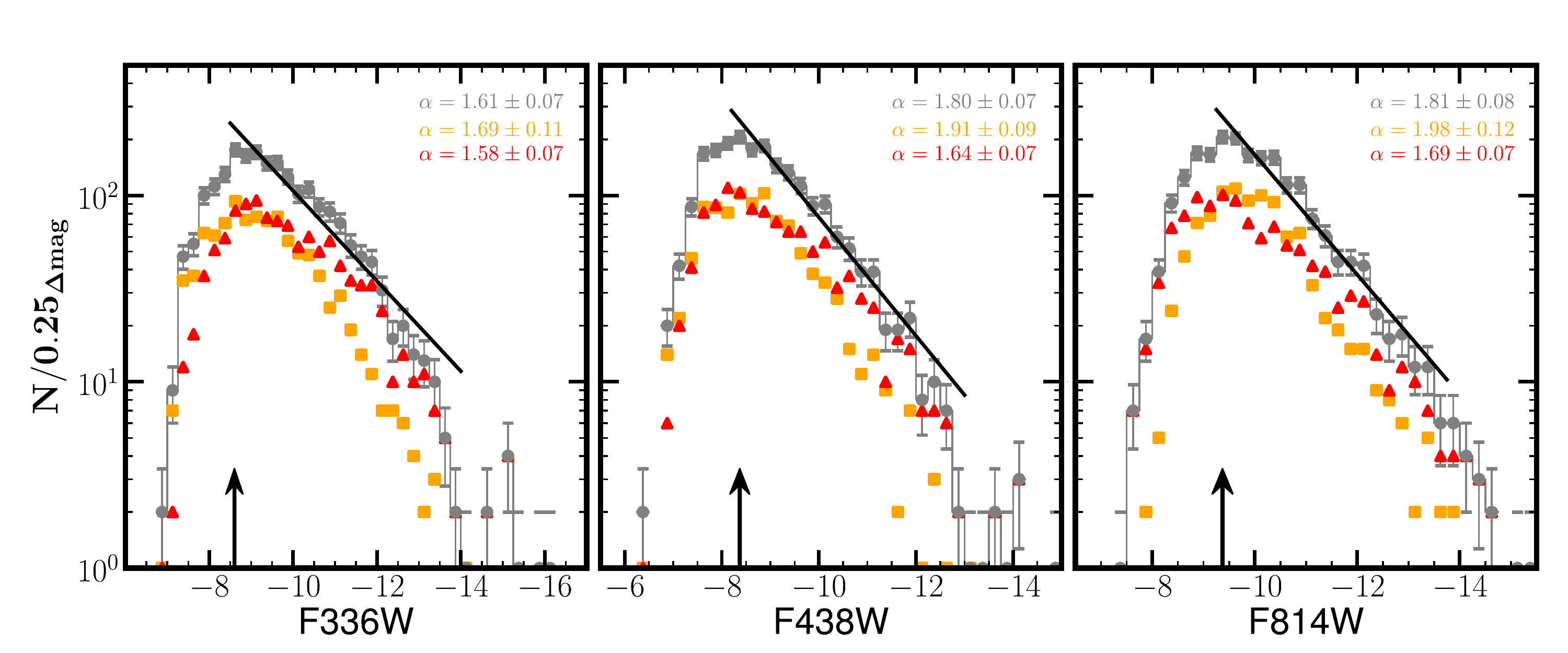}}}\\
\resizebox{.95\hsize}{!}{\rotatebox{0}{\includegraphics{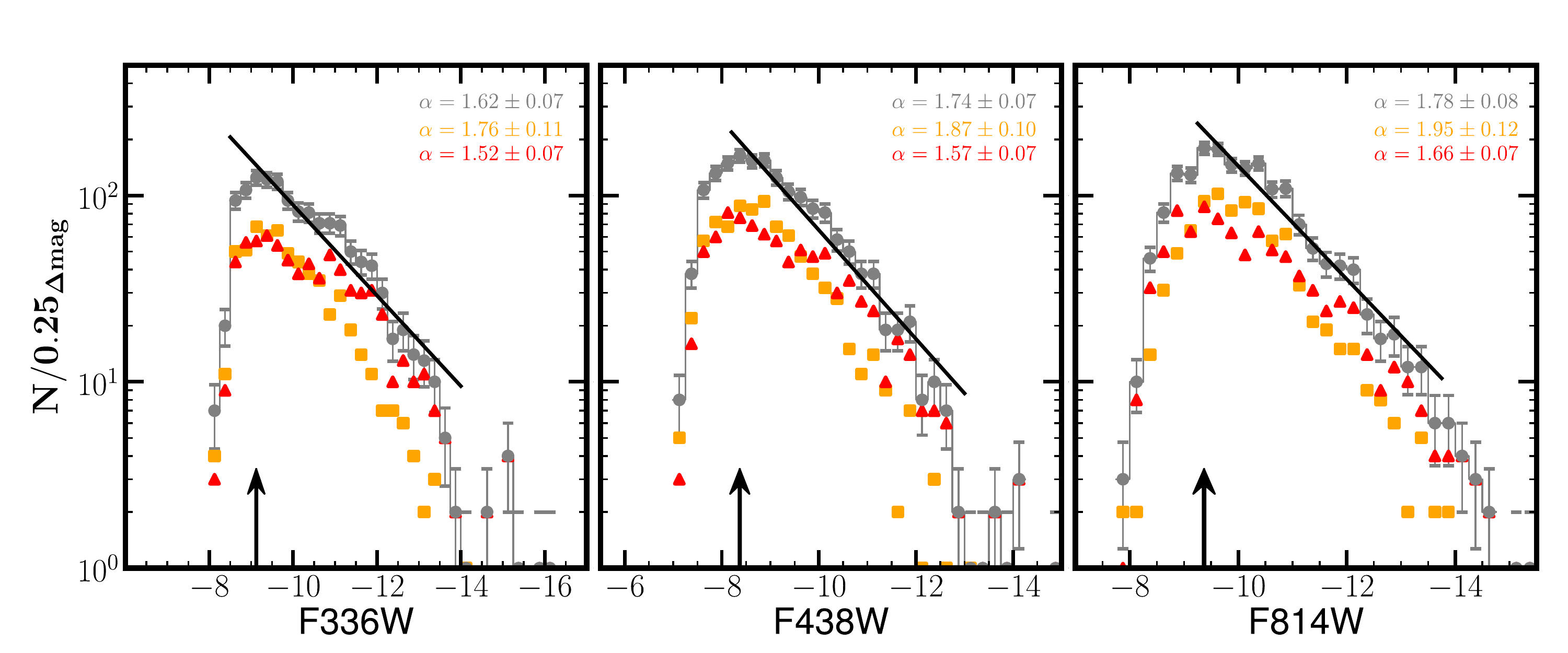}}}\\

\caption{\small {\em Top: }$U$- ({\em left}), $B$- ({\em middle}), $I$- ({\em right}) band CLFs of 
Arp\,299 and its galaxy components using the common sources in all three filters. The solid lines represent
the fit to a power-law distribution of the whole system LF \textcolor{black}{(grey circles)}. The vertical arrows indicate
\textcolor{black}{the critical magnitudes $M_{\rm cl,\,\lambda}$ until which we performed a power-law fit to the data. Orange squares and
red triangles respectively represent the CLFs of the east and the west components.} 
{\em Bottom: } Same CLFs as in the top panel but
using star clusters that were selected at each filter with $\sigma \leq 0.20$\,mag
and masses \textcolor{black}{${\rm M > 10^{3.8} M_{\odot}}$}. The power-law slopes $\alpha$ get steeper at 
redder filters for both components. The datasets of NGC\,3690W present
an underlying truncation at the faint end of the CLFs.  
In the case of NGC\,3690E, such a trend is weakly observed in the $I$-band LF.}

\label{fig:ssc-lfs}
\end{figure*}

\begin{figure*}
\centering
\resizebox{.95\hsize}{!}{\rotatebox{0}{\includegraphics{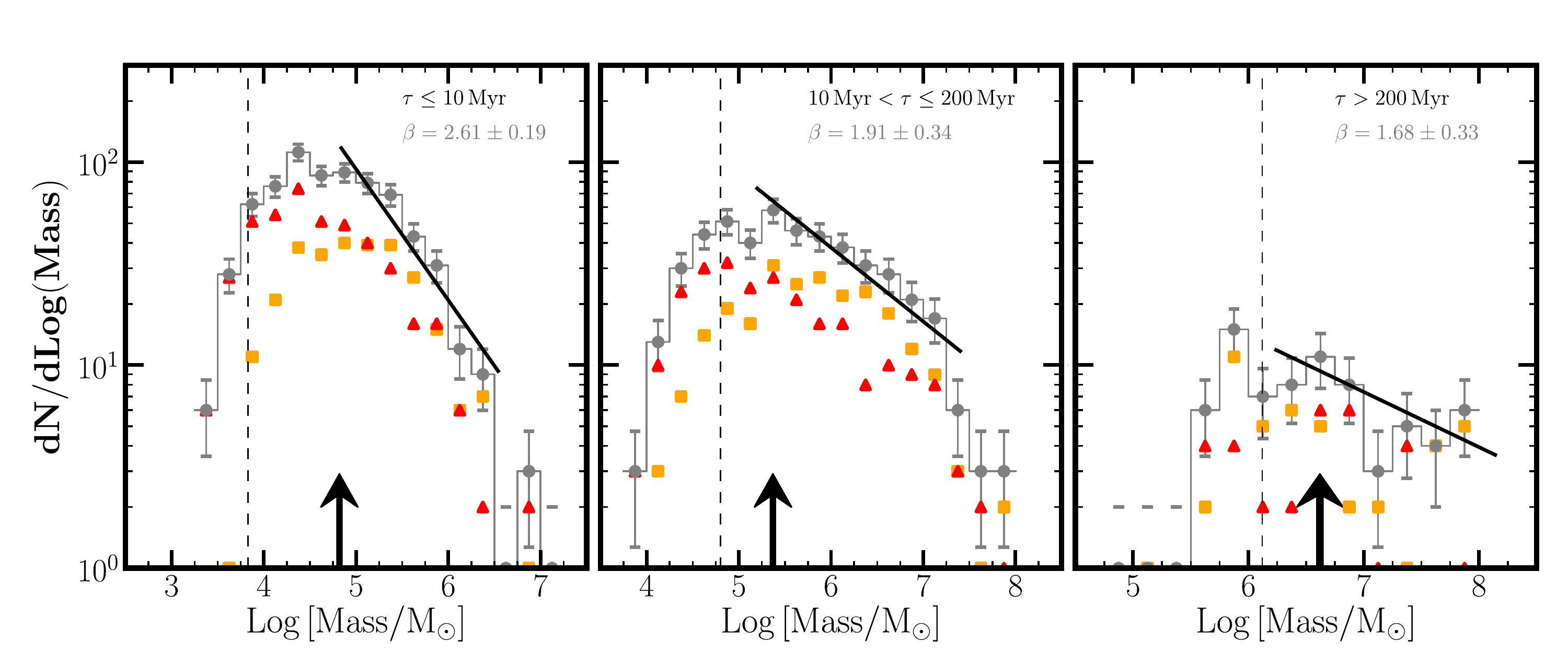}}}\\
\caption{\small The cluster mass functions of \arp\,(grey circles), NGC\,3690E
(orange squares), and NGC\,3690W (red triangles) for three different age
bins. Only SSCs with $\sigma \leq 0.20$\,mag are being considered to avoid any
photometric bias. The dashed lines show approximate values of the completeness limits
estimated from the age-mass plane. The solid lines represent the 
fit to a power-law distribution of \arp~MFs until a certain 
critical mass ${\rm M_{cl}}$ labelled as an arrow. The power-law slope $\beta$
flattens as we move to older age bins. CMFs of both components
show an underlying truncation/turnover, especially for 10\,Myr\,$ < \tau \leq$\,200\,Myr. 
}
\label{fig:ssc-mfs}
\end{figure*}

Before analysing the CMFs, we first derive the associated $UBI$-band CLFs
using a constant bin size. We fit a power-law function of the form ${\rm dN \sim  L^{-\alpha} dL}$
to the data, where $\alpha$ is the power-law slope. Figure\,\ref{fig:ssc-lfs} plots the 
derived $U$-band (left), $B$-band (middle) and $I$-band (right) CLFs of Arp\,299.
Top panels use all candidates in the cross-matched catalogue (i.e. 2182 sources), whereas
the bottom ones consider 
\textcolor{black}{1318\,($U$), 1680\,($B$), 1763\,($I$)} selected sources with $\sigma \leq 0.20$\,mag
and masses \textcolor{black}{${\rm M > 10^{3.8} M_{\odot}}$} in each filter. With such robust criteria, photometric biases
and stochastic fluctuations should not affect the second set of the luminosity functions. 
\textcolor{black}{Grey circles, orange squares, and red triangles} were used to draw the CLFs of \arp, NGC\,3690E and 
NGC\,3690W, respectively. The vertical arrows indicate a certain magnitude limit referred to as
\textcolor{black}{critical magnitude or $M_{\rm cl,\,\lambda}$ 
 where we only fit the clusters with bright luminosities up to 
that magnitude. Such a parameter corresponds to the bin 
before which the faint-end of the CLF starts to flatten or to exhibit a steep drop.} 
The fitted power-law functions of the interacting system are represented by 
the black solid lines.

In all three filters, the CLFs of \arp~generally follow a 
single power-law distribution with slopes below $\alpha = 2$, ranging between $1.61 - 1.81$. 
\textcolor{black}{Using a variable bin size, such a range becomes $1.52 - 1.68$ which 
implies that the choice of binning does not introduce bias.
By investigating the effect of binning, we also recorded different values of 
the critical magnitudes. A variation in the values of $M_{\rm cl,\,\lambda}$ only 
changes the slope $\alpha$ by $\sim\,0.10$.}
In addition, the values of
$\alpha$ from the two sets of CLFs (top vs. bottom panel of Fig.\,\ref{fig:ssc-lfs}) are similar within 
their uncertainties: e.g. $\alpha = 1.81 \pm 0.08$ 
(cross-matched sources) vs. \textcolor{black}{$\alpha = 1.78 \pm 0.08$} ($I$-band selected SSCs) in the $I$-band.
The interacting LIRG thus has flatter indices than the average slope associated with normal spiral galaxies
($\alpha \approx 2.4$, see e.g.\,\citealp{2014AJ....147...78W}). \citet{2013MNRAS.431..554R} 
ruled out resolution effects for galaxies with distances below 100\,Mpc. They suggested that a 
smaller value of the index $\alpha$ could result from a mass and/or environment-dependent disruption hapenning
within the galactic field of strongly star-forming galaxies such as LIRGs. Recent analyses by 
\citet{2016MNRAS.462.3766C} support such arguments; they reported a tight correlation between the galaxy environment and 
the power-law slopes of star-forming regions. 

Comparisons between the CLFs of the two galaxy components indicate that:
\begin{enumerate}

\item indices get steeper  with an increasing wavelength in both cases. 
\citet{2008A&A...487..937H}, \citet{2012ApJ...752...95J} and \citet{2014AJ....148...33R} among many others 
also found the same pattern in their work. Such a variation is due to 
the evolutionary fading of the star clusters with time \citep{2010ASPC..423..123G}.

\item regardless of the filter used, NGC\,3690W CLFs exhibit either a bend/truncation or a flattening below $M_{\rm cl,\,\lambda}$.
Such results are consistent with the steep  ${\rm dN/d{\tau}}$ age distribution in Section\,\ref{sec:Cfqcy}. 
Observational incompleteness alone cannot explain such a behaviour 
since the effects should be negligible on the data with magnitudes brighter than $M_{\rm cl,\,\lambda}$.
{\it Could it be an imprint of a mass/environment-dependent disruption?}

 \item NGC\,3690E is mostly fitted with a function of steeper slopes by $\approx 0.3$ in all filters. 
 One cannot entirely associate a size-of-sample effect to the trends since there are relatively more YMCs in NGC\,3690W
 than in the eastern component. Again, NGC\,3690W may be enduring stronger disruption mechanisms than its companion
 \citep{2008A&A...487..937H}.
\end{enumerate} 

Since clusters of different ages are binned together when constructing the LFs, analyses of the CMFs are
crucial to investigate the possibility of mass/environment-dependent disruption events.

\subsection{Cluster mass functions}\label{subsec:ssc-mfs}

\subsubsection{Properties of the very massive SSC candidates}\label{sec:massive-SSC}
Both shape and slope of the CMF are governed by the cluster mass range. 
It is therefore crucial to determine whether most of the very massive SSC candidates are 
complexes or that the environment of the interacting system favors their formation.
In the case of \arp, around 4\,percent of the cluster population have masses
${\rm M > 10^7\,M_{\odot}}$. They are mainly located in the nuclear regions 
with more than half of the population hosted by NGC\,3690E. Their
positions in the CMD are consistent with the results from SED fitting; their ages
are estimated to be older than 100\,Myr with an age median of $\sim\,200$\,Myr.
As already discussed in Section\,\ref{sec:distr-arp}, \arp~has the potential to form
massive SSCs in regions where dense molecular clouds are concentrated such as in the
galaxy centers. \citet{2006A&A...448..881B} and \citet{2017ApJ...843...91L}
also recorded star clusters with the same mass range. 
They suggested that the extreme star-forming environment offered by (interacting) LIRGs is the most likely interpretation.
This is consistent with the results from recent simulations of the formation of massive star clusters in major mergers
by \citet{2017ApJ...844..108M}.

We conclude that the concentration of very massive star clusters in the nuclear regions of \arp~is
evidence for {\it environmental dependence} of the cluster formation and that most of these YMCs
are less likely to be complexes. They could evolve to become present day globular clusters.
Nevertheless, we apply an upper mass cutoff of ${\rm M > 10^8\,M_{\odot}}$ to 
derive robust CMFs that should be free from blending effects. Three cluster candidates are excluded by setting
the above criteria.

\subsubsection{Constructing the CMFs}
Figure\,\ref{fig:ssc-mfs} shows the resulting CMFs of \arp\,(black points), NGC\,3690E (blue squares), 
and NGC\,3690W (green triangles) by considering only datapoints with $\sigma \leq 0.20$\,mag
and ${\rm M < 10^8\,M_{\odot}}$.
The SSC catalogue was split in subsets
as a function of the cluster ages for an optimal assessment of the mass distribution and hence a better understanding
of the cluster disruption mechanisms.  The age ranges are: $\tau \leq 10$\,Myr (left panel), 
$10\,{\rm Myr} < \tau \leq 200$\,Myr (middle), 
and $\tau > 200$\,Myr (right). The vertical lines mark approximate values of the cluster mass completeness
limits using the age-mass diagram in Fig.\,\ref{fig:age-mass}. A power-law distribution of the form
${\rm {dN} \sim M^{-\beta}\,dM}$ was fit to the CMF high-mass end until a critical mass 
${\rm M_{cl}}$, \textcolor{black}{ which is determined in a similar way as the critical magnitude in Section\,\ref{sec:CLF}.
The binning method defines the parameter ${\rm M_{cl}}$ and its value varies between 
log${(\rm M_{cl}/M_{\odot}}) \sim 5.24 - 5.63$ as we
adopt different bin sizes or apply a variable binning when constructing the CMF with $\beta = 1.75 - 1.91$.}
The completeness limit, the values of ${\rm M_{cl}}$ and the slope $\beta$ at each time interval are listed
in Table\,\ref{tab:mf-slopes}. 

The mass distribution as well as the critical mass shift toward higher masses as we move to older age ranges
because of evolutionary fading and size-of-sample effect \citep{2003AJ....126.1836H}. 
As for the power-law slopes, they become flatter with an increasing
time interval: from $2.61 \pm 0.19$, $1.91 \pm 0.34$ and $1.68 \pm 0.33$ in the case 
of \arp. Since  ${\rm M_{cl}}$ is always more massive than
the completeness limit, the flattening in the slope $\beta$ cannot be entirely associated with observational 
incompleteness; rapid dissolution of the less massive clusters (${\rm M \lesssim 10^5\,M_{\odot}}$) could be 
another reason \citep{2003MNRAS.338..717B,2005A&A...441..117L}.

Finally, CMFs of $\tau \leq 200$\,Myr are suggestive of an underlying truncation/turnover at high mass 
end and could be potentially fit by a broken power-law or a Schechter function. The mass distribution
of NGC\,3690W mostly deviates from a pure power-law function. In fact, the corresponding $I$-band CLFs 
(i.e.\,splitting the catalogue as a function of the cluster ages) in \textcolor{black}{Fig.\,\ref{fig:lfs-arp}}  
show similar trends well above the detection limit at $M_I = - 9.12\,{\rm mag}$. The slopes $\alpha$ also
decrease with older ages: from $\alpha = 2.08 \pm 0.08, 1.73 \pm 0.07, {\rm to}\,1.61 \pm 0.08$.
These results support the idea that a CLF is a mere reflection of the CMF \citep{2006A&A...450..129G}.

\begin{table*}
\begin{scriptsize}
\caption[The CMFs]{Characteristics of the star cluster mass functions in Arp\,299 and its two components.}
\label{tab:mf-slopes}
\begin{center}
\begin{tabular}{cccccccc}
\hline 
  \noalign{\smallskip}
  
Age range 	&  Comp.lim & Nb.SSCs & ${\rm M_{cl}}$ & ${\rm M_{\star}}$ &  $\beta$ &  $\beta_{{\rm A}}$ &  $\beta_{{\rm B+C}}$	\\
(Myr)	        &   log(${\rm M/M_{\odot}}$) & (${\rm SNR} > 5$)  &    log(${\rm M/M_{\odot}}$) & log(${\rm M/M_{\odot}}$)  &  &   &    \\
(1)	        &   (2)	   &	(3)	       & (4) & (5)	& (6) &  (7) & (8)	\\
  \noalign{\smallskip}
\hline \hline
  \noalign{\smallskip}
$\tau \leq 10$ & 3.83& 650 & 4.87 & \textcolor{black}{5.47}  &  2.61$\pm0.19$ & 2.31$\pm0.33$ &  2.45$\pm0.17$ \\
$10 - 200$     & 4.80& 377 & 5.37 & \textcolor{black}{6.19}  & 1.91$\pm0.34$ & 1.84$\pm0.23$ &  1.97$\pm0.22$  \\
$ \tau > 200$  & 6.12&  49 & 6.62 &  -    & 1.68$\pm0.33$ & 1.19$\pm0.21$ &  2.21$\pm0.20$  \\
\noalign{\smallskip}
\hline\\
\noalign{\smallskip}
\multicolumn{8}{@{} p{13.5cm} @{}}{\footnotesize{\textbf{Notes. }Column 1: SSC age
range to construct the CMF; Column 2: completeness mass limit 
estimated from the age-mass diagram; Column 3: number of SSCs with a magnitude error $\sigma \leq 0.20$\,mag
within the time interval and above the mass limit;  Column 4: critical cluster mass until 
which we performed the power-law fit; Column 5: Schechter characheristic/truncation mass; 
Columns $6 - 8$: single power-law slopes of \arp, NGC\,3690E and 
NGC\,3690W, respectively.}}

\end{tabular}
\end{center}
\end{scriptsize}
\end{table*}

Further analyses of the CMFs for the $10 - 200$\,Myr age range
are done in Section\,\ref{subsec:MF-mid} to check whether
the truncation is a mere result of a size-of-sample effect or does involve any physical explanation.
This time interval is very critical because the star
cluster's chance for a long-term survival after escaping infant mortality will be at play
during this period. Moreover, MFs of the younger ages ($\tau < 10\,{\rm Myr}$)
could be contaminated by unresolved unbound stellar associations \citep{2011MNRAS.410L...6G} while the ones with ages above 200\,Myr 
are likely affected by fading and evaporation \citep{2003AJ....126.1836H}.

\subsubsection{Applying a Schechter fit to the MF}\label{subsec:MF-mid}
 
\begin{figure}
\centering
\resizebox{1.\hsize}{!}{\rotatebox{0}{\includegraphics[trim= 0.cm 0cm 0cm 0cm, clip]{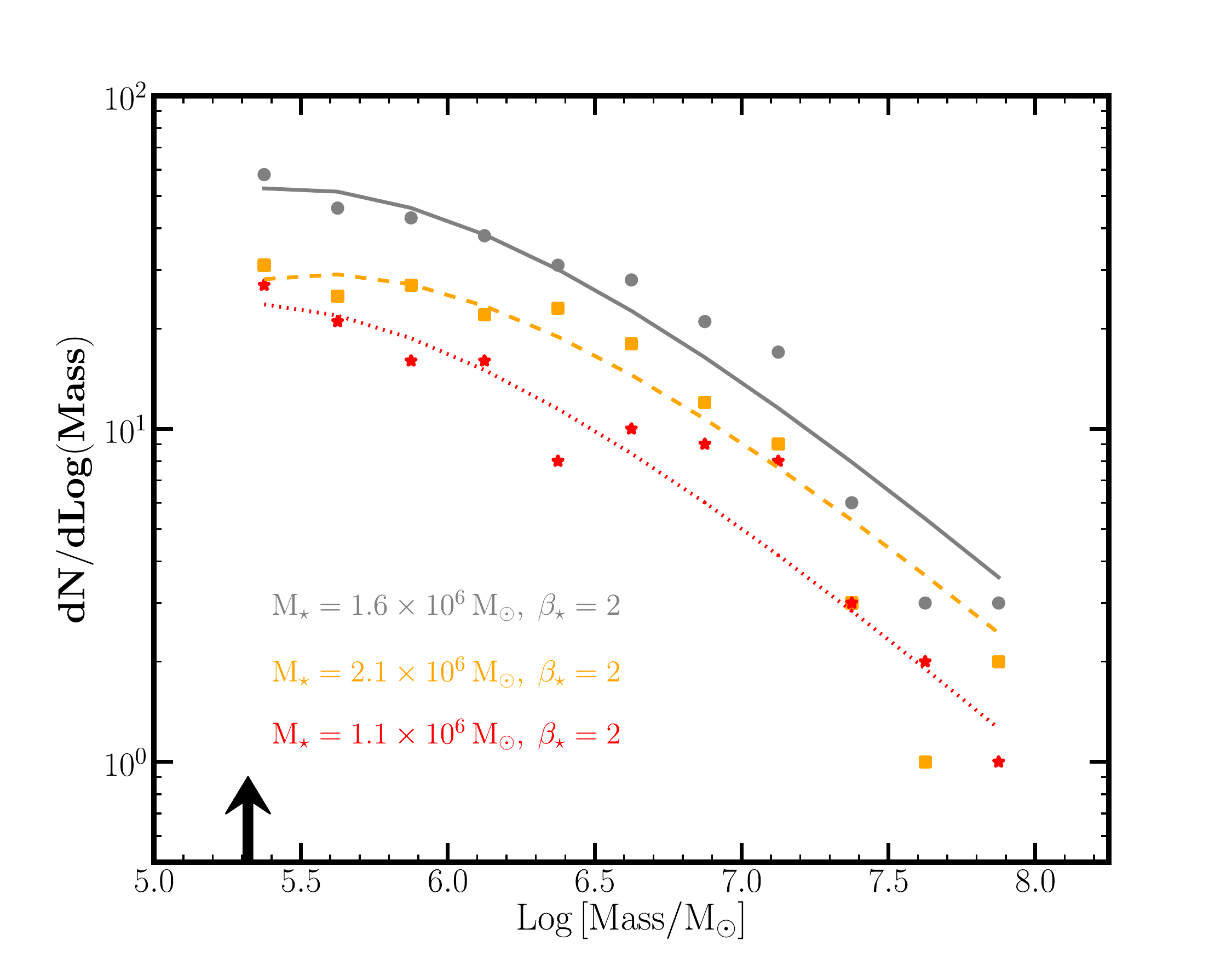}}}\\
\caption{\small The mass distribution of star clusters with ages between 10 - 200 Myr old. Same labels as in Fig.\,\ref{fig:ssc-mfs}. Schechter distributions are fit to 
the CMFs of \arp~(solid line) and its two components NGC\,3690E (dashed line) and NGC\,3690W (dotted line) 
\textcolor{black}{until the critical mass of ${\rm log(M_{cl}/M_{\odot}) = {5.37}}$. 
The fixed slopes ${\beta_{\star} = 2}$ and the derived characteristic masses ${\rm M_{\star}}$ are included as insets in the plot.}}
\label{fig:MF-schec}
\end{figure}

We also fit \textcolor{black}{a Schechter distribution of faint-end slope $\beta_{\star} = 2$
to the CMF with ages between $10 - 200$\,Myr old
and masses above the critical mass of ${\rm log(M_{cl}/M_{\odot}) = 5.37}$.}   
Figure\,\ref{fig:MF-schec} shows the resulted fits overplotted on top of the datapoints.
While the bend of a Schechter function occurs around ${\rm 2.1 \times 10^{5}\,\msun}$
for normal spiral galaxies \citep{2009A&A...494..539L}, a more massive truncation mass close to 
${\rm 10^{6}\,\msun}$ is required for starbursts and merger galaxies such as LIRGs 
\citep[e.g.][]{2008MNRAS.390..759B,2017ApJ...843...91L}.
This is consistent with the values we record for \arp\,(solid line) where 
\textcolor{black}{${\rm M_{\star} = 1.6 \times 10^{6}\,\msun}$}.
The varying truncation mass is thought to be governed by the local properties of the host environment
and \textcolor{black}{it has been reported} to be directly linked with the SFR surface density 
\citep{2015MNRAS.452..246A,2017ApJ...839...78J}.
Schechter characteristic masses are \textcolor{black}{${\rm 2.1 \times 10^{6}\,\msun}$ and 
${\rm 1.1 \times 10^{6}\,\msun}$} for NGC\,3690E (dashed) and NGC\,3690W (dotted), respectively.  
The relatively high ${\rm M_{\star}}$ of NGC\,3690E compared to that of the western component 
is consistent with this galaxy hosting more than half of the very massive clusters 
of the system, although it may also result from the large concentration
of high gas pressures in the environment of NGC\,3690E.

\subsubsection{Cluster mode of disruption in \arp}\label{subsec:MF-interp}

Features of the derived CMFs are in agreement with the cluster mass spatial distribution,
the age distribution, and the luminosity functions. These tools denote at least a strong signature of 
an environmental dependence of the cluster mode of disruption. Further investigations are 
reported in Section\,\ref{sec:env-dep}.
A mild mass-dependent disruption is also suggested. Such physical effects are indeed combined
with internal effects from the cluster evolutionary processes 
such as infant mortality, evolutionary fading and evaporation.

The value of the power-law slope $\beta$ decreases
 by $\sim 0.70$ moving from the youngest age range to the 10 - 200 Myr
time interval. The corresponding slopes of the $I$-band CLFs (Fig.\,\ref{fig:lfs-arp}) also differ
by $\sim 0.35$. \citet{2009MNRAS.394.2113G} interprets this decrease in the value of $\beta$ with an increasing age range
as a signature of a mass-dependent disruption mechanism. In addition, 
a mass-dependent disruption time
could also be the reason of the bend occuring at low masses of the CMFs since incompleteness bias is to be excluded.
In other words, the cluster
initial mass function defines the disruption time and hence the cluster survival chance. If not destroyed
during infant mortality at early ages, low mass star clusters are likely to be vulnerable by disrupting 
faster in such a scenario \citep[e.g.][]{2005A&A...441..117L}.

\subsection{The cluster formation efficiency}

In this work, a time interval  $\Delta t _1= 10 - 50$\,Myr was chosen over the $10 - 200$\,Myr range to estimate reliable cluster 
formation rate and  cluster formation efficiency $\Gamma$ of \arp ~(see Eq.\,\ref{eqn:CFE}). The inclusion of star clusters with ages
$\tau > 50$\,Myr old
certainly underestimates the value of $\Gamma$ because the YMCs could have already been affected by evolutionary processes. 
In addition, the star formation history over the past $\sim$\,50\,Myr is best represented by the current SFR, especially in the case of systems with ongoing starbursts such as LIRGs \citep{2016MNRAS.457L..24K}. Nonetheless, we also estimated $\Gamma$ by adopting $\Delta t_2 = 10 - 100$\,Myr for comparison. 
Only clusters with masses ${\rm M > 10^{3.83}\,M_{\odot}}$ were considered when computing the total mass of clusters with ages within $\Delta t$. 

Apart from the difficulty to accurately define regions that belong to each galaxy component of~\arp,
it is also not trivial to estimate their SFR without introducing non-negligible uncertainties.
Therefore, we decided to only derive the CFE of the whole system by considering 
\textcolor{black}{a ${\rm SFR = 86.1_{-18.4}^{+7.4} \,M_{\odot}\,yr^{-1}}$ retrieved from the galaxy
SED fitting performed by \citet{2017MNRAS.471.1634H}. For $\Delta t _1= 10 - 50$\,Myr, 
the value of $\Gamma = 19.4_{-1.5}^{+5.3}$\,percent. We did not use the SFR based on the galaxy infrared luminosity since 
both AGN and starburst activity contribute toward the total infrared luminosity 
and hence lead to an overestimate of the derived SFR.} Such a bias is however taken into
account during the SED modelling which output a SFR averaged over the past 50\,Myr. 
Had we considered $\Delta t_2 = 10 - 100$\,Myr, the value of 
\textcolor{black}{$\Gamma$ becomes $16.0_{-1.3}^{+4.4}$\,percent}. This slightly lower  
percentage arises because of the disruption effects endured by the 50 - 100\,Myr old clusters. 
Nevertheless, both values of $\Gamma$ are expected to slightly increase after correcting for biases such as incompleteness
due to the missing mass from low mass star clusters.

\begin{figure}
\centering
\resizebox{1.\hsize}{!}{\rotatebox{0}{\includegraphics{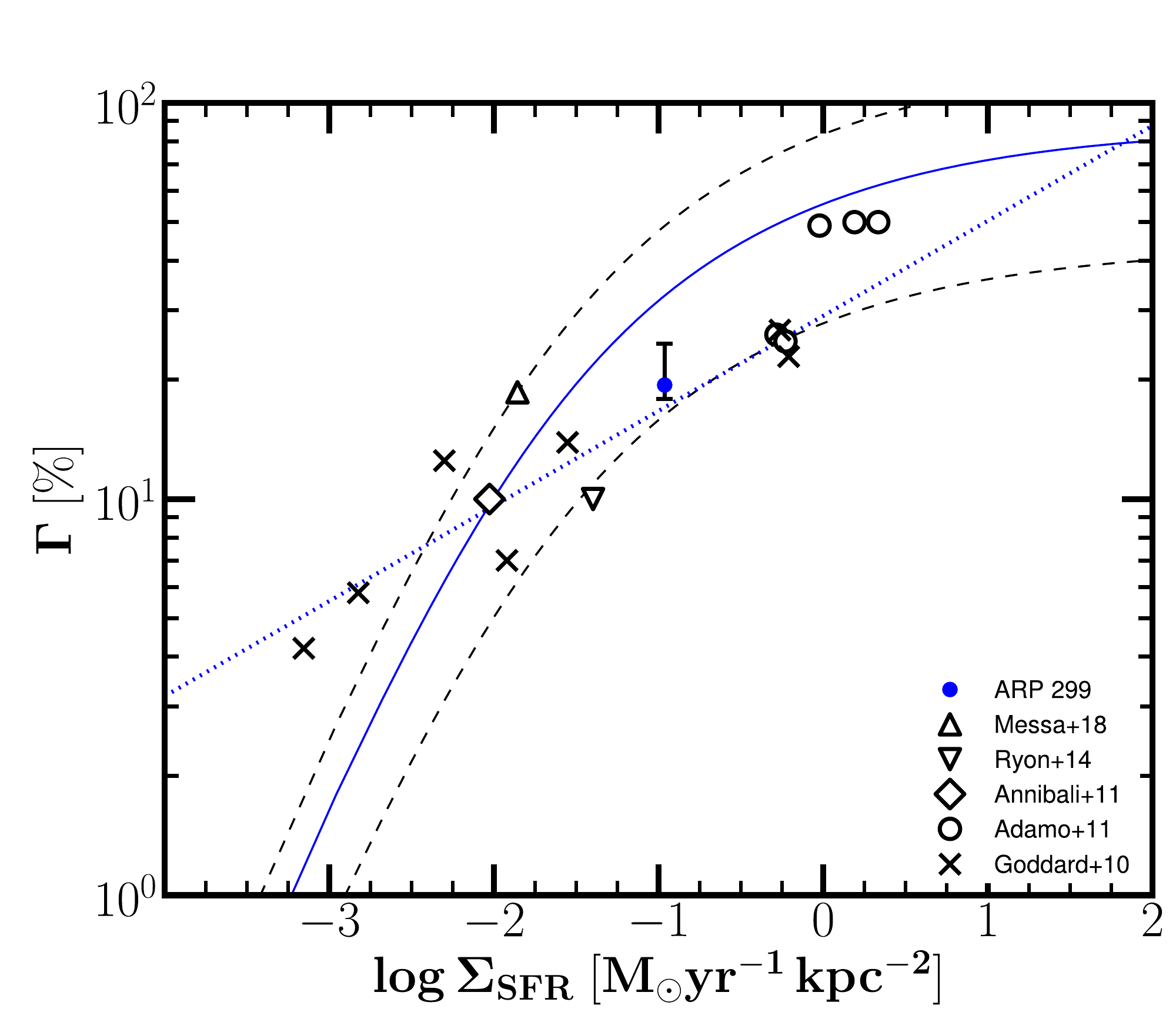}}}\\
\caption{\small The star formation occurring in bound clusters or $\Gamma$ plotted against the SFR density. 
\textcolor{black}{The blue point represents the derived value from this work using the time scale of $10 - 50$\,Myr
and a SFR estimated from galaxy SED fitting.}
The other datapoints (open markers) were retrieved from 
\citet{2010MNRAS.405..857G}, \citet{2011MNRAS.417.1904A}, \citet{2011AJ....142..129A},  \citet{2014AJ....148...33R} and \citet{2018MNRAS.473..996M}. The dotted line marks the best power-law fit of the relation according to \citet{2010MNRAS.405..857G}. The solid line represents the theoretical predictions from \citet{2012MNRAS.426.3008K} with the dashed lines within 3$\sigma$ uncertainty of the model.}
\label{fig:CFE-sfr}
\end{figure}

We also derived the SFR density ${\rm \Sigma_{SFR}}$ as a function of the galaxy SFR being used
and we get \textcolor{black}{${\rm \Sigma_{SFR} = 0.11\,M_{\odot}\,yr^{-1}\,kpc^{-2}}$.}
A similar approach to \citet{2011MNRAS.417.1904A} was adopted (plotting the galactic luminosity distribution as a
function of the galactocentric radius) to estimate the projected area of the starburst regions 
enclosed within a $\sim$\,\textcolor{black}{16}\,kpc radius.
A version of the CFE - ${\rm \Sigma_{SFR}}$ relation is presented in Fig.\,\ref{fig:CFE-sfr} where we plot results from this work 
(blue filled marker using  $\Delta t _1$) with data taken from the literature (open markers).  
Our results agree with the power-law model suggested by 
\citet{2010MNRAS.405..857G}. The blue point also lies within the 3$\sigma$ fiducial model of \citet{2012MNRAS.426.3008K}.

In the case of \arp, the fiducial model predicts a theoretical cluster formation efficiency $\Gamma_{th} \sim 33$\,percent for the given value
of ${\rm \Sigma_{SFR}}$. The incompleteness in the total mass contributes toward the discrepancy between the observed vs.\,theoretical
results.  Nevertheless, the observed CFE of \arp~is already $\sim 3 - 5$ times higher than the
typical value of $\Gamma$ in
gas-poor spiral galaxies such as NGC\,2997 \citep{2014AJ....148...33R} and NGC\,4395 \citep{2011A&A...529A..25S}.
In fact, it is comparable with the mean CFE $\sim 24.6 \pm 4.1$\,percent of the LIRG NGC\,3256 and two blue compact 
galaxies with slightly higher $\Sigma_{{\rm SFR}}$  \citep{2010MNRAS.405..857G,2011MNRAS.417.1904A}. These are not surprising since extreme conditions necessary to form dense GMCs are easily met in the star-forming regions of interacting systems such as \arp~\citep[e.g.][]{2017ApJ...844..108M}. Their high-pressure environments favor star formation to happen in bound stellar clusters. Our results thus provide further evidence
that the CFE is a parameter dependent on the galactic environments \citep{2010MNRAS.405..857G, 2012MNRAS.426.3008K, 2018ASSL..424...91A}. 

\subsection{Further analyses on the star cluster formation and disruption mechanisms}\label{sec:env-dep}

\subsubsection{SSC properties in inner vs.\,outer field}\label{subsec:var-reg}

\begin{figure*}
\centering
\resizebox{1.\hsize}{!}{\rotatebox{0}{\includegraphics[trim= 0.cm 0cm 0cm 0cm, clip]{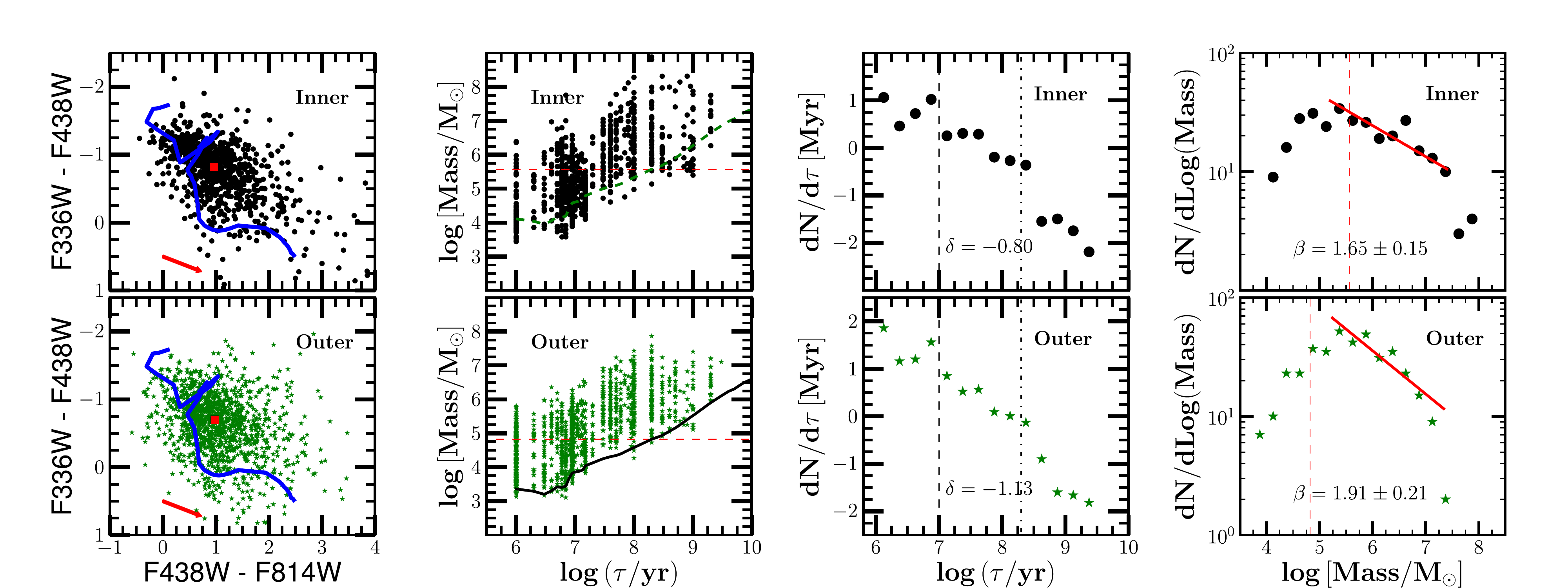}}}\\
\caption{\small {\em Top: } Properties of the SSC candidates located in the inner 
regions of  \arp ~(black points). The first column overplots the star cluster color-color diagram with
the {\tt Yggdrasil} SSP model (blue solid line).
The arrow and the red square indicate a reddening of $E(B-V) = 0.25$ and the median colors, respectively. 
The second column shows the cluster age-mass plane where the horizontal line marks the mass 
cutoff used to define a complete sample 
for clusters younger than 200\,Myr and the dashed line the 80 percent photometric detection limit.
The cluster age distribution is shown in the third column where the vertical lines mark the ages at 10 and 200\,Myr.
The cluster mass function for the $10 - 200$\,Myr age range is 
displayed in the last column where the red solid line represents a single power-law fit of the MF bright end 
\textcolor{black}{and the dashed line an approximate value of the completeness limit
estimated from the age-mass plane.}
{\em Bottom: }  Similar to the top panel, but for the outer regions (green stars) where the black
solid line corresponds to the 80 percent photometric detection limit.
A peak is observed around log($\tau) \approx$\,7.7 in both age distributions.
The inner field has a shallower CMF slope than the outer regions.}
\label{fig:prop-reg}
\end{figure*}

Figure\,\ref{fig:prop-reg} shows different properties of the SSCs hosted by the inner (upper panel, black circles)
and outer regions (lower panel, green stars) of the interacting system \arp. The color-color diagram (first column),
the age-mass plane (second),  the cluster 
age distribution (third) as well as the cluster mass function for the $10 - 200$\,Myr age range
(fourth column) are plotted for each field. 
We notice that the distributions
are not the same but (slightly) change as a function of the varying background, 
especially in the case of the age-mass plane and the CMF:
\begin{enumerate}
 \item {\it the color diagrams:} the ranges of the $U - B$ and $B - I$ colors are generally similar for both regions. 
The median color labelled as a red square only shifts by $\sim$\,0.13\,mag. Based on the CCDs, we cannot conclude whether the two regions 
exhibit a different age distribution over the past few tens of Myr as observed by \citet{2013AJ....145..137K}.
 
 \item {\it the age-mass diagrams:} the applied mass cutoff to draw a mass-limited 
sample for clusters younger than 200\,Myr changes by $\sim$\,1\,dex, i.e.\,from ${\rm 10^{5.5}\,M_{\odot}}$ down to 
${\rm 10^{4.8}\,M_{\odot}}$, when moving 
away from the nuclear regions. This is because of the relatively high-luminosity background of the inner 
field which challenges the detection of faint low-mass clusters or highly-extinguished sources in that region.
The inner field hosts 68 massive star clusters with log${\rm (M/M_{\odot}) > 7}$\,(8\,percent of its population) against 31\,(2\,percent)
candidates in the outer regions. There are however similar numbers of young and old populations in both regions:
around 65\,percent of the SSCs are younger than 30\,Myr and 14\,percent older than 200\,Myr.
 
 \item {\it the age distributions: } the dN/d$\tau$ shapes generally look similar for both regions. 
 The age distributions of the $1 - 10$\,Myr time interval are consistent with the ones plotted in Fig.\,\ref{fig:CFH-arp}:
 an increase of the cluster formation rate (though bias from unbound systems and the chimney around 10\,Myr should
be accounted for). A peak is  also seen around log$(\tau) \sim$\,7.7\,($\approx$\,50\,Myr)
which may indicate a past starburst activity. A power-law fit for the 10 - 200\,Myr age range results to $\delta = -0.80$ and
$-1.13$ for the inner and outer regions, respectively. We do not interpret these values since we applied different mass limits
to draw the two age distributions (see the horizontal line in the age-mass plane).

 \item {\it the mass functions:} we get slopes $\beta = 1.65 \pm 0.15$ (inner) and $\beta = 1.91 \pm 0.21$ (outer) by fitting a
 single power-law to the high end of the mass distributions.  Alternatively, Schechter or broken power-law fits can also 
 represent the CMF of the nuclear regions with an underlying truncation around ${\rm 3.9 \times 10^{6}\,\msun}$.
 
\end{enumerate}

\begin{figure*}
\centering
\resizebox{1.\hsize}{!}{\rotatebox{0}{\includegraphics[trim= 0.cm 0cm 0cm 0cm, clip]{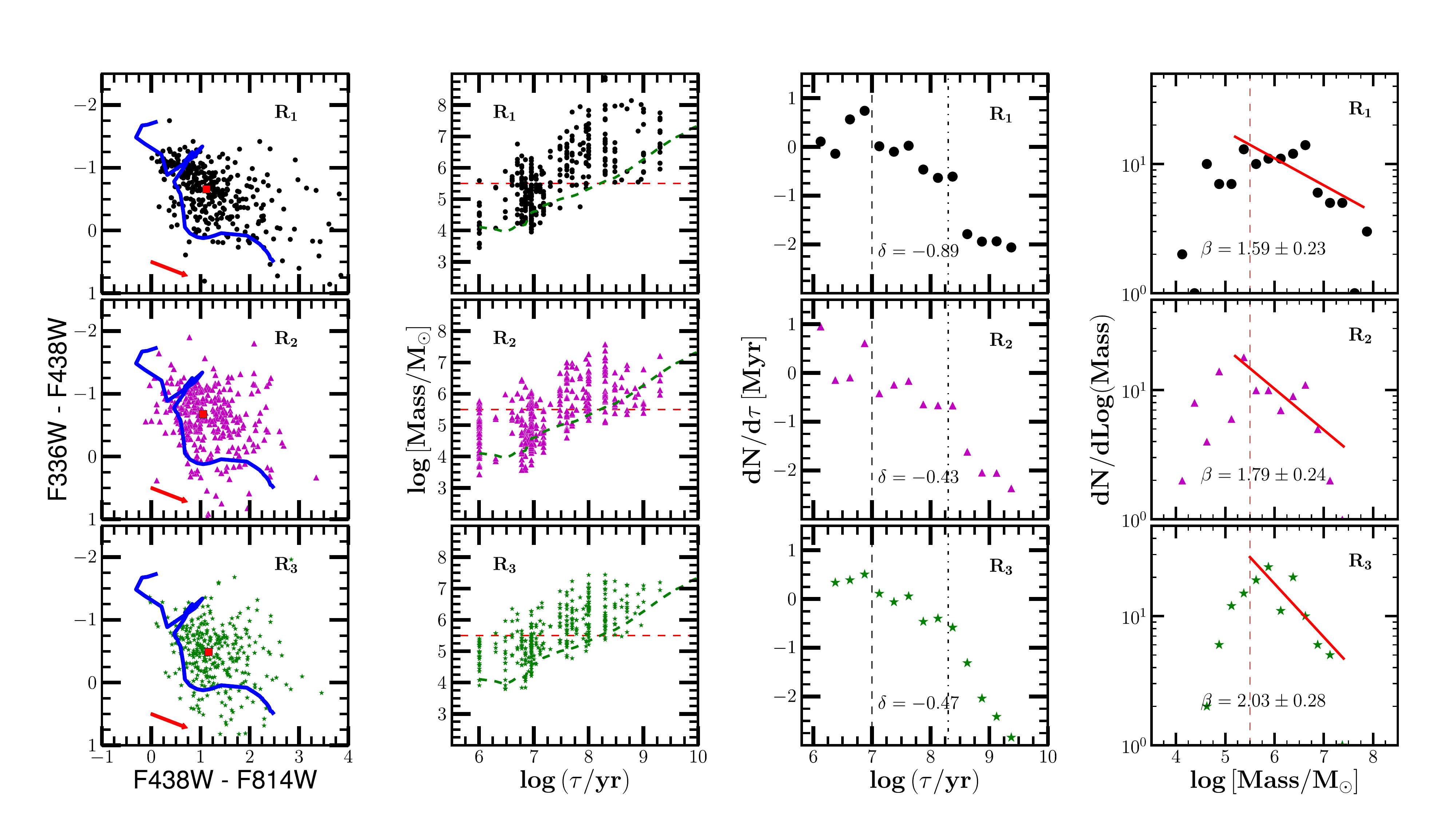}}}\\
\caption{\small Similar to Fig.\,\ref{fig:prop-reg}, \textcolor{black}{but using a fixed mass 
cutoff of ${\rm 10^{5.5}\,M_{\odot}}$ and} splitting the
SSC population of NGC\,3690E in 
three groups (with equal number) as a function of the cluster galactocentric radius: from 0 to 1.38\,kpc (R$_1$, top panel - black points), 
1.38 - 2.25\,kpc (R$_2$, middle - purple triangles), 2.25\,kpc and beyond (R$_3$, bottom - green stars). 
A peak is observed around log($\tau) \approx$\,7.7 in all three age distributions. CMFs of the R$_1$ and R$_2$
regions clearly exhibit a truncation at high mass end.}
\label{fig:prop-rad}
\end{figure*}

The outer field hosts more YMCs than the nuclear regions: 1272 against 836 candidates. 
A size-of-sample effect is therefore a possibility to explain the discrepancies in the distributions.
However, the truncation at the high mass end and the relatively shallower slope of the inner field CMF can also be 
an evidence for environmentally-dependent cluster disruption. This could confirm the previously suggested scenario
in the field of e.g.\,M83 \citep{2011MNRAS.417L...6B, 2012MNRAS.419.2606B} or NGC\,4041 \citep{2013AJ....145..137K}.
Nevertheless, results from other works
emphasize a quasi-universal disruption mechanism  independent of the galaxy host environment \citep[e.g.][]{2016ApJ...826...32M}.

\begin{table*}
\begin{scriptsize}
\caption[The CMFs]{Properties of the star cluster population of NGC\,3690E in three separate radial bins with an equal source number.}
\label{tab:ic-radial}
\begin{center}
\begin{tabular}{cccccccccc}
\hline 
  \noalign{\smallskip}

Region name & Annulus & Nb.SSCs & \textcolor{black}{Nb.SSCs} & Nb.SSCs  & Nb.SSCs &  Nb.SSCs & ${\rm M^{i^{th}}_{max}}$ & $\delta$ &  $\beta$ 	\\
        &   (kpc) & (${\rm SNR} > 3$)  & \textcolor{black}{($\tau < 10$\,Myr)} & ($\tau < 30$\,Myr)    & ($\tau > 200$\,Myr)  & (${\rm M > 10^7\,M_{\odot}}$) &log(${\rm M_{max}/M_{\odot}}$)&  &    \\
(1)	        &   (2)	   &	(3)	       & \textcolor{black}{(4)} & (5)	& (6) &  (7) & (8) & (9) & (10)	\\
  \noalign{\smallskip}
\hline \hline
  \noalign{\smallskip}
R$_1$ & $0 - 1.38$   & 329 & 45\,\% (147) & 58\,\% (191) & 19\,\% (64)  & 4\,\% (14) &8.88, 8.14, 7.99 &  $-0.89$ &  1.59$\pm0.23$ \\
R$_2$ & $1.38 - 2.25$& 331 & 50\,\% (164) & 59\,\% (194) & 18\,\% (60)  & 1\,\% (3)  &7.58, 7.27, 7.17 &  $-0.43$ &  1.79$\pm0.24$  \\
R$_3$ & $> 2.25$     & 328 & 42\,\% (137) & 50\,\% (163) & 18\,\% (60)  & 2\,\% (6)    &7.44, 7.24, 7.23 &  $-0.47$ &  2.03$\pm0.28$   \\
\noalign{\smallskip}
\hline\\
\noalign{\smallskip}
\multicolumn{10}{@{} p{18cm} @{}}{\footnotesize{\textbf{Notes. }
Columns 1 \& 2:  name and size of the radial bin;
Column 3: number of SSCs hosted by the region within the radial bin;
Columns 4 \& 5: number of SSCs \textcolor{black}{younger than 10\,Myr} and 30\,Myr;
Column 6: number of SSCs older than 200\,Myr;
Column 7: number of  SSCs more massive than ${\rm 10^{7}\,M_{\odot}}$; 
Column 8: masses of the first, the third and the fifth most massive star clusters; 
Columns 9 \& 10: power-law slopes  of the age distribution and the CMF.}}

\end{tabular}
\end{center}
\end{scriptsize}
\end{table*}

\subsubsection{Radial binning of the clusters hosted by NGC\,3690E}\label{subsec:var-rad}

Three radial bins were defined in the field of NGC\,3690E to split its SSC catalogue 
into three equal numbers (see Section\,\ref{select-arp}). The same distributions
as shown in Fig.\,\ref{fig:prop-reg} were also
plotted for each subpopulation in Fig.\,\ref{fig:prop-rad}:  SSCs within the innermost
bin R$_1$ labelled as black points (top panel), those hosted by the middle annulus R$_2$ as purple triangles (middle),
and the ones located in the outermost bin R$_3$ as green stars (bottom). 
Table\,\ref{tab:ic-radial} summarizes the star cluster properties of each radial bin derived from the different distributions.
\textcolor{black}{There is no obvious difference in the color-color diagrams and each radial
bin hosts approximately the same amount of old clusters ($\sim$\,18\,percent for $\tau > 200$\,Myr). A peak is
also  observed around log($\tau) \approx$\,7.7 in all three age distributions. However, we record a variation as a function of galactocentric distance
in the properties of the massive cluster candidates, the cluster age distributions and the CMFs,}
especially by comparing the SSC physical paramaters of R$_1$ to 
those of R$_2$ or R$_3$: {\it i)}  a relatively high percentage of massive star clusters in the innermost regions compared to those of
the other annuli ($2 - 4$ times higher); {\it ii)} masses of the first, the third, and the fifth most massive SSC candidates generally
decrease with an increasing distance as listed
in the 8th column of Table\,\ref{tab:ic-radial}. \textcolor{black}{\citet{2013MNRAS.435.2604P} and \citet{2016ApJ...816....9S}
also reported a similar trend in their work. Because the masses of the most massive star clusters are less likely to endure strong
stochastic IMF sampling effects, interpreting such a correlation helps in establishing the properties of the CIMF}; {\it iii)} the power-law index of the ${\rm dN/d\tau}$ $10 - 200$\,Myr age
range declines with an increasing radius (9th column, a fixed lower mass limit of ${\rm 10^{5.5}M_{\odot}}$ was applied to all 
three functions); {\it iv)} the power-law slope $\beta$ flattens as we approach the galaxy centre (from  2.03$\pm0.28$ going down
to 1.79$\pm0.24$ and then 1.59$\pm0.23$) 
\textcolor{black}{and a truncation at high masses is observed in the CMFs of the two inner annuli}. 
Note, however, that a power-law function does not seem to best represent the
CMFs of R$_1$ and R$_2$. One should thus be careful when interpreting such results. Nevertheless, given that there are
equal SSC numbers in each bin, the above results support the theory of an environmentally-dependent cluster formation 
and disruption as already suggested earlier in this work and the literature 
\citep[e.g.][]{2014ApJ...786..117F, 2015MNRAS.452..246A, 2016MNRAS.460.2087H}.

\subsubsection{The cluster mass-galactocentric radius relation}\label{sec:mass-Rc}

\begin{figure}
\centering
\resizebox{1.\hsize}{!}{\rotatebox{0}{\includegraphics[trim= 0.cm 0cm 0cm 0cm, clip]{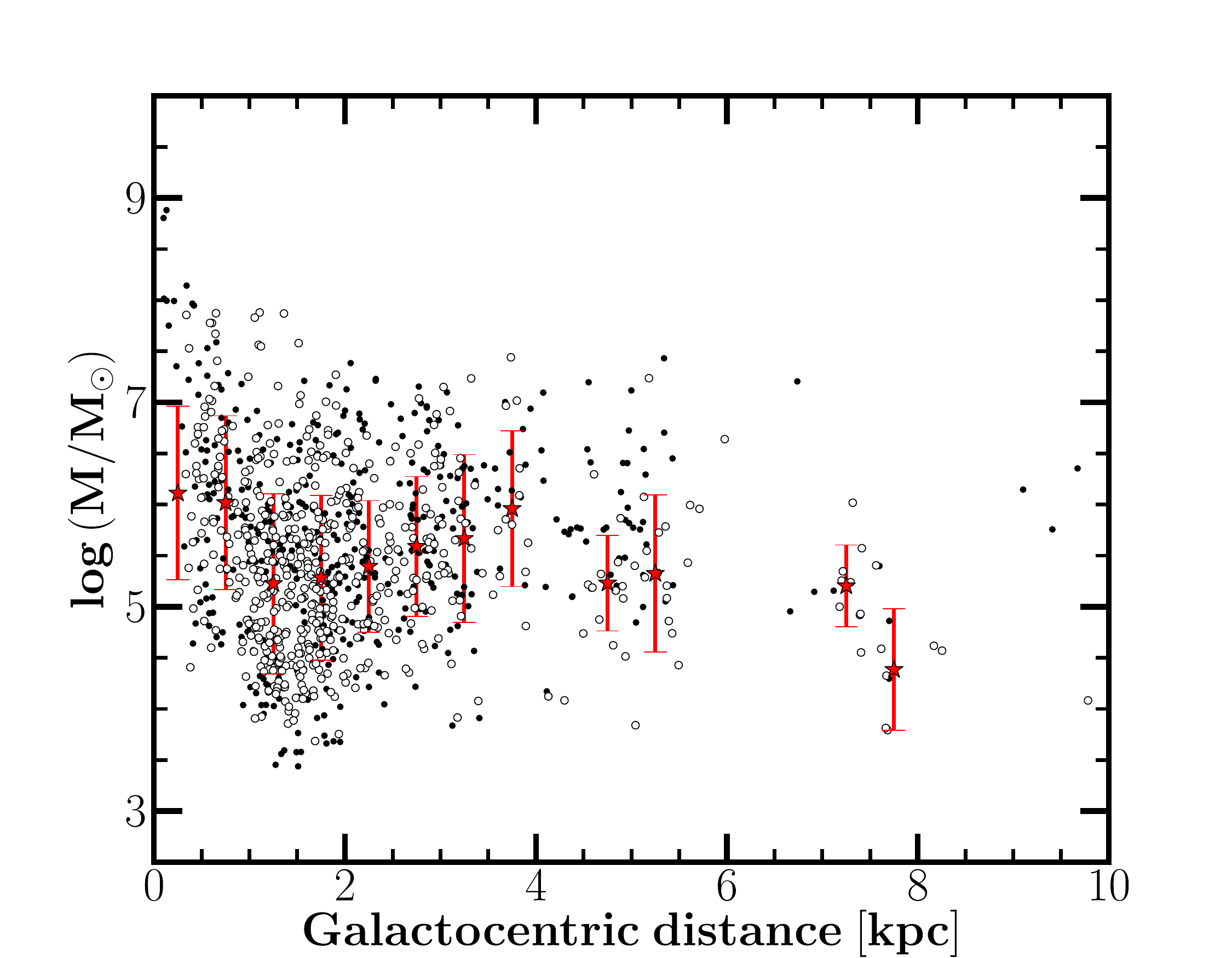}}}\\
\caption{\small The star cluster mass-galactocentric radius relation of NGC\,3690E shown as black points  and where open circles
represent high-confidence photometric cluster candidates. The red points plotted with their $1\sigma$ uncertainties
are the mean cluster mass values at each radius step of 0.5\,kpc.}
\label{fig:mass-Rc}
\end{figure}

Because of its symmetrical morphological feature, the star cluster population of NGC\,3690E was also used to check the 
tightness of the cluster mass-galactocentric radius relation. {\it Does it only apply to the first, the third, and the fifth most 
massive clusters of each radial bin?} Figure\,\ref{fig:mass-Rc} shows the star cluster mass plotted against its galactocentric
distance. The cluster mass decreases with an increasing galactocentric radius up to 
$\approx 2$\,kpc away from the galaxy nucleus, but after which it slowly increases until $\approx 4$\,kpc. 
While the high gas density in the nuclear regions of NGC\,3690E likely explains the behaviour within 2\,kpc, the steady increase beyond this
point arises because of the extreme environments provided by the ongoing merging of the eastern component with its companion NGC\,3690W.
In fact, the merging point of the system is estimated to be located $\approx 2$\,kpc away from the nucleus of NGC\,3690E.
The star cluster mass-galactocentric radius relation thus suggests that the formation of star clusters in \arp, especially in
the eastern galaxy, is not only
governed by stochastic sampling. Despite the possibility of any size-of-sample effect, the YMC birthsites should have played an important
role  \citep{2015MNRAS.452..246A, 2017ApJ...834...69L}. This is consistent with the results from our radial binning analyses. 
Note, however, that other studies such as \citet{2016ApJ...816....9S} rule out any environment-dependence scenario and favor
a stochastic cluster formation process. 

\begin{figure*}
\centering
 \begin{tabular}{c}
 \resizebox{1.\hsize}{!}{\rotatebox{0}{\includegraphics{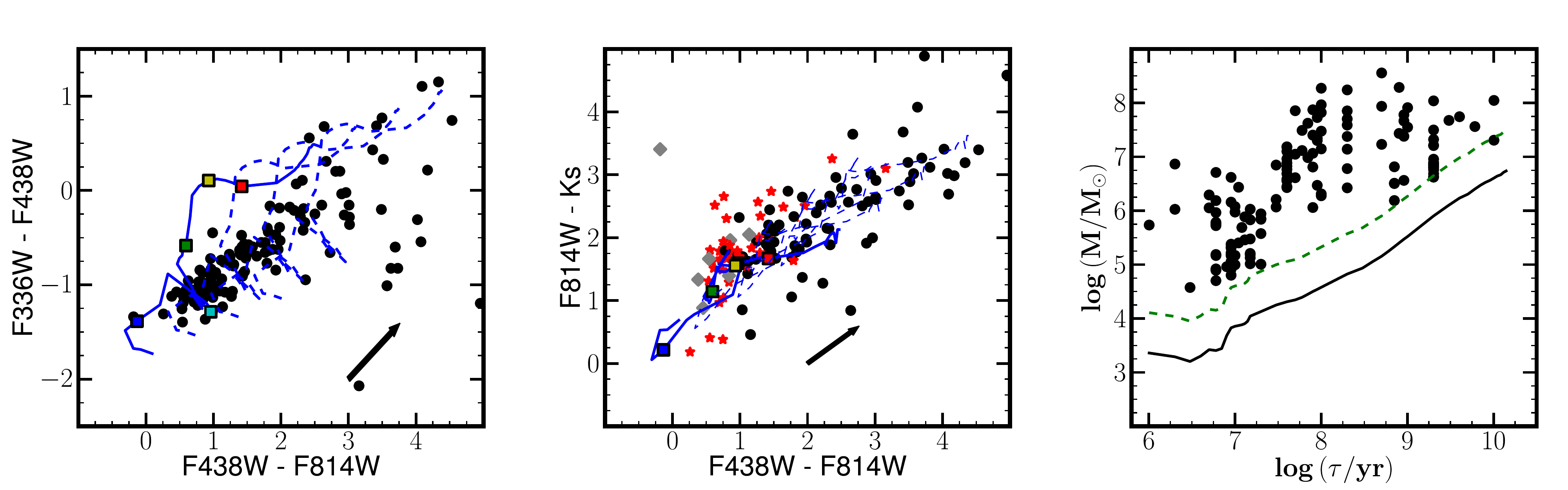}}}
 \end{tabular}
\caption{\small Photometric properties of the $K$-band selected SSC candidates. Left and 
middle panels show $U -B$ versus $B - I$ colors and $I - K$ versus $B - I$ colors, 
respectively. The blue solid and dashed lines respectively represent {\tt Yggdrasil} SSP models
for $A_V = 0$ and $A_V > 0$ and the arrows indicate 
a redenning of $E(B - V) = 0.25$. In the $BIK$ CCD, datapoints with $\tau\,>$\,30\,Myr are labelled as black circles while
those with $\tau\,<$\,30\,Myr are labelled as a function of the cluster 
masses: grey diamonds for ${\rm log(M/M_{\odot})\,\leq}$\,5, else red circles.
The right panel shows the resulting age-mass plane. Solid and dashed 
lines denote photometric detection limits for the outer and inner regions, respectively.}
\label{fig:ssc-excess}
\end{figure*} 

\subsection{The $K$-band selected SSC candidates}\label{subsec:K-selected}
It is important to have a separate analysis of the NIR sources since datasets of the ongoing SUNBIRD survey are mainly
based on $K$-band observations only \citep[e.g.][]{2013ApJ...775L..38R,2015PhDT.......214R}. The WFC3/UVIS images were thus resampled using
{\tt GEOTRAN}  and {\tt GEOMAP\,IRAF} tasks to match spatially with the NIR AO data pixel scale and orientation. Photometry of bright
non-saturated sources in the field
of both original and resampled HST images were derived and the measurements indicated that the values of the
visual magnitudes are similar within their photometric errors. Hence, we did not
apply any correction from the resampling and any offset slightly higher than 0.5\,pixels could simply be due
to relative extinction effects. Not all the $K$-selected sources were detected in the optical images of the system. 
Those located in the dust-obscured regions of the galaxy ($\sim$\,25\,percent of the NIR sources) were 
apparently being missed while running {\tt SExtractor} in Section\,\ref{detect-sec}. 
Figure\,\ref{fig:ssc-excess} overplots the {\tt Yggdrasil} SSP models with the $K$-band datapoints
in $U - B$ vs.\,$B - I$ (left panel) and $I - K$ vs.\,$B - I$ (middle)
color-color diagrams. Both distributions suggest that the NIR star clusters  
occupy a wide range of extinction and age.
Their positions in the $BIK$-band CCD generally follow the predicted colors of the SSP
models. In other words, most of the $K$-selected star cluster candidates of \arp~do not show any NIR excess. 

We also combined the SSC $K$-band magnitudes with the optical data 
to run two new sets of the $\chi^2$ minimization (using $UBIK$- and $BIK$-filters) other than the adopted $UBI$-band SED fitting.
In all three sets, we find that the star cluster masses 
generally lie in the range of ${\rm 10^{3} - 10^8 \msun}$. 
Although some of the datapoints with a derived mass 
${\rm M \approx 10^8\,{\rm \msun}}$ could be star cluster complexes or sources with strong NIR excess, 
the interacting system has also the potential to host very massive star clusters due to its extreme 
environment as already discussed in Section\,\ref{sec:massive-SSC}. Apart from an overestimate of the star cluster mass, outputs from $BIK$-fit 
converge to an older age range compared with the resulting cluster ages of the other fits. This is expected because
of the influence from the age-extinction degeneracy once excluding the UV-data. On the other hand, $UBI$-fit performs
better (with the smallest reduced $\chi^2$ values) than the two other sets. Such results confirm the reliability of using 
$UBI$-fit outputs to investigate the star cluster properties of \arp~in this work. 
The resulting age-mass plane of the $K$-band selected clusters (from $UBI$-fit) is presented in the right panel of Fig.\,\ref{fig:ssc-excess}. 
The lower mass limits, in the order of ${\rm 10^{4.5}\,\msun}$, are higher than the ones of the 
optically-selected clusters. Such a difference cannot be associated with an overestimate in the NIR cluster mass
since the $K$-band filter was already excluded during the SED modelling. 

\begin{figure}
\centering
 \begin{tabular}{c}
\resizebox{.87\hsize}{!}{\rotatebox{0}{\includegraphics[trim= 0.cm 0cm 0.cm 0cm, clip]{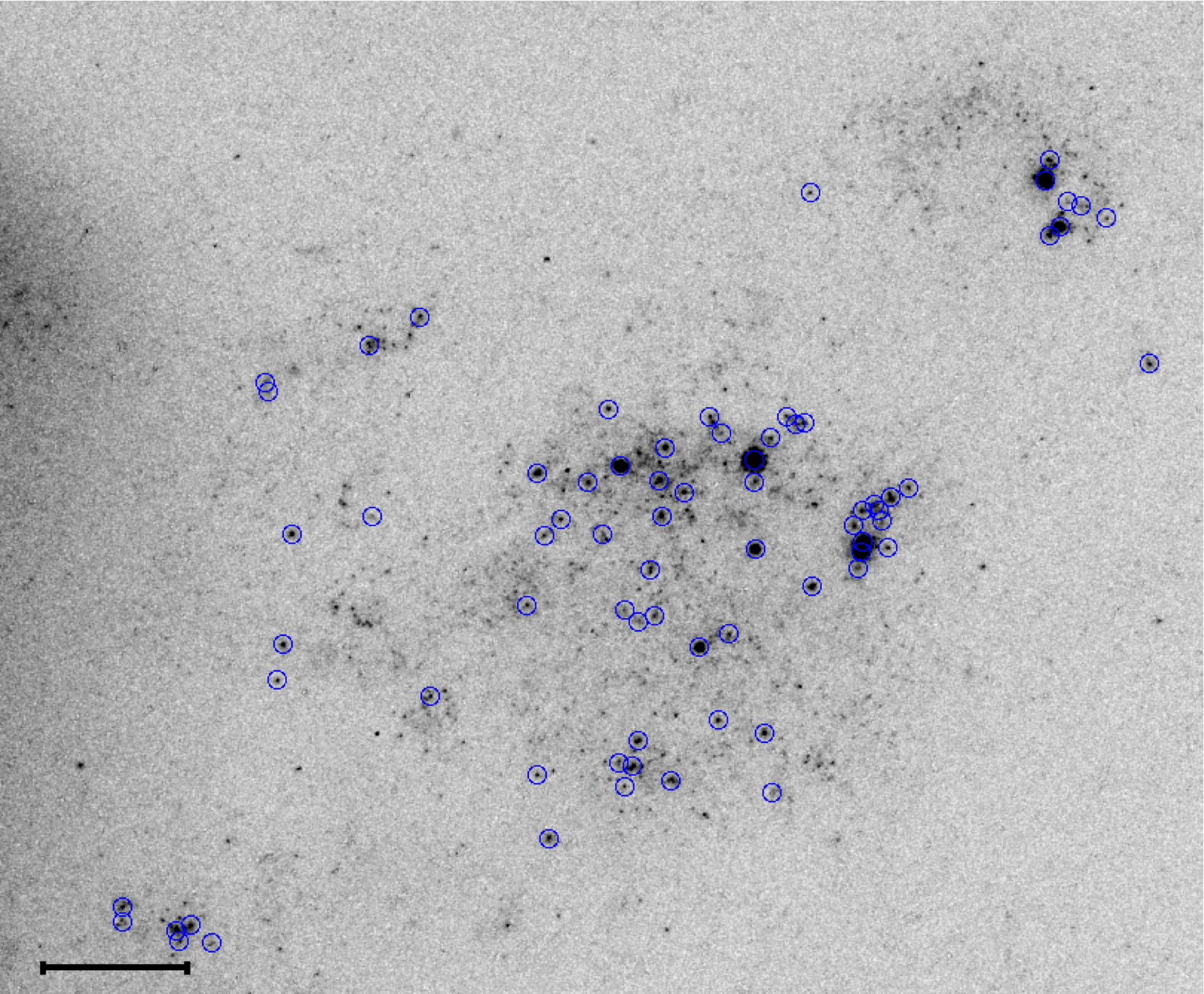}}}\\
\resizebox{1.\hsize}{!}{\rotatebox{0}{\includegraphics[trim= 1cm 0cm 1cm 0cm, clip]{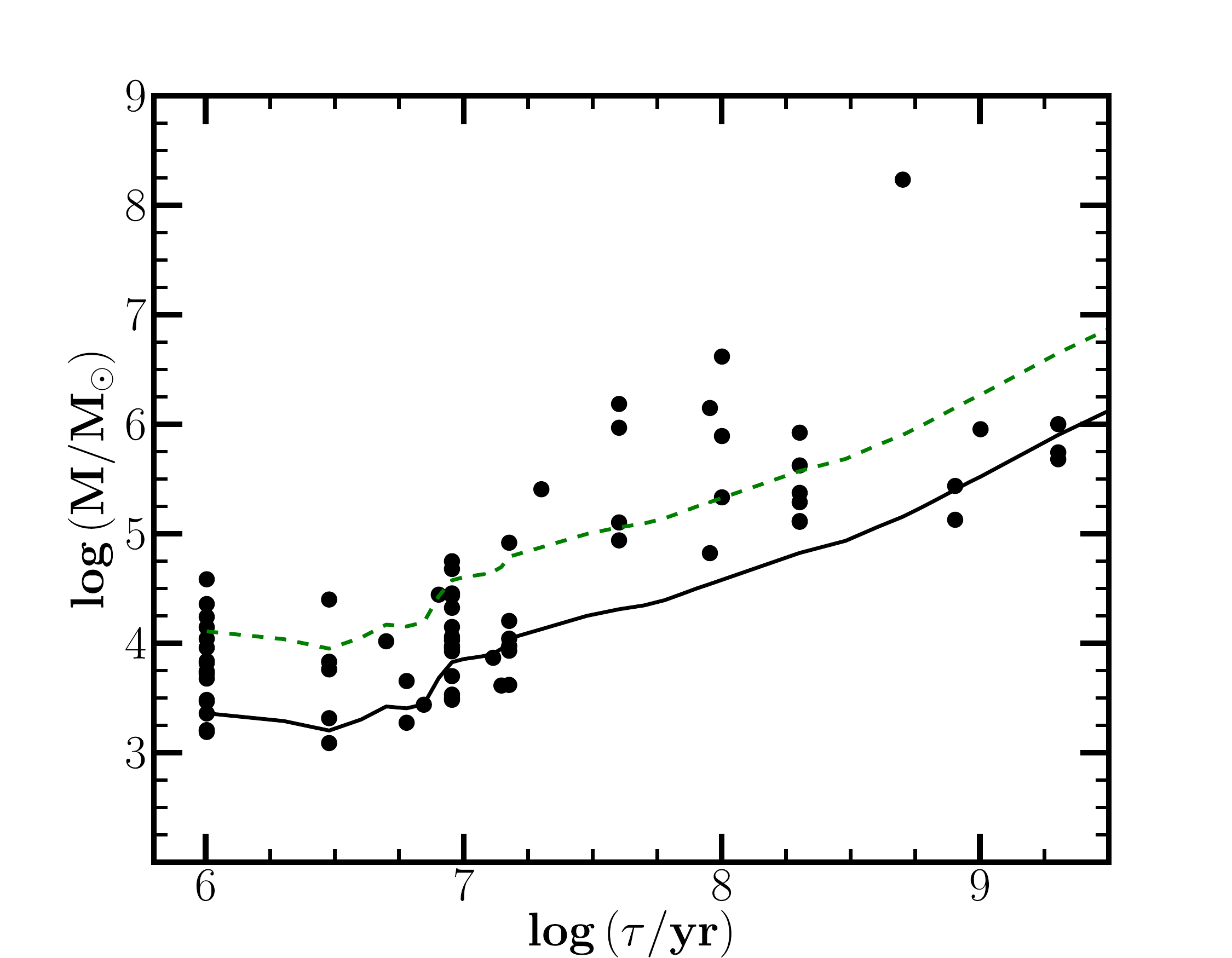}}}
\end{tabular}
\caption{\small {\it Top:} A snapshot ($7 \times 6$\,kpc) from the $I$-band image of the northwest star-forming region. Open circles represent
the selected YMC candidates in the field. The horizontal line marks a 1\,kpc scale.
{\it Bottom:} Distribution of the clusters in the age-mass plane.
The solid and dashed lines respectively denote the 80\,percent photometric detection limits for the outer
and inner regions as defined in Fig.\,\ref{fig:comp-frac}.}
\label{fig:NWreg-arp}
\end{figure}

The $BIK$ color-color plot has a few outliers that unexpectedly strech to redder colors. 
Such a dispersion cannot be associated to high photometric uncertainties as we have already excluded sources with $\sigma > $\,0.35\,mag. 
These outliers appear to show a NIR excess in their fluxes. 
Apart from stochastic sampling
effects and/or resolution effects, the inclusion of the surrounding nebular emission in the photometric
measurements could be another reason
behind such an excess \citep[e.g.][]{2010MNRAS.407..870A,2013MNRAS.431.2917D,2014MNRAS.444.3829B}. 
If there is stochasticity, massive SSCs are
expected to follow the distribution of the SSP models in the CCD, whereas low mass star clusters (${\rm M < 10^{5}\,\msun}$)
are distributed over a wide range of the $I - K$ color because of the dominant NIR flux of RSGs in their stellar population.
This is not, however, the case when looking at the mass distribution of young clusters ($\tau \lesssim 30$\,Myr)
of the current dataset: no record of any significant scatter in the distribution
of the low mass star clusters compared to that of the massive ones.
Stochasticity is thus believed to only partly affect the data and other causes
of the excess such as blending effects or nebular emission should also be considered.
In fact, \citet{2014MNRAS.444.3829B} recently reported that the intensity of the NIR excess
strongly depends on the spatial resolution of the NIR data. 
Since we have applied a smaller aperture radius of $\sim$\,0.1\,arcsec to derive the star cluster magnitudes and have already
shown that blending effects are insignificant at distances below 100\,Mpc
\citep{2013MNRAS.431..554R}, we conclude that
NIR excess of the young clusters also arises due to the emission included inside the aperture measurements.

\subsection{YMCs in the northwest star-forming regions}\label{sec:SSCs-NWreg}
As can be seen in Fig.\,\ref{arp299-rgb}, a star-forming region $\sim$\,10\,kpc northwest of the
interacting system was also imaged by the HST/WFC3 camera. The region could either be a low-metallicity blue 
dwarf companion or a metal-rich tidal tail forming stars on its own after gas has
been thrown out from the advanced merging galaxies. A snapshot of the region from the $I$-band image
clearly reveals that young massive clusters are present in the field (see top panel of Fig.\,\ref{fig:NWreg-arp}). Object detection and
selection followed the same procedures
as those of \arp~and resulted in the extraction of 74 YMC candidates (labelled as open circles). Note, however, that
visual inspection of the HST images suggests that there could be more YMCs than the recorded number in this region. 

YMCs of the northwest star-forming region are mainly very young, relatively less massive,
and with low extinction compared to the star cluster population hosted by \arp.
The bottom panel of Fig.\,\ref{fig:NWreg-arp} presents the YMC age-mass diagram:
58\,percent of the clusters have ages $\tau < 10$\,Myr with 23 candidates
younger than 5\,Myr. More than 70\,percent of the clusters 
have masses ${\rm M < 10^5\,M_{\odot}}$ and only one candidate that could also be identified as a supergiant star 
(an outlier in the CMD/CCD) is more massive than ${\rm 10^8\,M_{\odot}}$.
In terms of visual extinction, the median value is $A_V = 0.44$\,mag and more than 75\,percent of the
star cluster population have $A_V < 0.8$\,mag. We did not see any trend in the spatial distribution
plotted as a function of the cluster age and extinction (see bottom-right panels of 
Fig.\,\ref{fig:age-spatial-distr} and Fig.\,\ref{fig:Ebv-spatial-distr}, respectively). However, there seems
to exist at least a weak pattern when
labelling the YMC candidates with respect to their masses (bottom-right panel of Fig.\,\ref{fig:mass-spatial-distr}).
Regions with a relatively low-luminosity background are the ones to host the least massive star clusters 
(grey stars, ${\rm M \leq 10^{3.8}\,M_{\odot}}$).
{\it Could it be related with the distribution of the gas reservoir in the field?} Kinematic studies
are required to trace the (star) cluster formation history of the region. These are relevant
to check whether the extreme environments of \arp~affect the evolution of the neighbouring young star clusters in any way
(e.g. could there be any gas outflow from the interaction?).

\section{Summary and conclusions}\label{end-sec}

We performed a photometric study of young massive star clusters in the LIRG system of \arp:
an ongoing merger composed of NGC\,3690E\,(east) and NGC\,3690W\,(west). The work mainly used 
archival datasets from the HST WFC3/UVIS camera to understand the formation, evolution and disruption mechanisms
of the star clusters. The final $UBI$-band catalogue contains 2182 sources common to all three filters with magnitude
errors below $\sigma_m = 0.35$\,mag. This includes 1323 high confidence photometric sources and 74 SSCs hosted by a star-forming region 
$\sim$\,10\,kpc to the northwest of the interacting system. 
{\tt Yggdrasil} SSP models were used as a reference in deriving the age, mass and extinction
of the cluster population. Each individual cluster had its own
range of extinction that was constrained from an $U - I$-based extinction map while performing the SED fitting. The use of 
$U$-band data helped to optimally reduce the age-extinction degeneracy. The major findings 
of this work are outlined as follows:

\begin{enumerate}

\item \arp~hosts a very young star cluster population with 62\,percent of 
them having an age less than 15\,Myr old. They are massive with a lower
mass limit of $\approx$\,$10^4\,{\rm \msun}$ for a subsample that is complete up to 10\,Myr. 
Their visual extinctions $A_V$ range between 0 and 4.5\,mag. 
Apart from hosting more luminous star clusters, a significant number of the YMC population of NGC\,3690W
have a low extinction value and ages younger than 10\,Myr compared to the candidates hosted by NGC\,3690E. 

\item The star clusters are generally randomly distributed in terms of their ages. Nevertheless, the disk overlap region and the southern spiral
arm of NGC\,3690E mainly host cluster candidates younger than 50\,Myr old. There are supergiant HII regions located in those star-forming regions.  On the 
other hand, the very massive clusters ${\rm M > 10^7\,\msun}$ are spatially associated with the galaxy nuclear regions. Such a distribution could be an imprint of the displacement of large quantities of dense molecular gas toward the nuclear regions as the two gas-rich
galaxies collide.

\item The derived cluster age distribution suggests that the cluster formation rate has increased
over the past 10\,Myr and will still be increasing in the future. This is a highly probable scenario
given the ubiquity of extreme starburst activities induced by the violent merging of the two galaxy
components of \arp. The fitted age distribution of a mass-limited sample over the $10 - 200$\,Myr time
interval results in a power-law slope $\delta$ equals to $-0.67$. Such a value becomes $-0.59$ and $-0.84$ in
the case of NGC\,3690E and NGC\,3690W, respectively. The difference in the slope suggests that
YMCs of NGC\,3690W undergo stronger disruption than those located in the eastern component. 
 Finally, a bump observed around 50\,Myr could possibly trace an intense starburst activity which is already 
 reported by other studies.

\item Fitting a single power-law distribution to the $UBI$-band CLFs of \arp~results in slopes 
ranging between $\alpha = 1.61 - 1.81$. Unlike normal spiral galaxies, intensely star-forming 
galaxies such as LIRGs are found to be associated with shallower slopes. The values of $\alpha$
get steeper with redder filter because of cluster evolutionary fading. With indices shallower 
by $\approx\,0.3$ dex than those of NGC\,3690E, the CLFs of NGC\,3690W also display clear truncations at high 
luminosities and a flat distribution in the faint end magnitudes. Such a trend is likely a signature
of stronger dissolution of the star clusters hosted by this component compared to the disruption 
occuring in NGC\,3690E. 

\item The final catalogue was split in three age bins to better analyse the cluster mass distribution. Apart from fitting
a power-law function to the CMFs, a Schechter fit was also applied to the $10 - 200$\,Myr mass-limited data. 
The value of the power-law slope $\beta$ decreases with an increasing age bin (from 2.61 down to 1.68). The critical mass until which
we performed the power-law fit shifts toward higher masses. The mass functions at younger ages exhibit an underlying
truncation at the high mass end. The Schechter characteristic mass \textcolor{black}{${\rm M_{\star} = 1.6 \times 10^6\,\msun}$} is consistent with the
values reported in the literature in the case of starbursts and merger galaxies. The eastern component has a more massive
\textcolor{black}{${\rm M_{\star} = 2.1 \times 10^6\,\msun}$} than its companion which is consistent with that galaxy hosting more than half of the very 
massive YMCs in \arp. Although statistical effects and/or incompleteness may contribute to these trends, they
are not enough to fully explain the CMF peculiar behaviours. Both environmental and mass-dependence of the cluster mode
of disruption should also be considered. 

\item Properties of the YMCs as a function of the varying background also provide evidence for environmentally-dependent 
cluster disruption. We found that the CMF associated with the nuclear regions has a relatively shallower slope and a distinct
truncation at the high mass end. Furthemore, radial binning in equal number of the star clusters hosted by NGC\,3690E strengthens the
idea of some physical effects (partly) constraining the cluster formation and disruption. In particular, we found that, within
$\sim$\,2\,kpc radius of this galaxy nucleus, the cluster mass decreases with an increasing galactocentric radius.

\item We considered the time interval of $10 - 50$\,Myr to derive the cluster formation rate and the cluster formation efficiency
$\Gamma \sim 19$\,percent. Such a value is underestimated because of incompleteness in low mass star clusters. Nevertheless, the 
extreme environments of \arp~enable the system to have $\sim 3 - 5$\,times more star formation occuring in bound clusters than in 
the case of gas-poor spiral galaxies. With a SFR density of \textcolor{black}{$0.11\,{\rm M_{\odot}\,yr^{-1}\,kpc^{-2}}$},
\arp~generally follows the 
${\rm CFE - \Sigma_{SFR}}$ relation established from observational and theoretical predictions. 

\item YMCs of the $K$-band catalogue are more massive than the optically-selected ones.
They do not have specific age and mass ranges and they are uniformly distributed in the cluster age-mass plane.
Only a few cases present NIR excess in their fluxes. 
Besides stochasticity, contamination from nebular emission encircled within the aperture 
photometry could also explain the excess.

\item The star-forming region northwest of the interacting LIRG mainly hosts young and relatively less massive YMCs 
(${\rm \tau < 50\,Myr,\,M < 10^{5}\,\msun}$) of low visual extinction (with a median of ${\rm A_V = 0.44\,mag}$). The lack
of spectroscopic data, however, limited our investigation on the effects of the ongoing merging toward the cluster 
formation history of the region.

\end{enumerate}

In summary, the star clusters of NGC\,3690W are more
vulnerable to disruption while those hosted by the eastern component stand
a higher chance to have masses ${\rm M > 10^{7}\,\msun}$. These results strongly indicate that
the extreme environments of \arp~influence both formation and evolutionary processes of the star cluster population
while a mass-dependent scenario is also a possibility. An upcoming paper will use the binary population synthesis model
{\tt BPASS} \citep{2008MNRAS.384.1109E, 2017PASA...34...58E}
to estimate the ages and masses of the young massive star clusters. The objectives of such analyses are: {\it i)} to evaluate
the effects of the inclusion of binary stars into the chosen SSP models; {\tt ii)} and to predict the total X-ray flux from 
high-mass X-ray binaries (HMXBs) after combining the derived star cluster properties with Chandra observations of Arp\,299
\citep{2016MNRAS.460.3570A}. HMXBs are believed to be progenitors of CCSNe.

\section*{Acknowledgements}
We thank the referee for helpful comments and suggestions to improve this article.
ZR acknowledges financial support from the South African Radio Astronomical Observatory,
which is a facility of the National Research Foundation (NRF). PV is also grateful to the NRF.
Based on observations
made with the NASA/ESA Hubble Space Telescope, and obtained from 
the Hubble Legacy Archive, which is a collaboration between the Space Telescope Science Institute (STScI/NASA), 
the Space Telescope European Coordinating Facility (ST-ECF/ESA) and the Canadian Astronomy Data Centre (CADC/NRC/CSA).
Based in part on observations obtained at the Gemini Observatory, which is operated by the Association of Universities
for Research in Astronomy, Inc., under a cooperative agreement with the NSF on behalf of the Gemini partnership: the National
Science Foundation (United States), the National Research Council (Canada), CONICYT (Chile), Ministerio de Ciencia, Tecnolog\'ia
e Innovaci\'on Productiva (Argentina), and Minist\'erio da Ci\^encia, Tecnologia e Inova\c c\~ao (Brazil).  

\bibliographystyle{apj}
\small


\appendix
\renewcommand\thefigure{\thesection.\arabic{figure}}
\section{Appendix}
\setcounter{figure}{0}

\begin{figure}
 \begin{tabular}{cc}
\resizebox{.55\hsize}{!}{\rotatebox{0}{\includegraphics{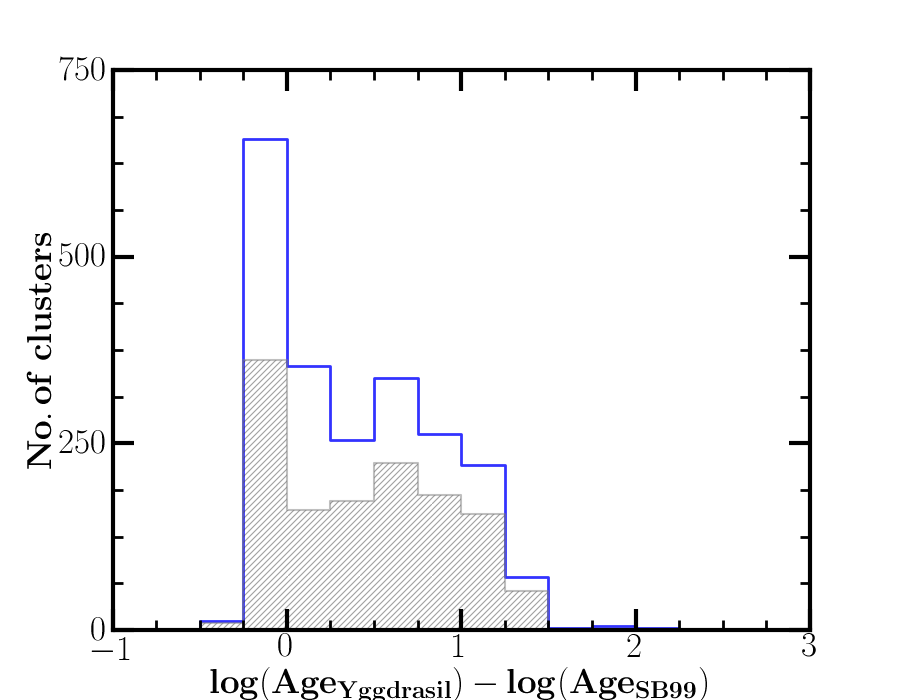}}} 
\resizebox{.55\hsize}{!}{\rotatebox{0}{\includegraphics{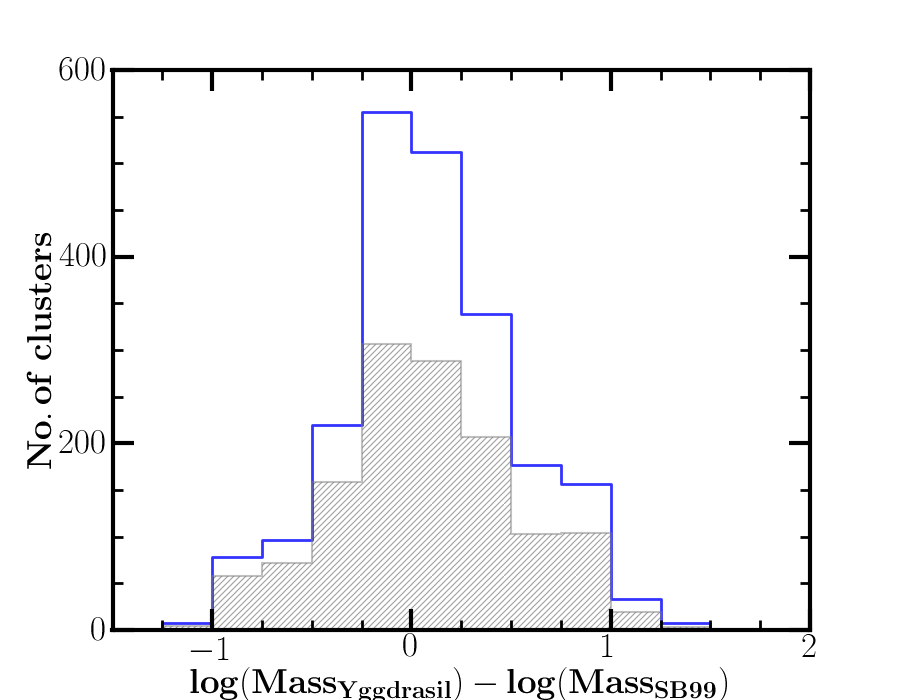}}}\\ 

\end{tabular}
\caption{Comparing the derived star cluster ages ({\it left}) and masses ({\it right}) using two different SSP models: {\tt Yggdrasil} versus
{\tt SB99} models. The solid lines consider all selected candidates while the hatched
distributions only include YMCs with $\sigma \leq$\,0.20\,mag.
}
\label{fig:comp-SB99}
\end{figure}

\begin{figure}
\centering
 \begin{tabular}{cc}
\resizebox{.55\hsize}{!}{\rotatebox{0}{\includegraphics{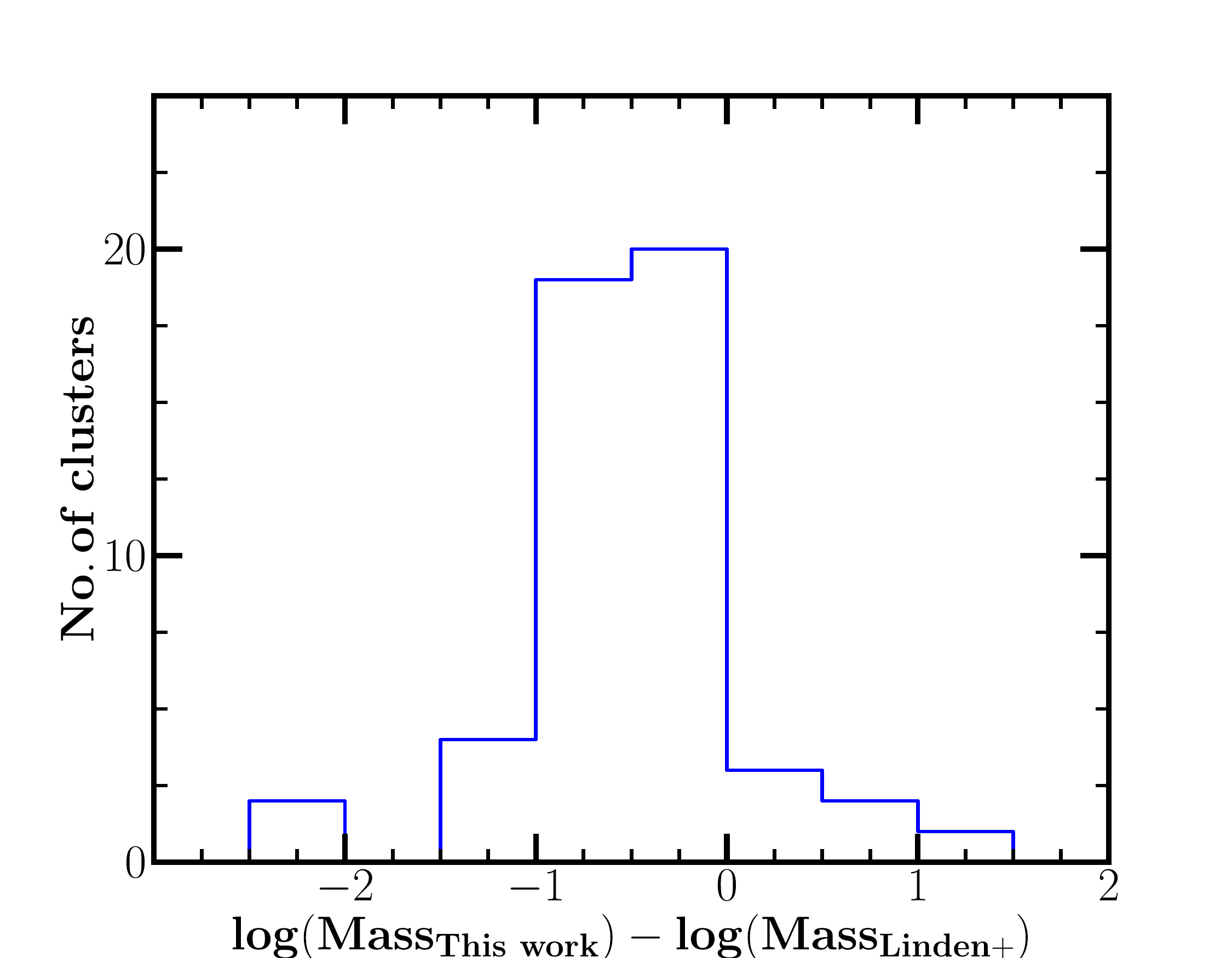}}} 
\resizebox{.55\hsize}{!}{\rotatebox{0}{\includegraphics{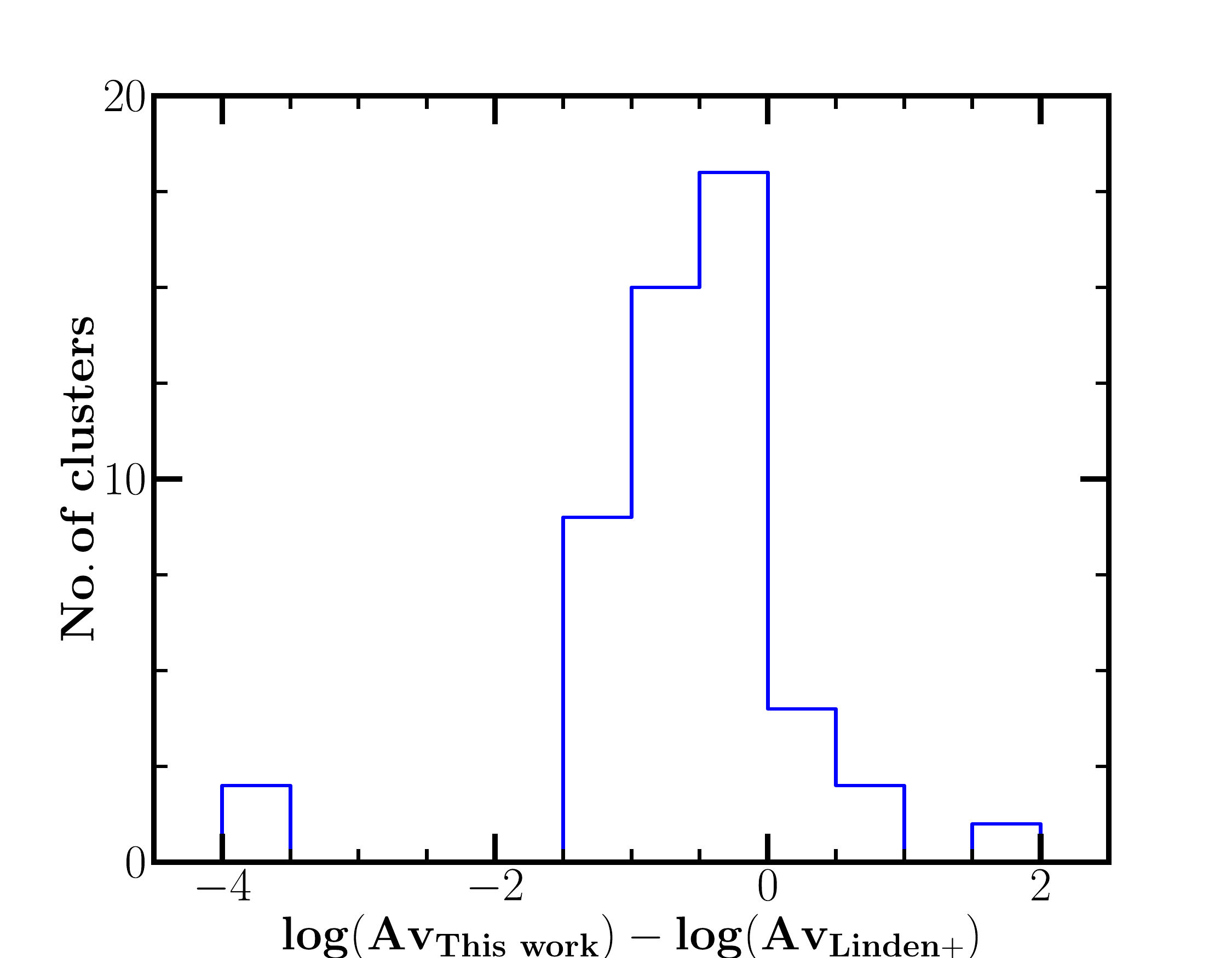}}}\\ 

\end{tabular}
\caption{Comparing our derived star cluster masses ({\it left}) and extinctions ({\it right}) with the ones
estimated by \citet{2017ApJ...843...91L}.
}
\label{fig:comp-Linden}
\end{figure}

\begin{figure}
\centering
\resizebox{1.\hsize}{!}{\rotatebox{0}{\includegraphics{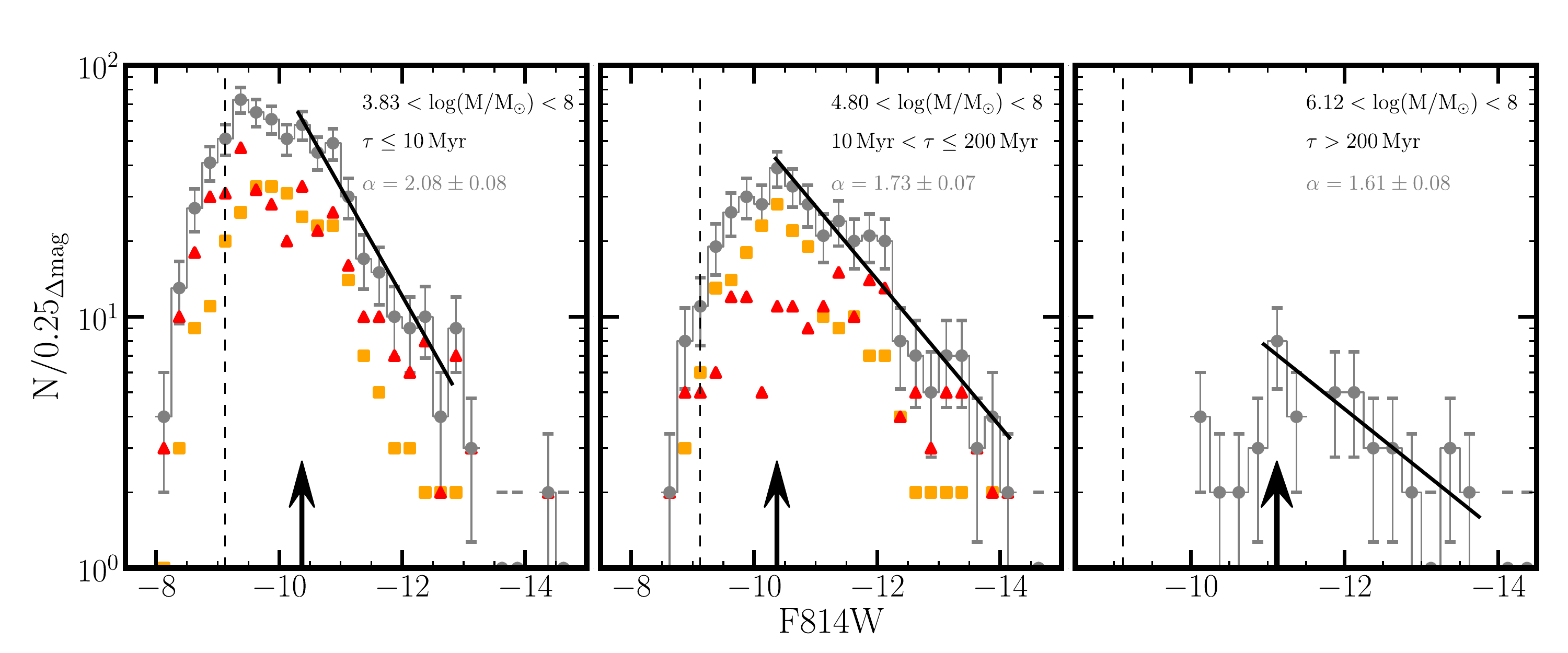}}}\\
\caption{\small $I$-band LFs of the optically-selected YMCs with $\sigma \leq 0.20$\,mag for three
different age bins. Same labels as in Fig.\,\ref{fig:ssc-mfs}. Different mass limits were set
at each age range to account for completeness.
The dashed lines at $M_I = - 9.12\,{\rm mag}$ mark $\sim$\,80\,percent photometric detection limits.
The solid lines represent the 
fit to a power-law distribution of \arp~LF bright end and the vertical arrows indicate
the critical magnitudes $M_{\rm cl,\,\lambda}$. The CLFs of the $10 - 200$\,Myr age range
present a prominent underlying truncation at the faint end of the distribution, 
especially in the case of NGC\,3690W.}
\label{fig:lfs-arp}
\end{figure}

\end{document}